\pdfoutput=1
\documentclass[aps,prb,reprint,superscriptaddress,longbibliography]{revtex4-2}

\usepackage{cellspace}
\usepackage{amsmath}
\usepackage{amsfonts}
\usepackage{amssymb}
\usepackage{graphicx}
\usepackage{svg}
\usepackage{bbding}
\usepackage{mathtools}
\usepackage{centernot} % not parallel, etc.
\usepackage{amsthm} % theorems
\usepackage{cmap}
\usepackage{xcolor}
\usepackage{bigints}
\usepackage{dsfont} %mathbb 1
\usepackage{esint} % beatiful integrals
\usepackage{physics}
\usepackage{natbib}
\usepackage{nicefrac} % nice fraction	
\usepackage{cancel}
\usepackage{bm}
\usepackage{upref}
\usepackage{multirow}
\usepackage{slashed}
\usepackage[compat=1.1.0]{tikz-feynman}
\tikzfeynmanset{
  warn luatex=false,
  every medium@@/.style={/tikz/node distance=1.5cm},
}
\usepackage{paralist}
\usepackage{varioref} % references to show page
\usepackage[]{hyperref} %pagebackref
\usepackage{url}
\usepackage[capitalise]{cleveref} 

\newcommand{\transp}{\top}

\usepackage{wrapfig}
\usepackage[protrusion=true,tracking=true,expansion,final]{microtype}      % microtypography %kerning=true,
\usepackage{xcolor} %colors
\usepackage{booktabs} % for professional tables
\usepackage{adjustbox}

\renewcommand{\cite}[1]{\citep{#1}}

\usepackage[T1]{fontenc}
\vbadness=\maxdimen
\definecolor{citecolor}{HTML}{2980b9}
\definecolor{mydarkblue}{rgb}{0,0.08,0.45}
\definecolor{urlcolor}{rgb}{0,.145,.698}
\definecolor{linkcolor}{rgb}{.71,0.21,0.01}

\hypersetup{ %
	pdftitle={Topological phase transitions between bosonic and fermionic quantum Hall states near even-denominator filling factors},
	pdfauthor={Evgenii Zheltonozhskii, Ady Stern, Netanel H. Lindner},
	pdfsubject={},
	pdfkeywords={},
	pdfborder=0 0 0,
	pdfpagemode=UseOutlines,
	bookmarksopen=true,
	breaklinks=true,  % so long urls are correctly broken across lines
	colorlinks=true,
	urlcolor=urlcolor,
	linkcolor=linkcolor,
	citecolor=citecolor,
	filecolor=mydarkblue,
	pdfview=FitH}

\DeclareFontFamily{OMX}{lmex}{}
\DeclareFontShape{OMX}{lmex}{m}{n}{<-> lmex10}{}

\begin{document}

%\preprint{APS/123-QED}

%opening
\title{Topological phase transitions between bosonic and fermionic quantum Hall states near even-denominator filling factors}% Force line breaks with \\

\author{Evgenii Zheltonozhskii}
\affiliation{\mbox{Physics Department, Technion, 320003 Haifa, Israel}}
 \email{evgeniizh@campus.technion.ac.il}

\author{Ady Stern}
\affiliation{\mbox{Department of Condensed Matter Physics, Weizmann Institute of Science, Rehovot 7610001, Israel}}

\author{Netanel H. Lindner}
\affiliation{\mbox{Physics Department, Technion, 320003 Haifa, Israel}}


\date{\today}% It is always \today, today,

\begin{abstract}
We study the quantum critical point between the fermionic $\nu=8$ quantum Hall state and the bosonic $\nu=2$ quantum Hall state of Cooper pairs. Our study is motivated by the composite fermion construction for the daughter states of even-denominator fractional quantum Hall states and the experimentally observed transition between the daughter and the Jain states at the same filling. We show that this transition is equivalent to the transition between a neutral invertible $E_8$ state and a topologically trivial state. These transitions can be described in a partonic framework as a cascade of mass changes of four neutral Dirac fermions coupled to multiple Abelian Chern--Simons $U(1)$ gauge fields. In the absence of fine-tuning, the transition is split into a series of at least eight distinct transitions, with at least seven distinct intermediate topologically ordered phases that host neutral anyons. 
\end{abstract}
\maketitle

\section{Introduction}
Quantum Hall states are typically characterized by their Landau-level filling factor $\nu$, or, equivalently, the quantized Hall conductance $\sigma_{\mathrm{xy}}=(e^2/h)\nu$. However, $\sigma_{\mathrm{xy}}$ alone is not enough to distinguish between different topological orders. Identifying these orders requires additional probes \cite{ma2022fractional,li2017denominator,zucker2016phpfaffian,sharma2023compositefermion,spnsltt2019topological,schiller2021extracting,stern2005proposed,bonderson2005detecting,feldman2006detecting,lee2022nonAbelian,yutushui2021identifying,hein2022thermal,manna2022classification,cooper2008observable,yang2009thermopower,haldane2021graviton,mross2017theory,wang2017topological,simon2020pfaph,hsin2020eftfqh,lotric2025reconstructionnu52,rosenow2010neutral,takei2011neutral,spnsltt2020conductance, park2020noise,das2022current,park2024fingerprints,vinjamuri2025shotnoise,spnsltt2026bosonization}, such as thermal Hall measurements, shot noise, or quasiparticle interferometry. 
For example, even-denominator fillings, such as $\nu=5/2$ \cite{willet1987nu52}, admit multiple Abelian and non-Abelian candidate topological orders \cite{mooreread1991nonabelions,read1999paired,nayak2007nonabelian,levin2007antipfaffian,son2015composite,hansson2016quantum,kane2017pairing,ma2022fractional}. 
One way to identify the specific topological order is to study the fractional states that occur near an even-denominator ``parent'' filling. These nearby states can be viewed as the result of the quasiparticles in the parent state forming a new quantum Hall liquid, generating a hierarchy of ``daughter'' states \cite{khveshchenko2006composite,bonderson2007fractional,levin2008collective,hermanns2009condensing}. The parent topological order uniquely determines the sequence of fillings in this hierarchy \cite{levin2008collective,yutushui2024paired,zheltonozhskii2024identifying,zhang2024hierarchy}. Such daughter states have been observed in GaAs \cite{kumar2010nonconventional,singh2023topological} and graphene \cite{zibrov2016robust,huang2021valley,assouline2023energy,hu2024studying,kumar2024quarter,chanda2025denominator}, providing constraints on the nature of the even-denominator states in these systems.

Conventional Jain states \cite{jain1989cf} constitute a competing topological order appearing at the same filling as the daughter states, e.g., $\nu=8/17$ and $\nu=7/13$. 
The quantized response that distinguishes between the daughter and the Jain states is the thermal Hall conductance $\kappa_{\mathrm{xy}}$~\cite{kane1997thermal}. This response measures the net chirality of gapless edge modes (corresponding to a topological invariant called the chiral central charge) and, for a gapped bulk in the low-temperature limit with thermally equilibrated edges, is quantized in units of $\kappa_0 = \frac{\pi^2 k_B^2 T}{3h}$.

By studying the longitudinal resistance  $R_{\mathrm{xx}}$ as a function of carrier density and an in‑plane magnetic field,
\citet{singh2023topological} reported evidence for two distinct phases with the same $\sigma_{\mathrm{xy}}$  at $\nu=8/17$ and $\nu=7/13$. In wide‑quantum‑well devices they used, the electron wavefunction can become effectively ``bilayer-like''; tuning these parameters reduces interlayer tunneling $\Delta_{\mathrm{SAS}}$, thereby favoring the daughter phase due to pairing of composite fermions, as explained below. Similar phenomena were observed in bilayer graphene \cite{kumar2024quarter}, where at high displacement fields, the single-electron wavefunction is localized at a single layer; as the displacement field decreases, the wavefunction spreads across both layers, and the daughter state becomes more favorable.
These observations raise a fundamental question: what is the nature of a phase transition between topologically distinct phases at fixed filling?

In this work, we address this question. We find that the Jain state can be represented as a tensor product of the daughter state and an additional neutral quantum Hall state: specifically, the Jain phase is equivalent to the daughter phase stacked with a neutral bosonic $E_8$ state. This allows us to analyze the transition based on the low-energy description of the neutral sector.

The bosonic $E_8$ state is a short‑range‑entangled \cite{chen2011symmetry,kapustin2014bosonic,kong2014braided,freed2014shortrange,wen2017colloquium,freed2021reflection,barkeshli2021classification,tantivasadakarn2022hierarchy} chiral phase with $\kappa_{\mathrm{xy}}=8$ (in units of $\kappa_0$). It supports no anyons (no intrinsic topological order). Since, in our settings, it is electrically neutral, its stacking with the daughter state leaves $\sigma_{\mathrm{xy}}$ and $\nu$ unchanged. Thus, the Jain--daughter transition maps to
\begin{equation*}
\text{trivial neutral sector} \;\longleftrightarrow\; E_8\ \text{neutral sector},
\end{equation*}
within a low-energy description in which the charged excitations remain gapped. We focus on $\nu=8/17$, but other cases, in which the bosons form $\nu=2m$ states with additional fermionic modes, can be analyzed similarly. Numerically accessing these phases \cite{furukawa2013integer,wu2013quantum} is challenging because the relevant fractions have large denominators and, in the composite‑fermion description, correspond to high effective integer fillings.

We formulate a parton critical theory in which the direct transition corresponds to the simultaneous change of the mass sign of eight neutral Dirac fermions coupled to multiple Abelian Chern--Simons $U(1)$ gauge fields. Physically, the mass terms correspond to the energy gaps of the emergent neutral excitations. Eight flavors arise naturally because the change in thermal conductance is $\Delta \kappa_{\mathrm{xy}}=8$. 

Our large-$N$ renormalization-group (RG) analysis indicates that mass perturbations destabilize this multicritical point, so reaching it requires fine-tuning. A generic path traverses a cascade of intermediate phases where the neutral sector becomes topologically ordered, changing $\kappa_{\mathrm{xy}}$ by increments of 1, following the principle of minimal allowed change in topological invariants~\cite{lee1993unified,ostrovsky2007theory,mross2017theory}. Consequently, between the trivial ($\kappa_{\mathrm{xy}}=0$) and $E_8$ ($\kappa_{\mathrm{xy}}=8$) limits, one expects intermediate gapped fractional quantum Hall (FQH) phases with quantized integer-valued thermal Hall conductance $\kappa_{\mathrm{xy}}$.
%\cref{fig:pd} shows a schematic phase diagram in the space of fermion masses. 
We prove that the minimal torus ground-state degeneracy of the neutral sector is $\min(\kappa_{\mathrm{xy}}+1, 9-\kappa_{\mathrm{xy}})$. While there are multiple critical points between the two limits that can be described within our approach, we focus on an explicit example that attains these values.

Thus, the experimentally inferred transition should, in general, split into several intermediate phases. While a direct $\kappa_{\mathrm{xy}}$ measurement is challenging, improved sample quality should resolve what at present appears to be a single transition \cite{singh2023topological,kumar2024quarter} into multiple transitions. Moreover, the intermediate phases host neutral anyons detectable via interferometry \cite{wei2023thermalinterferometry,han2024interferometry} or neutral-mode noise measurements \cite{bid2009shotnoise,takei2011neutral}. Finally, our findings suggest that the daughter states provide a possible platform for creating and studying the $E_8$ phase.

In the context of FQH, the quantum phase transitions~\cite{carr2010understanding,sachdev2011qpt} are topological, meaning there is no local order parameter. While plateau transitions between states at different fillings have been extensively explored both theoretically \cite{aoki1985localization,pruisken1988singlarities,kivelson1992globalphasediagram,huckestein1995scaling,pruisken2005theta,kumar2020interaction,sbierski2020criticality,pu2021anderson,andrews2023localization} and experimentally \cite{koch1991scaling,hohls2002scaling,dodooamoo2014nonuniversality,madathil2023delocalization,kaur2024qhtransition}, examples at constant filling remain limited. The prominent example is $\nu=2/3$ \cite{eisenstein1990transition,mcdonald1995topological,kronmueller1998huge,kraus2002logtitudinal,wu2012microscopic}, which, in addition to first-order spin transition \cite{mcdonald1995topological,smet2001morphology,verdene2007firstorder,geraedts2015competing}, is predicted to host non-Abelian states \cite{geraedts2015competing,peterson2015abelian,liu2015nonabelian}. Recently, evidence for a transition between Abelian and non-Abelian states also emerged at $\nu=1/2$~\cite{singh2025nonabeliantransition}. Additional candidate transitions were proposed, described by anyon condensation \cite{wen1999continuous,barkeshli2010anyon,burnell2017anyoncondensation} and effective Chern--Simons field theories \cite{barkeshli2012continuous,senthil2012integer,grover2012quantum,lee2018emergent,ma2020emergent,lotri2025paired}.

We now review the physical origin of the Jain and daughter phases within composite fermion theory \cite{jain1989cf,jain2007cfbook}. At even-denominator filling, composite fermions experience zero effective magnetic field and can form a superconducting state. The nature of this state depends on the pairing channel, characterized by the integer-valued chiral central charge $\mathcal{C}$ (the number of chiral neutral Majorana edge modes). The $\mathcal{C}=16$ state is equivalent to the $E_8$ state with no topological order, and thus only sixteen distinct topological orders exist at even denominator filling~\cite{kitaev2006anyons,kane2017pairing}.

Away from the even-denominator filling, composite fermions experience a nonzero effective field. If it destroys the superconducting order, an integer quantum Hall (IQH) state at filling $\nu_{\text{CF}}$ is formed: the standard Jain sequence at $\nu = \frac{\nu_{\text{CF}}}{p\nu_{\text{CF}} \pm 1}$. An alternative, which may be favored when the effective magnetic field is small, is the vortex lattice melting \emph{without the breaking of the composite-fermion Cooper pairs} \cite{sinova2002quantum,kwasigroch2012quantum,kumar2018symmetry,nguyen2023quantum,du2023quantum}, and the subsequent formation of the daughter state: a bosonic IQH (BIQH) state \cite{lu2012theory,senthil2012integer,grover2012quantum,lu2012quantum} of the Cooper pairs of composite fermions. 
The charge-conserving BIQH phases have no topological order, carry even filling $\nu$, and have $\kappa_{\mathrm{xy}} = 0$.  All other known bosonic states without topological order are constructed from the charge-conserving BIQH and the $E_8$ state that has $\kappa_{\mathrm{xy}} = 8$. As such, the bosonic state without topological order has $\kappa_{\mathrm{xy}} \equiv 0 \mod 8$.

\section{Critical theory}
The Abelian states we study can be described within the $K$-matrix formalism \cite{wenzee1992structures,wenzee1992classification,wen1995edge,hansson2016quantum}. We construct $K$-matrices from four building blocks: non-chiral bosonic $\sigma_x= \begin{pmatrix} 0 & 1 \\ 1 & 0 \end{pmatrix}$ and fermionic $\sigma_z = \begin{pmatrix} 1 & 0 \\ 0 & -1 \end{pmatrix}$ matrices, which carry zero $\kappa_{xy}$,  and maximally chiral $8\times 8$ matrices: the identity matrix $I_8$ (fermionic) and the $K$-matrix of the $E_8$ state $K_{E_8}$ (bosonic). In the absence of symmetries, the $K$-matrix with determinant 1 and zero signature defines a topologically trivial state \cite{belov2005classification}; for example, both $\sigma_x$ and $\sigma_z$ describe trivial phases. With $U(1)$ symmetry, in the topologically trivial phase, $\sigma_{\mathrm{xy}}$ must also vanish. Therefore, $\sigma_z$ with equally charged modes describes a topologically trivial phase. Therefore, $K$ and $K \oplus \sigma_z$ describe topologically equivalent phases, representing the addition of a gappable pair of fermionic modes \cite{cano2013bulkedge}.
 
The fermionic $\nu=8$ state is described by $K = I_8$ with charge vector entries $t_i=1$. The bosonic $\nu_{\text{pair}}=2$ state is described by $K = \sigma_x $ with charge vector entries $t_i=2$. To show the relation between the two, we start from an augmented $K$-matrix for the fermionic state:  a $12 \times 12$ matrix $K_{\text{f}} = I_8 \oplus \sigma_z \oplus \sigma_z$ with charge vector $t_i=1$.

A key insight comes from applying a $\mathrm{SL}(12, \mathbb{Z})$ transformation $W$ (given explicitly in \cref{app:matrices}) to the $K_{\text{f}}$ such that
\begin{align}
    K'_{\text{f}} = W^\transp K_{\text{f}} W &= \sigma_x \oplus K_{E_8} \oplus \sigma_z, \label{eq:maptoe8}
\end{align}
where the transformed charge vector $t'=W^\transp t$ has entries $t'_i=2$ for the $\sigma_x$ block, $t'_i=0$ for the $K_{E_8}$ block, and $t'_i=1$ for the $\sigma_z$ block. The transformation is merely a change of basis; it does not change the state. The last $\sigma_z$ block represents a trivial phase both before and after the transformation, in which single-fermion excitations are gapped. Beyond this block, the fermionic $\nu=8$ state ($I_8$) is topologically equivalent to the bosonic $\nu_{\text{pair}}=2$ state ($\sigma_x$) stacked with a neutral $E_8$ state. Hence, the two phases differ in their neutral sector: a chiral $E_8$ state for the fermionic $\nu=8$ state and a trivial state for the bosonic $\nu_{\text{pair}}=2$.

 To describe the critical point between the neutral $E_8$
 and topologically trivial states, 
 we employ a parton construction  \cite{jain1989partons,grover2012quantum,balram2018anomalfqhe}. 
We express both theories using a single parton construction; they differ only in the signs of certain parton Chern numbers. In this representation, the transition is a mass sign change of the partonic Dirac fermions coupled to the gauge fields. We then perform an RG analysis of the critical point to demonstrate the emergence of intermediate phases.

We begin with a general description of this approach.
Parton theory describes physical states using constituent fermions $f_j$ (partons), each represented by a bosonic field $\beta_j$, such that the parton currents are $j = \frac{1}{2\pi} \curl{\beta}$. When the partons are gapped, their low-energy dynamics follow an action with kinetic and topological Chern--Simons   terms subject to constraints on the parton density imposed by the emergent gauge fields $a_k$:
 \begin{align}
 \mathcal{L} &= \sum_{j=1}^{N_{\text{p}}}\qty(\frac{\mathcal{K}_{jj}}{4\pi} \beta_j \partial \beta_j + \frac{t_j}{2\pi}\beta_j \partial A) \nonumber\\ &\quad + \mathcal{L}_{\text{constraint}} + \mathcal{L}_{\text{kinetic}} , \label{eq:parton_lagrangian}
 \end{align}
 where spacetime indices are omitted for brevity, $\beta \partial \beta$ is a shorthand for  $\epsilon^{\mu\nu\rho} \beta_\mu \partial_\nu \beta_\rho$, $ \mathcal{L}_{\text{constraint}}$ is the constraint Lagrangian defined below, $\mathcal{L}_{\text{kinetic}}$ is the kinetic term of form $\beta M \beta$ with $M$ being a diagonal matrix both in parton and spacetime indices, $A$ is an external gauge field, and $t$ is the charge vector. In the first term, $\mathcal{K}_{jj} \in \qty{\pm 1, 0}$ is the Chern number of the parton $f_j$ (we use $K$ for the $K$-matrix for the physical degrees of freedom and calligraphic $\mathcal{K}$ otherwise), reflecting the partons being either in an IQH state or in a trivial insulator state. The constraint forces parton densities to be identified: 
\begin{align}
 \mathcal{L}_{\text{constraint}} &= \frac{1}{2\pi}\sum_{j=1}^{N_{\text{g}}} a_j \partial \sum_{k=1}^{N_{\text{p}}} Q_{kj} \beta_k, \label{eq:constraint}
 \end{align}
 where $a_j$ is the set of $N_{\text{g}}$ gauge fields  and $Q_{kj}$ is the charge of the parton $f_k$ with respect to the gauge field $a_j$. Solving the linear constraints (\ref{eq:constraint}) yields a set of physical fields, whose Lagrangian, in turn, can represent various Chern--Simons theories depending on $Q$. For example, three partons constrained to have equal densities generate $\nu=1/3$ FQH Lagrangian \cite{jain1989partons}. 

We now use a parton theory to construct an electronic state defined by the $K$-matrix $K_{\text{e}}$ involving $N_{\text{e}}$ physical fields  $\psi_i$. These fields are constructed from $N_{\text{p}}$ partons $f_j$ via $\psi_i = \prod_j f_j^{P_{ij}}$.  The matrix $P$ is an $N_{\text{e}}\times N_{\text{p}} $ full-rank 
 matrix, and spans the null space of the coupling of the partons to the gauge fields $a_k$ in \cref{eq:constraint}: $Q$ is an $N_{\text{p}} \times N_{\text{g}}$ matrix satisfying $PQ=0$. 
 If all parton Chern numbers are non-zero, integrating out the gauge fields in \cref{eq:parton_lagrangian} yields a Lagrangian for the gauge-invariant physical fields with the $K$-matrix $K_{\text{e}} =  P \mathcal{K} P^\transp$; 
 the charge vector of the physical fields is $t_{\text{e}} =  P t$.
 We restrict $P$ to have values in the set $\qty{-1,0,1}$, meaning each parton appears at most once in the definition of $\psi_i$. 

If a parton has zero Chern number ($\mathcal{K}_{jj}=0$), it is a trivial band insulator. Its current response starts at $\order{\omega^2}$, yielding a finite transverse term in the inverse kernel of $\beta_j$. After imposing $\beta=P^\transp\varphi$, the physical inverse kernel at zero spatial momentum has the form
\begin{align}
 D_{\mathrm{phys}}^{mn}(\omega)
 &= (PMP^\transp)\delta^{mn}
 +\frac{i\omega}{2\pi}(P\mathcal{K}P^\transp)\epsilon^{mn}
 +\order{\omega^2}, \label{eq:physical_kernel}
\end{align}
where $M$ has nonzero entries for the trivial partons. Integrating out the fields on which $PMP^\transp$ is nonzero leaves
\begin{align}
 K_{\mathrm e}&=LP\mathcal{K}P^\transp L^\transp,
 &t_{\mathrm e}&=LPt, \label{eq:projected_k}
\end{align}
with the rows of $L$ spanning $\ker(PMP^\transp)$. The eliminated fields contribute only higher-derivative corrections to this sector; the full derivation is given in \cref{app:c0}. A non-singular $K_{\mathrm e}$ describes a gapped topological sector within this effective theory; otherwise the theory requires further analysis \cite{barkeshli2012continuous,grover2012quantum}.

There are multiple possible partonic constructions (i.e.,  choices of $P$ and $\mathcal{K}$, possibly with different $N_{\text{p}}$) for each topological state defined by $K_e$. In what follows, we point out which of our conclusions depend on the particular construction choice and which are universal.

We now use a specific construction for the neutral $E_8$ and trivial states that appear in \cref{eq:maptoe8} using $N_{\text{p}}=12$ partons. 
The trivial state is realized with $\mathcal{K}_1 = I_6 \oplus (-I_6)$ and the $E_8$ state with $\mathcal{K}_2 = I_{10} \oplus (-I_2)$, using the $8 \times 12$ projection matrix $P$ and $12 \times 4$ coupling matrix $Q$; the charge vector is $0$ since the state is neutral. These matrices (and an alternative partonic construction) are given in \cref{app:matrices}. 

In this system, each change of a parton's Chern number changes $\kappa_{\mathrm{xy}}$ by 1. If we consider three phases differing only by the value of $\mathcal{K}_{jj}\in \qty{-1, 0,1}$, their physical $K$-matrices $K_e$ differ by a rank-1 matrix proportional to $P_j P_j^\transp$, which changes the signature $\kappa_{\mathrm{xy}}$ by at most 2~\cite{golub1973matrixreview}. Since $\kappa_{\mathrm{xy}}$ drops from 8 to 0 across four partonic Chern number sign flips, the difference in $\kappa_{\mathrm{xy}}$ between phases with $\mathcal{K}_{jj} = \pm 1$ is exactly 2. Moreover, since in the $\mathcal{K}_{jj}=0$ phase the $\kappa_{\mathrm{xy}}$ is odd, each transition changes $\kappa_{\mathrm{xy}}$ by 1.

%The phase transition between the two states closes the gap, requiring a description that includes partons. 
A transition in which the sign of $\mathcal{K}_{jj}$ is flipped can be described by a pair of Dirac fermions $f_{j,\pm}$ whose masses change signs. Therefore, the critical point describing a direct transition between the $E_8$ and trivial states is described by $N_{\text{f}}=4$ such pairs. As all the other partons are gapped in both phases, we can integrate them out. 
%and include only the partons that change mass sign in the Lagrangian. 
The Lagrangian describing the low-energy physics near the critical point is
\begin{align}
    \mathcal{L} &= \sum_{i=1}^{N_{\text{f}}}\sum_{s=\pm}
 \bar f_{i,s}\qty(\gamma_\mu D_i^\mu+m_{i,s})f_{i,s} \nonumber\\
 &\quad +\frac{i}{4\pi}\sum_{i,j=1}^{N_{\text{g}}}\mathcal{K}^{\text{eff}}_{ij}a_i\partial a_j
 +\mathcal{L}_{\text{Maxwell}}, \label{eq:Ltwophases}
\end{align}
where we use Euclidean signature, $D_i^\mu=\partial^\mu-i\sum_{j=1}^{N_{\text{g}}}R_{ij}a_j^\mu$, $\mathcal{K}^{\text{eff}}_{ij} = \sum_k Q_{ki} \mathcal{K}_{kk}^{-1} Q_{kj}$ with $k$ running over partons we integrate out, $R$ is a $N_{\text{f}} \times N_{\text{g}}$ matrix of rows of $Q$ corresponding to the remaining partons, and $\mathcal{L}_{\text{Maxwell}}$ includes non-universal higher-order terms.
Far from the transition, the Lagrangians defined in  \cref{eq:parton_lagrangian,eq:Ltwophases} are equivalent after integrating out the fields $\beta$ in \cref{eq:parton_lagrangian} and the Dirac fermions in \cref{eq:Ltwophases}. The phase in question is determined by the signs of $\mathcal{K}_{jj}$ in \cref{eq:parton_lagrangian}, or, equivalently, by the signs of $m_{i,\pm}$ in \cref{eq:Ltwophases}.

In our case, the matrices $\mathcal{K}^{\text{eff}}$  and $R$ are given by (\cref{app:rg}):
\begin{align}
    \mathcal{K}^{\text{eff}} &=
\begin{pmatrix}
 0 & 0 & 1 & 1 \\
 0 & -1 & 2 & 0 \\
 1 & 2 & 0 & -1 \\
 1 & 0 & -1 & -1 \\
\end{pmatrix} \label{eq:Keff}\\
R &= \begin{pmatrix} 
0 & -1 & 0 & 0 \\ 
0 & 1 & 0 & 0 \\ 
0 & 0 & -1 & 0 \\ 
0 & 0 & 0 & 1 \end{pmatrix}. \label{eq:Rcrit}
\end{align}

A direct transition from the $E_8$ to the trivial state requires the $2N_{\text{f}}=8$ masses to vanish simultaneously. Yet, a generic path connecting two phases passes through multiple intermediate phases; the simultaneous vanishing of all masses should not happen generically. While it may be imposed by a symmetry, the system in question lacks a suitable microscopic symmetry. 
Another possibility is the emergence of a larger gauge group \cite{ma2020emergent} if there is a symmetry between the partons, i.e., the linear dependence between the columns of $P$. However, since the change in $\kappa_{\mathrm{xy}}$ is bounded by twice the rank of the $P(\mathcal{K}_1-\mathcal{K}_2)P^\transp$, in a transition between $E_8$ and trivial states, there are at least four linearly independent partons.  

Alternatively, all eight masses can vanish simultaneously if, at low energies, the mass difference operators are irrelevant in the RG sense, resulting in an emergent symmetry. %For example, in the Jain state edges with multiple modes \cite{kane1995impurity}.
 
The relevance of an operator is determined by its scaling during the RG flow. The coefficient $h$ of the operator $\mathcal{M}$ flows as 
\begin{align}
    \dv{h}{\ell} &= (d+1-\Delta_{\mathcal{M}})h, \label{eq:rg_scaling}
\end{align}
where $d+1$ is spacetime dimension, $\ell$ is the RG time and $\Delta_{\mathcal{M}}$ is the scaling dimension of $\mathcal{M}$: $h$ decreases if $\mathcal{M}$ is irrelevant, i.e., $d+1-\Delta_{\mathcal{M}}<0$ \cite{cardy1996scaling,difrancesco1997cft}.

Thus, we want to check whether near the critical point of \cref{eq:Ltwophases} there are relevant operators beyond the one that controls the phase transition. If any such mass-like operator $\mathcal{M} = \sum_{ij} \bar{\psi}_i m_{ij} \psi_j$ is relevant, there is no direct transition between the two phases. 
In a non-interacting 2+1D fermionic theory, the bilinear $\mathcal{M}$ has scaling dimension $\Delta^{(0)}_\mathcal{M} = 2$ (twice the fermion’s), hence it is relevant.

We treat interactions via a large-$N$ expansion \cite{hermele2005algebraic,hermele2005algebraicerratum,chester2016anomalous,thomson2017qed3,lee2018emergent} by rewriting the Lagrangian (\ref{eq:Ltwophases}) as 
\begin{align}
    \mathcal{L} &= \sum_{k=1}^{N}\sum_{i=1}^{N_{\text{f}}}\sum_{s=\pm}
 \bar f_{i,k,s}\qty(\gamma_\mu D_i^\mu+m_{i,k,s})f_{i,k,s} \nonumber\\
 &\quad +\frac{iN}{4\pi}\sum_{i,j=1}^{N_{\text{g}}}\mathcal{K}^{\text{eff}}_{ij}a_i\partial a_j, \label{eq:LtwophasesNf}
\end{align}
In this Lagrangian, there are still $N_{\text{g}}$ gauge fields, but $2 N_{\text{f}} \cdot N$ fermionic species, and $2N$ species with the same $i$ are coupled to the gauge fields in the same way. Moreover, the Chern--Simons coupling between the gauge fields is also linear in $N$. The original Lagrangian corresponds to $N=1$, and we can calculate observables using perturbation theory in $1/N$. 
 
While applying the large-$N$ limit for $N=1$ is not rigorously justified, it yields corrections consistent with other results \cite{witten1979baryons,altshuler1994lowenergy,nikolic2007renormalization,gonzales1012higherorder}, such as the $\epsilon$-expansion for QED \cite{dipientro2017epsilonqed} even for $N=1$. Note that the $\epsilon$-expansion is inapplicable in our case due to the Chern--Simons term.

Due to the mixing between the scalar mass terms $\mathcal{M}^\alpha(x)=\frac{1}{\sqrt{2N}}\sum_{k=1}^N\sum_{s=\pm}\bar f_{\alpha,k,s}f_{\alpha,k,s}$, the operators with well-defined scaling at $\order{1/N}$ are certain linear combinations of $\mathcal{M}^\alpha(x)$ \cite{wilson1970ope,constable2002operator,chester2016anomalous,benvenuti2019easyplane}. To find the corrections to the tree-level scaling dimension  $\Delta^{(0)}=2$ we calculate the leading-order correction to the correlator 
\begin{align}
    \expval{\mathcal{M}^\alpha(x)\mathcal{M}^\beta(0)} = \frac{\delta_{\alpha\beta} - M_{\alpha\beta}\log(\abs{x}^2 \Lambda^2) + \order{\frac{1}{N^2}}}{\abs{x}^{2\Delta^{(0)}}}. \label{eq:corr_log}
\end{align}
Here $\alpha=1,\ldots,4$ labels the Dirac pairs, the tree-level correlator is normalized to unity, and $M=\order{1/N}$. 
The operators with well-defined scaling are given by the eigenvectors $\mu_{\alpha \beta}$ of $M_{\alpha \beta}$ via $\tilde{\mathcal{M}}^\alpha =\mu_{\alpha \beta} \mathcal{M}^\beta$. The scaling dimension can then be extracted using \cref{eq:corr_log} as an expansion of $\expval{\tilde{\mathcal{M}}^\alpha (x)\tilde{\mathcal{M}}^\alpha(0)} =\abs{x}^{-2\Delta_\alpha}$ around $\Delta_{\alpha}=\Delta^{(0)}$.
While off-diagonal mass terms can, in principle, appear in the action, such terms are gauge-invariant only if two partons $f_i$ and $f_j$ have the same charge under all gauge fields. 

To evaluate \cref{eq:corr_log}, we first calculate the photon propagator by integrating out the Dirac fermions at the critical point of \cref{eq:LtwophasesNf}, resulting in a momentum-space Lagrangian
\begin{align}
    \mathcal{L}_{\text{crit}}(p) = a_{i\mu}(-p)\Big[&\frac{\abs{p}}{32}G^{ij}
    \qty(\delta^{\mu\nu}-\frac{p^\mu p^\nu}{p^2})\nonumber\\
    &+\frac{1}{4\pi}\epsilon^{\mu\nu\rho}p_\rho
    \qty(\mathcal{K}^{\text{eff}})^{ij}\Big]a_{j\nu}(p).
\end{align}
Here $G=2N R^\transp R$, and $\mathcal{K}^{\text{eff}}$ is redefined to incorporate the factor $N$. The photon propagator $(D_{\mu\nu})_{ij}=\expval{a_{i\mu}(p)a_{j\nu}(-p)}$ is then an $N_{\text{g}}\times N_{\text{g}}$ matrix. When multiplied by the charge matrices, it is given by
\begin{align}
    \tilde{D}_{\mu\nu}(p)&=R D_{\mu\nu}(p)R^\transp
    =\frac{\tilde{A}}{\abs{p}}\qty(\delta_{\mu\nu}-\frac{p_\mu p_\nu}{\abs{p}^2})
    +\tilde{C}\frac{p^\rho}{p^2}\epsilon_{\rho\mu\nu},\label{eq:final_propagator}\\
    V&=\qty[\frac{\pi}{128}G{\mathcal{K}^{\text{eff}}}^{-1}G
    +\frac{1}{2\pi}\mathcal{K}^{\text{eff}}]^{-1},\\
    \tilde{A}&=\frac{\pi}{8}R{\mathcal{K}^{\text{eff}}}^{-1}G V R^\transp,\\
    \tilde{C}&=-R V R^\transp,
\end{align}
where we use transverse (Landau) gauge ($a_\mu p^\mu=0$), and the matrix multiplication is over gauge-field indices $i,j$. The elements of $\tilde{A}$ and $\tilde{C}$ are $\order{1/N}$.

The corrections to the scaling dimensions depend on the parton construction and are given by the eigenvalues of the matrix
\begin{align}
    M_{\alpha\beta}=\frac{-16\tilde{A}_{\alpha\beta}\delta_{\alpha\beta}
    +6N\qty[(\tilde{A}_{\alpha\beta})^2-(\tilde{C}_{\alpha\beta})^2]}{12\pi^2}.
    \label{eq:finalm}
\end{align}
Here the squares are elementwise, and each Dirac pair contains $2N$ species.
For the parton construction corresponding to \cref{eq:Keff,eq:Rcrit}, all eight diagonal mass perturbations remain relevant at $\order{1/N}$, although one scaling-dimension correction is positive. As a result, the fixed point is unstable to mass-like perturbations, meaning that a direct transition within this critical theory requires fine-tuning and intermediate phases appear generically. The values of $\Delta_k$ are given in \cref{app:rg}.

These results, as expected, are independent of choice of $Q$ in \cref{eq:constraint}: if $Q' = Q \cdot O$ for some (invertible) matrix $O$, then $\mathcal{K}^{\text{eff}}\mapsto O^\transp \mathcal{K}^{\text{eff}}O$, $G \mapsto O^\transp G O$, and $R\mapsto R O$ so that \cref{eq:final_propagator} is unaltered.

We conclude that generically there is no direct transition between the $E_8$ and the trivial bosonic states, and at least seven intermediate phases with $\kappa_{\mathrm{xy}} = 1\dots 7$ should appear. 
Since integer bosonic phases have $\kappa_{\mathrm{xy}} \equiv 0 \mod 8$, any intermediate state must be fractional~\cite{lu2012theory,kaidi2021higher}. 

\section{Discussion}
We now discuss the considerations for our particular choice of the parton construction. Different parton constructions, defined by the choice of $P$, $\mathcal{K}_1$, and $\mathcal{K}_2$, describe different critical theories, resulting in intermediate phases with distinct topological orders.
%and show that in this construction, the intermediate phases host the minimal possible number of anyon species.  
Even for a specific parton construction, the phase diagram contains multiple distinct phases with the same value of $\kappa_{\mathrm{xy}}$, with different anyon content, defined by the different mass sign assignments in \cref{eq:Ltwophases}. As we now argue, our choice of parton construction supports a sequence of intermediate phases that have the minimal possible number of anyon species.  

For the transition between the daughter and the Jain states at filling $\nu=\nicefrac{p}{q}$, we expect intermediate phases with ground-state degeneracy $r\cdot q$, where $r$ is the degeneracy of the fractionalized neutral sector.
For an Abelian neutral sector, $r=\abs{\det K_{\mathrm e}}$ counts the anyon types and the torus ground-state degeneracy. When all parton Chern numbers are $\pm1$, changing their signs preserves $K_{\mathrm e}=P\mathcal{K}P^\transp$ modulo 2 and hence preserves its odd determinant. This restricts the even-$\kappa_{\mathrm{xy}}$ phases along the successive-flip path. Together with the classification of Abelian bosonic anyon theories \cite{wang2020abelian}, it gives $r\geqslant\min(\kappa_{\mathrm{xy}}+1,9-\kappa_{\mathrm{xy}})$ for this class of paths \cite{tao2012topics} as shown in \cref{app:intermediate}.
 In that sense, our construction is ``minimal'' \cite{musser2024fractionalization,cheng2025ordering}, generating a single-parameter path between the $E_8$ and trivial phases passing through the minimal-degeneracy phases with $\kappa_{\mathrm{xy}} =1\dots 8$.

Our results suggest that the daughter states can serve as a platform for engineering various bosonic quantum Hall states that have not been observed experimentally, including the $E_8$ state and more complicated interacting integer states. For instance, a bilayer system with opposite-filling states could gap out the charge sector, thereby isolating the neutral bosonic state. Alternatively, the proximity-coupled counter-propagating edges of the daughter and Jain states are a simpler setup for probing the edge physics of the bosonic states. Thermal Hall measurements can verify our predictions for $\kappa_{\mathrm{xy}}$ of the intermediate phases. Probing the quasiparticle braiding or the ground-state degeneracy is necessary to fully verify our predictions. Finally, our approach can be extended to more general daughter states \cite{zhang2024hierarchy} beyond even-denominator parent states.

%  Acknowledgments
\begin{acknowledgments}
We thank Yarden Sheffer and Siddharth Singh for helpful discussions.
EZ is supported by the Adams Fellowships Program of the Israel Academy of Sciences and Humanities. AS was supported by grants from the ERC under the European Union’s Horizon 2020 research and innovation programme (Grant Agreements LEGOTOP No. 788715), the DFG (CRC/Transregio 183, EI 519/71), and by the ISF Quantum Science and Technology (2074/19). NL acknowledges support from ISF-MAFAT Quantum Science and Technology Grant no. 2478/24.
\end{acknowledgments}

%\clearpage
\bibliography{main}

@article{wilson1970ope,
  title = {Operator-Product Expansions and Anomalous Dimensions in the {Thirtring} Model},
  author = {Wilson, Kenneth G.},
  year = 1970,
  month = oct,
  journal = {Physical Review D},
  publisher = {American Physical Society},
  volume = 2,
  pages = {1473--1477},
  issue = 8,
  numpages = {0},
  doi = {10.1103/PhysRevD.2.1473},
  url = {https://link.aps.org/doi/10.1103/PhysRevD.2.1473},
}

@article{golub1973matrixreview,
  title = {Some Modified Matrix Eigenvalue Problems},
  author = {Golub, Gene H.},
  year = 1973,
  journal = {SIAM Review},
  volume = 15,
  number = 2,
  pages = {318--334},
  doi = {10.1137/1015032},
  url = {https://doi.org/10.1137/1015032},
}

@article{witten1979baryons,
  title = {Baryons in the {$1/N$} expansion},
  author = {Edward Witten},
  year = 1979,
  month = nov,
  journal = {Nuclear Physics B},
  volume = 160,
  number = 1,
  pages = {57--115},
  doi = {https://doi.org/10.1016/0550-3213(79)90232-3},
  issn = {0550-3213},
  url = {https://www.sciencedirect.com/science/article/pii/0550321379902323},
}

@article{aoki1985localization,
  title = {Critical localization in two-dimensional {Landau} quantization},
  author = {Aoki, Hideo and Ando, Tsuneya},
  year = 1985,
  month = {Feb},
  journal = {Physical Review Letters},
  publisher = {American Physical Society},
  volume = 54,
  pages = {831--834},
  issue = 8,
  numpages = {0},
  doi = {10.1103/PhysRevLett.54.831},
  url = {https://link.aps.org/doi/10.1103/PhysRevLett.54.831},
}

@article{willet1987nu52,
  title = {Observation of an even-denominator quantum number in the fractional quantum {Hall} effect},
  author = {Willett, Robert and Eisenstein, James P. and St\"ormer, Horst L. and Tsui, Daniel C. and Gossard, Arthur C. and English, John H.},
  year = 1987,
  month = oct,
  journal = {Physical Review Letters},
  publisher = {American Physical Society},
  volume = 59,
  pages = {1776--1779},
  issue = 15,
  numpages = {0},
  doi = {10.1103/PhysRevLett.59.1776},
  url = {https://link.aps.org/doi/10.1103/PhysRevLett.59.1776},
}

@article{pruisken1988singlarities,
  title = {Universal Singularities in the Integral Quantum {Hall} Effect},
  author = {Pruisken, Adrianus M. M.},
  year = 1988,
  month = sep,
  journal = {Physical Review Letters},
  publisher = {American Physical Society},
  volume = 61,
  pages = {1297--1300},
  issue = 11,
  numpages = {0},
  doi = {10.1103/PhysRevLett.61.1297},
  url = {https://link.aps.org/doi/10.1103/PhysRevLett.61.1297},
}

@article{jain1989cf,
  title = {Composite-fermion approach for the fractional quantum {Hall} effect},
  author = {Jain, Jainendra K.},
  year = 1989,
  month = jul,
  journal = {Physical Review Letters},
  publisher = {American Physical Society},
  volume = 63,
  pages = {199--202},
  issue = 2,
  doi = {10.1103/PhysRevLett.63.199},
  url = {https://link.aps.org/doi/10.1103/PhysRevLett.63.199},
}

@article{jain1989partons,
  title = {Incompressible quantum {Hall} states},
  author = {Jain, Jainendra K.},
  year = 1989,
  month = oct,
  journal = {Physical Review B},
  publisher = {American Physical Society},
  volume = 40,
  pages = {8079--8082},
  issue = 11,
  numpages = {0},
  doi = {10.1103/PhysRevB.40.8079},
  url = {https://link.aps.org/doi/10.1103/PhysRevB.40.8079},
}

@article{eisenstein1990transition,
  title = {Evidence for a spin transition in the $\nu=2/3$ fractional quantum {Hall} effect},
  author = {Eisenstein, James P. and Stormer, Horst L. and Pfeiffer, Loren N. and West, Kirk W.},
  year = 1990,
  month = apr,
  journal = {Physical Review B},
  publisher = {American Physical Society},
  volume = 41,
  pages = {7910--7913},
  issue = 11,
  numpages = {0},
  doi = {10.1103/PhysRevB.41.7910},
  url = {https://link.aps.org/doi/10.1103/PhysRevB.41.7910},
}

@article{mooreread1991nonabelions,
  title = {Nonabelions in the fractional quantum {Hall} effect},
  author = {Gregory Moore and Nicholas Read},
  year = 1991,
  journal = {Nuclear Physics B},
  volume = 360,
  number = 2,
  pages = {362--396},
  doi = {https://doi.org/10.1016/0550-3213(91)90407-O},
  issn = {0550-3213},
  url = {https://www.sciencedirect.com/science/article/pii/055032139190407O},
}

@article{koch1991scaling,
  title = {Experiments on scaling in {$\mathrm{Al}_{\mathit{x}}\mathrm{Ga}_{1-\mathit{x}}$As/GaAs} heterostructures under quantum {Hall} conditions},
  author = {Koch, S. and Haug, Rolf J. and von Klitzing, Klaus and Ploog, K.},
  year = 1991,
  month = mar,
  journal = {Physical Review B},
  publisher = {American Physical Society},
  volume = 43,
  pages = {6828--6831},
  issue = 8,
  numpages = {0},
  doi = {10.1103/PhysRevB.43.6828},
  url = {https://link.aps.org/doi/10.1103/PhysRevB.43.6828},
}

@article{wenzee1992structures,
  title = {Topological structures, universality classes, and statistics screening in the anyon superfluid},
  author = {Wen, Xiao-Gang and Zee, Anthony},
  year = 1991,
  month = jul,
  journal = {Physical Review B},
  publisher = {American Physical Society},
  volume = 44,
  pages = {274--284},
  issue = 1,
  numpages = {0},
  doi = {10.1103/PhysRevB.44.274},
  url = {https://link.aps.org/doi/10.1103/PhysRevB.44.274},
}

@article{kivelson1992globalphasediagram,
  title = {Global phase diagram in the quantum {Hall} effect},
  author = {Kivelson, Steven and Lee, Dung-Hai and Zhang, Shou-Cheng},
  year = 1992,
  month = jul,
  journal = {Physical Review B},
  publisher = {American Physical Society},
  volume = 46,
  pages = {2223--2238},
  issue = 4,
  numpages = {0},
  doi = {10.1103/PhysRevB.46.2223},
  url = {https://link.aps.org/doi/10.1103/PhysRevB.46.2223},
}

@article{wenzee1992classification,
  title = {Classification of {Abelian} quantum {Hall} states and matrix formulation of topological fluids},
  author = {Wen, Xiao-Gang and Zee, Anthony},
  year = 1992,
  month = jul,
  journal = {Physical Review B},
  publisher = {American Physical Society},
  volume = 46,
  pages = {2290--2301},
  issue = 4,
  numpages = {0},
  doi = {10.1103/PhysRevB.46.2290},
  url = {https://link.aps.org/doi/10.1103/PhysRevB.46.2290},
}

@article{lee1993unified,
  title = {Unified model for two localization problems: Electron states in spin-degenerate {Landau} levels and in a random magnetic field},
  author = {Lee, Derek K. K. and Chalker, John T.},
  year = 1994,
  month = mar,
  journal = {Physical Review Letters},
  publisher = {American Physical Society},
  volume = 72,
  pages = {1510--1513},
  issue = 10,
  numpages = {0},
  doi = {10.1103/PhysRevLett.72.1510},
  url = {https://arxiv.org/abs/cond-mat/9311050},
  eprint = {cond-mat/9311050},
  archiveprefix = {arXiv},
  primaryclass = {cond-mat},
}

@article{altshuler1994lowenergy,
  title = {Low-energy properties of fermions with singular interactions},
  author = {Altshuler, Boris L. and Ioffe, Lev B. and Millis, Andrew J.},
  year = 1994,
  month = nov,
  journal = {Physical Review B},
  publisher = {American Physical Society},
  volume = 50,
  pages = {14048--14064},
  issue = 19,
  numpages = {0},
  doi = {10.1103/PhysRevB.50.14048},
  url = {https://arxiv.org/abs/cond-mat/9406024},
  eprint = {cond-mat/9406024},
  archiveprefix = {arXiv},
  primaryclass = {cond-mat},
}

@article{huckestein1995scaling,
  title = {Scaling theory of the integer quantum {Hall} effect},
  author = {Huckestein, Bodo},
  year = 1995,
  month = apr,
  journal = {Reviews of Modern Physics},
  publisher = {American Physical Society},
  volume = 67,
  pages = {357--396},
  issue = 2,
  numpages = {0},
  doi = {10.1103/RevModPhys.67.357},
  url = {https://arxiv.org/abs/cond-mat/9501106},
  eprint = {cond-mat/9501106},
  archiveprefix = {arXiv},
  primaryclass = {cond-mat},
}

@article{wen1995edge,
  title = {Topological orders and edge excitations in fractional quantum {Hall} states},
  author = {Xiao-Gang Wen},
  year = 1995,
  month = sep,
  journal = {Advances in Physics},
  publisher = {Taylor & Francis},
  volume = 44,
  number = 5,
  pages = {405--473},
  doi = {10.1080/00018739500101566},
  url = {https://arxiv.org/abs/cond-mat/9506066},
  eprint = {cond-mat/9506066},
  archiveprefix = {arXiv},
  primaryclass = {cond-mat},
}

@article{mcdonald1995topological,
  title = {Topological phase transition in the $\nu=2/3$ quantum {Hall} effect},
  author = {McDonald, Ian A. and Haldane, F. Duncan M.},
  year = 1996,
  month = jun,
  journal = {Physical Review B},
  publisher = {American Physical Society},
  volume = 53,
  pages = {15845--15855},
  issue = 23,
  numpages = {0},
  doi = {10.1103/PhysRevB.53.15845},
  url = {https://arxiv.org/abs/cond-mat/9511061},
  eprint = {cond-mat/9511061},
  archiveprefix = {arXiv},
  primaryclass = {cond-mat},
}

@article{kane1997thermal,
  title = {Quantized thermal transport in the fractional quantum {Hall} effect},
  author = {Kane, Charles L. and Fisher, Matthew P. A.},
  year = 1997,
  month = jun,
  journal = {Physical Review B},
  publisher = {American Physical Society},
  volume = 55,
  pages = {15832--15837},
  issue = 23,
  numpages = {0},
  doi = {10.1103/PhysRevB.55.15832},
  url = {https://arxiv.org/abs/cond-mat/9603118},
  eprint = {cond-mat/9603118},
  archiveprefix = {arXiv},
  primaryclass = {cond-mat},
}

@article{kronmueller1998huge,
  title = {New Resistance Maxima in the Fractional Quantum {Hall} Effect Regime},
  author = {Kronm\"uller, Silvia and Dietsche, Werner and Weis, Jürgen and von Klitzing, Klaus and Wegscheider, Werner and Bichler, Max},
  year = 1998,
  month = sep,
  journal = {Physical Review Letters},
  publisher = {American Physical Society},
  volume = 81,
  pages = {2526--2529},
  issue = 12,
  numpages = {0},
  doi = {10.1103/PhysRevLett.81.2526},
  url = {https://arxiv.org/abs/cond-mat/9804283},
  eprint = {cond-mat/9804283},
  archiveprefix = {arXiv},
  primaryclass = {cond-mat.mes-hall},
}

@article{read1999paired,
  title = {Paired states of fermions in two dimensions with breaking of parity and time-reversal symmetries and the fractional quantum {Hall} effect},
  author = {Read, Nicholas and Green, Dmitry},
  year = 2000,
  month = apr,
  journal = {Physical Review B},
  publisher = {American Physical Society},
  volume = 61,
  pages = {10267--10297},
  issue = 15,
  numpages = {0},
  doi = {10.1103/PhysRevB.61.10267},
  url = {https://arxiv.org/abs/cond-mat/9906453},
  eprint = {cond-mat/9906453},
  archiveprefix = {arXiv},
  primaryclass = {cond-mat.mes-hall},
}

@article{wen1999continuous,
  title = {Continuous Topological Phase Transitions between Clean Quantum {Hall} States},
  author = {Wen, Xiao-Gang},
  year = 2000,
  month = apr,
  journal = {Physical Review Letters},
  publisher = {American Physical Society},
  volume = 84,
  pages = {3950--3953},
  issue = 17,
  numpages = {0},
  doi = {10.1103/PhysRevLett.84.3950},
  url = {https://arxiv.org/abs/cond-mat/9908394},
  eprint = {cond-mat/9908394},
  archiveprefix = {arXiv},
  primaryclass = {cond-mat.mes-hall},
}

@article{smet2001morphology,
  title = {Ising Ferromagnetism and Domain Morphology in the Fractional Quantum {Hall} Regime},
  author = {Smet, Jurgen H. and Deutschmann, Rainer A. and Wegscheider, Werner and Abstreiter, Gerhard and von Klitzing, Klaus},
  year = 2001,
  month = mar,
  journal = {Physical Review Letters},
  publisher = {American Physical Society},
  volume = 86,
  pages = {2412--2415},
  issue = 11,
  numpages = {0},
  doi = {10.1103/PhysRevLett.86.2412},
  url = {https://link.aps.org/doi/10.1103/PhysRevLett.86.2412},
}

@article{hohls2002scaling,
  title = {Hopping Conductivity in the Quantum {Hall} Effect: Revival of Universal Scaling},
  author = {Hohls, Frank and Zeitler, Ulrich and Haug, Rolf J.},
  year = 2002,
  month = jan,
  journal = {Physical Review Letters},
  publisher = {American Physical Society},
  volume = 88,
  pages = {036802},
  issue = 3,
  numpages = 4,
  doi = {10.1103/PhysRevLett.88.036802},
  url = {https://arxiv.org/abs/cond-mat/0107412},
  eprint = {cond-mat/0107412},
  archiveprefix = {arXiv},
  primaryclass = {cond-mat.mes-hall},
}

@article{sinova2002quantum,
  title = {Quantum Melting and Absence of {Bose}-{Einstein} Condensation in Two-Dimensional Vortex Matter},
  author = {Sinova, Jairo and Hanna, Charles B. and MacDonald, Allan H.},
  year = 2002,
  month = jun,
  journal = {Physical Review Letters},
  publisher = {American Physical Society},
  volume = 89,
  pages = {030403},
  issue = 3,
  numpages = 4,
  doi = {10.1103/PhysRevLett.89.030403},
  url = {https://arxiv.org/abs/cond-mat/0201020},
  eprint = {cond-mat/0201020},
  archiveprefix = {arXiv},
  primaryclass = {cond-mat},
}

@article{kraus2002logtitudinal,
  title = {From Quantum {Hall} Ferromagnetism to Huge Longitudinal Resistance at the $2/3$ Fractional Quantum {Hall} State},
  author = {Kraus, Stefan and Stern, O. and Lok, Jaap G. S. and Dietsche, Werner and von Klitzing, Klaus and Bichler, Max and Schuh, Dieter and Wegscheider, Werner},
  year = 2002,
  month = dec,
  journal = {Physical Review Letters},
  publisher = {American Physical Society},
  volume = 89,
  pages = 266801,
  issue = 26,
  numpages = 4,
  doi = {10.1103/PhysRevLett.89.266801},
  url = {https://link.aps.org/doi/10.1103/PhysRevLett.89.266801},
}

@article{constable2002operator,
  title = {Operator Mixing and the {BMN} Correspondence},
  author = {Neil R. Constable and Daniel Z. Freedman and Matthew Headrick and Shiraz Minwalla},
  year = 2002,
  month = dec,
  journal = {Journal of High Energy Physics},
  volume = 2002,
  number = 10,
  pages = {068},
  doi = {10.1088/1126-6708/2002/10/068},
  url = {https://dx.doi.org/10.1088/1126-6708/2002/10/068},
  eprint = {hep-th/0209002},
  archiveprefix = {arXiv},
  primaryclass = {hep-th},
}

@article{belov2005classification,
  title = {Classification of abelian spin {Chern}-{Simons} theories},
  author = {Dmitriy Belov and Gregory W. Moore},
  year = 2005,
  month = may,
  journal = {arXiv preprint},
  url = {https://arxiv.org/abs/hhep-th/0505235},
  eprint = {hep-th/0505235},
  archiveprefix = {arXiv},
  primaryclass = {hep-th},
}

@article{hermele2005algebraic,
  title = {Algebraic spin liquid as the mother of many competing orders},
  author = {Hermele, Michael and Senthil, Todadri and Fisher, Matthew P. A.},
  year = 2005,
  month = sep,
  journal = {Physical Review B},
  publisher = {American Physical Society},
  volume = 72,
  pages = 104404,
  issue = 10,
  numpages = 16,
  doi = {10.1103/PhysRevB.72.104404},
  url = {https://arxiv.org/abs/cond-mat/0502215},
  eprint = {cond-mat/0502215},
  archiveprefix = {arXiv},
  primaryclass = {cond-mat.str-el},
}

@article{kitaev2006anyons,
  title = {Anyons in an exactly solved model and beyond},
  author = {Alexei {\relax Yu}. Kitaev},
  year = 2006,
  month = jan,
  journal = {Annals of Physics},
  volume = 321,
  number = 1,
  pages = {2--111},
  doi = {https://doi.org/10.1016/j.aop.2005.10.005},
  issn = {0003-4916},
  url = {https://arxiv.org/abs/cond-mat/0506438},
  eprint = {cond-mat/0506438},
  archiveprefix = {arXiv},
  primaryclass = {cond-mat.mes-hall},
  note = {{January} Special Issue},
}

@article{bonderson2005detecting,
  title = {Detecting Non-{Abelian} Statistics in the $\nu=5/2$ Fractional Quantum {Hall} State},
  author = {Bonderson, Parsa and Kitaev, Alexei and Shtengel, Kirill},
  year = 2006,
  month = jan,
  journal = {Physical Review Letters},
  publisher = {American Physical Society},
  volume = 96,
  pages = {016803},
  issue = 1,
  numpages = 4,
  doi = {10.1103/PhysRevLett.96.016803},
  url = {https://arxiv.org/abs/cond-mat/0508616},
  eprint = {cond-mat/0508616},
  archiveprefix = {arXiv},
  primaryclass = {cond-mat.mes-hall},
}

@article{stern2005proposed,
  title = {Proposed Experiments to Probe the Non-{Abelian} $\nu=5/2$ Quantum {Hall} State},
  author = {Stern, Ady and Halperin, Bertrand I.},
  year = 2006,
  month = jan,
  journal = {Physical Review Letters},
  publisher = {American Physical Society},
  volume = 96,
  pages = {016802},
  issue = 1,
  numpages = 4,
  doi = {10.1103/PhysRevLett.96.016802},
  url = {https://arxiv.org/abs/cond-mat/0508447},
  eprint = {cond-mat/0508447},
  archiveprefix = {arXiv},
  primaryclass = {cond-mat.mes-hall},
}

@article{feldman2006detecting,
  title = {Detecting Non-{Abelian} Statistics with an Electronic {Mach}-{Zehnder} Interferometer},
  author = {Feldman, Dima E. and Kitaev, Alexei},
  year = 2006,
  month = nov,
  journal = {Physical Review Letters},
  publisher = {American Physical Society},
  volume = 97,
  pages = 186803,
  issue = 18,
  numpages = 4,
  doi = {10.1103/PhysRevLett.97.186803},
  url = {https://arxiv.org/abs/cond-mat/0607541},
  eprint = {cond-mat/0607541},
  archiveprefix = {arXiv},
  primaryclass = {cond-mat.mes-hall},
}

@article{nikolic2007renormalization,
  title = {Renormalization-group fixed points, universal phase diagram, and {$1/N$} expansion for quantum liquids with interactions near the unitarity limit},
  author = {Nikoli\'{c}, Predrag and Sachdev, Subir},
  year = 2007,
  month = mar,
  journal = {Physical Review A},
  publisher = {American Physical Society},
  volume = 75,
  pages = {033608},
  issue = 3,
  numpages = 14,
  doi = {10.1103/PhysRevA.75.033608},
  url = {https://arxiv.org/abs/cond-mat/0609106},
  eprint = {cond-mat/0609106},
  archiveprefix = {arXiv},
  primaryclass = {cond-mat.supr-con},
}

@article{khveshchenko2006composite,
  title = {Composite {Dirac} fermions in graphene},
  author = {Khveshchenko, Dmitri V.},
  year = 2007,
  month = apr,
  journal = {Physical Review B},
  publisher = {American Physical Society},
  volume = 75,
  pages = 153405,
  issue = 15,
  numpages = 4,
  doi = {10.1103/PhysRevB.75.153405},
  url = {https://arxiv.org/abs/cond-mat/0607174},
  eprint = {cond-mat/0607174},
  archiveprefix = {arXiv},
  primaryclass = {cond-mat.mes-hall},
}

@article{verdene2007firstorder,
  title = {Microscopic manifestation of the spin phase transition at filling factor 2/3},
  author = {Verdene, Basile and Martin, Jens and Gamez, Gerardo and Smet, Jurgen and von Klitzing, Klaus and Mahalu, Diana and Schuh, Dieter and Abstreiter, Gerhard and Yacoby, Amir},
  year = 2007,
  month = apr,
  journal = {Nature Physics},
  volume = 3,
  number = 6,
  pages = {392--396},
  doi = {10.1038/nphys588},
  isbn = {1745-2481},
  url = {https://doi.org/10.1038/nphys588},
}

@article{pruisken2005theta,
  title = {$\theta$ renormalization, electron–electron interactions and super universality in the quantum {Hall} regime},
  author = {Adrianus M. M. Pruisken and Igor S. Burmistrov},
  year = 2007,
  month = jun,
  journal = {Annals of Physics},
  volume = 322,
  number = 6,
  pages = {1265--1334},
  doi = {https://doi.org/10.1016/j.aop.2006.11.007},
  issn = {0003-4916},
  eprint = {cond-mat/0502488},
  archiveprefix = {arXiv},
  primaryclass = {cond-mat.mes-hall},
}

@article{hermele2005algebraicerratum,
  title = {Erratum: Algebraic spin liquid as the mother of many competing orders [{Phys. Rev. B} 72, 104404 (2005)]},
  author = {Hermele, Michael and Senthil, T. and Fisher, Matthew P. A.},
  year = 2007,
  month = oct,
  journal = {Physical Review B},
  publisher = {American Physical Society},
  volume = 76,
  pages = 149906,
  issue = 14,
  numpages = 2,
  doi = {10.1103/PhysRevB.76.149906},
  url = {https://link.aps.org/doi/10.1103/PhysRevB.76.149906},
}

@article{levin2007antipfaffian,
  title = {Particle-Hole Symmetry and the {Pfaffian} State},
  author = {Levin, Michael and Halperin, Bertrand I. and Rosenow, Bernd},
  year = 2007,
  month = dec,
  journal = {Physical Review Letters},
  publisher = {American Physical Society},
  volume = 99,
  pages = 236806,
  issue = 23,
  numpages = 4,
  doi = {10.1103/PhysRevLett.99.236806},
  url = {https://arxiv.org/abs/0707.0483},
  eprint = {0707.0483},
  archiveprefix = {arXiv},
  primaryclass = {cond-mat.mes-hall},
}

@article{ostrovsky2007theory,
  title = {Theory of anomalous quantum {Hall} effects in graphene},
  author = {Ostrovsky, Pavel M. and Gornyi, Igor V. and Mirlin, Alexander D.},
  year = 2008,
  month = may,
  journal = {Physical Review B},
  publisher = {American Physical Society},
  volume = 77,
  pages = 195430,
  issue = 19,
  numpages = 13,
  doi = {10.1103/PhysRevB.77.195430},
  url = {https://link.aps.org/doi/10.1103/PhysRevB.77.195430},
  eprint = {0712.0597},
  archiveprefix = {arXiv},
  primaryclass = {cond-mat.mes-hall},
}

@article{bonderson2007fractional,
  title = {Fractional quantum {Hall} hierarchy and the second {Landau} level},
  author = {Bonderson, Parsa and Slingerland, Joost K.},
  year = 2008,
  month = sep,
  journal = {Physical Review B},
  publisher = {American Physical Society},
  volume = 78,
  pages = 125323,
  issue = 12,
  numpages = 10,
  doi = {10.1103/PhysRevB.78.125323},
  url = {https://arxiv.org/abs/0711.3204},
  eprint = {0711.3204},
  archiveprefix = {arXiv},
  primaryclass = {cond-mat.mes-hall},
}

@article{nayak2007nonabelian,
  title = {Non-{Abelian} anyons and topological quantum computation},
  author = {Nayak, Chetan and Simon, Steven H. and Stern, Ady and Freedman, Michael and Das Sarma, Sankar},
  year = 2008,
  month = sep,
  journal = {Reviews of Modern Physics},
  publisher = {American Physical Society},
  volume = 80,
  pages = {1083--1159},
  issue = 3,
  numpages = {0},
  doi = {10.1103/RevModPhys.80.1083},
  url = {https://arxiv.org/abs/0707.1889},
  eprint = {0707.1889},
  archiveprefix = {arXiv},
  primaryclass = {cond-mat.str-el},
}

@article{yang2009thermopower,
  title = {Thermopower as a possible probe of non-{Abelian} quasiparticle statistics in fractional quantum {Hall} liquids},
  author = {Yang, Kun and Halperin, Bertrand I.},
  year = 2009,
  month = mar,
  journal = {Physical Review B},
  publisher = {American Physical Society},
  volume = 79,
  pages = 115317,
  issue = 11,
  numpages = 5,
  doi = {10.1103/PhysRevB.79.115317},
  url = {https://arxiv.org/abs/0901.1429},
  eprint = {0901.1429},
  archiveprefix = {arXiv},
  primaryclass = {cond-mat.mes-hall},
}

@article{cooper2008observable,
  title = {Observable Bulk Signatures of Non-{Abelian} Quantum {Hall} States},
  author = {Cooper, Nigel R. and Stern, Ady},
  year = 2009,
  month = apr,
  journal = {Physical Review Letters},
  publisher = {American Physical Society},
  volume = 102,
  pages = 176807,
  issue = 17,
  numpages = 4,
  doi = {10.1103/PhysRevLett.102.176807},
  url = {https://arxiv.org/abs/0812.3387},
  eprint = {0812.3387},
  archiveprefix = {arXiv},
  primaryclass = {cond-mat.mes-hall},
}

@article{levin2008collective,
  title = {Collective states of non-{Abelian} quasiparticles in a magnetic field},
  author = {Levin, Michael and Halperin, Bertrand I.},
  year = 2009,
  month = may,
  journal = {Physical Review B},
  publisher = {American Physical Society},
  volume = 79,
  pages = 205301,
  issue = 20,
  numpages = 11,
  doi = {10.1103/PhysRevB.79.205301},
  url = {https://arxiv.org/abs/0812.0381},
  eprint = {0812.0381},
  archiveprefix = {arXiv},
  primaryclass = {cond-mat.mes-hall},
}

@article{bid2009shotnoise,
  title = {Shot Noise and Charge at the $2/3$ Composite Fractional Quantum {Hall} State},
  author = {Bid, Aveek and Ofek, Nissim and Heiblum, Moty and Umansky, Vladimir and Mahalu, Diana},
  year = 2009,
  month = dec,
  journal = {Physical Review Letters},
  publisher = {American Physical Society},
  volume = 103,
  pages = 236802,
  issue = 23,
  numpages = 4,
  doi = {10.1103/PhysRevLett.103.236802},
  url = {https://link.aps.org/doi/10.1103/PhysRevLett.103.236802},
}

@article{hermanns2009condensing,
  title = {Condensing Non-{Abelian} Quasiparticles},
  author = {Hermanns, Maria},
  year = 2010,
  month = feb,
  journal = {Physical Review Letters},
  publisher = {American Physical Society},
  volume = 104,
  pages = {056803},
  issue = 5,
  numpages = 4,
  doi = {10.1103/PhysRevLett.104.056803},
  url = {https://arxiv.org/abs/0906.2073},
  eprint = {0906.2073},
  archiveprefix = {arXiv},
  primaryclass = {cond-mat.str-el},
}

@article{barkeshli2010anyon,
  title = {Anyon Condensation and Continuous Topological Phase Transitions in Non-{Abelian} Fractional Quantum {Hall} States},
  author = {Barkeshli, Maissam and Wen, Xiao-Gang},
  year = 2010,
  month = nov,
  journal = {Physical Review Letters},
  publisher = {American Physical Society},
  volume = 105,
  pages = 216804,
  issue = 21,
  numpages = 4,
  doi = {10.1103/PhysRevLett.105.216804},
  url = {https://arxiv.org/abs/1007.2030},
  eprint = {1007.2030},
  archiveprefix = {arXiv},
  primaryclass = {cond-mat.mes-hall},
}

@article{kumar2010nonconventional,
  title = {Nonconventional Odd-Denominator Fractional Quantum {Hall} States in the Second {Landau} Level},
  author = {Kumar, Ashwani and Cs\'athy, Gábor A. and Manfra, Michael J. and Pfeiffer, Loren N. and West, Kirk W.},
  year = 2010,
  month = dec,
  journal = {Physical Review Letters},
  publisher = {American Physical Society},
  volume = 105,
  pages = 246808,
  issue = 24,
  numpages = 4,
  doi = {10.1103/PhysRevLett.105.246808},
  url = {https://arxiv.org/abs/1009.0237},
  eprint = {1009.0237},
  archiveprefix = {arXiv},
  primaryclass = {cond-mat.mes-hall},
}

@article{gonzales1012higherorder,
  title = {Higher-order renormalization of graphene many-body theory},
  author = {Gonz{\'a}lez, Jose},
  year = 2012,
  month = aug,
  journal = {Journal of High Energy Physics},
  volume = 2012,
  number = 8,
  pages = 27,
  doi = {10.1007/JHEP08(2012)027},
  isbn = {1029-8479},
  url = {https://arxiv.org/abs/1204.4673},
  eprint = {1204.4673},
  archiveprefix = {arXiv},
  primaryclass = {cond-mat.supr-con},
}

@article{lu2012theory,
  title = {Theory and classification of interacting integer topological phases in two dimensions: A {Chern-Simons} approach},
  author = {Lu, Yuan-Ming and Vishwanath, Ashvin},
  year = 2012,
  month = sep,
  journal = {Physical Review B},
  publisher = {American Physical Society},
  volume = 86,
  pages = 125119,
  issue = 12,
  numpages = 28,
  doi = {10.1103/PhysRevB.86.125119},
  url = {https://link.aps.org/doi/10.1103/PhysRevB.86.125119},
  eprint = {1205.3156},
  archiveprefix = {arXiv},
  primaryclass = {cond-mat.str-el},
}

@article{wu2012microscopic,
  title = {Microscopic study of edge excitations of spin-polarized and spin-unpolarized $\nu=2/3$ fractional quantum {Hall} effect},
  author = {Wu, Ying-Hai and Sreejith, Ganesh Jaya and Jain, Jainendra K.},
  year = 2012,
  month = sep,
  journal = {Physical Review B},
  publisher = {American Physical Society},
  volume = 86,
  pages = 115127,
  issue = 11,
  numpages = 11,
  doi = {10.1103/PhysRevB.86.115127},
  url = {https://arxiv.org/abs/1207.3566},
  eprint = {1207.3566},
  archiveprefix = {arXiv},
  primaryclass = {cond-mat.mes-hall},
}

@article{kwasigroch2012quantum,
  title = {Quantum fluctuations of vortex lattices in ultracold gases},
  author = {Kwasigroch, Michal P. and Cooper, Nigel R.},
  year = 2012,
  month = dec,
  journal = {Physical Review A},
  publisher = {American Physical Society},
  volume = 86,
  pages = {063618},
  issue = 6,
  numpages = 9,
  doi = {10.1103/PhysRevA.86.063618},
  url = {https://arxiv.org/abs/1211.0245},
  eprint = {1211.0245},
  archiveprefix = {arXiv},
  primaryclass = {cond-mat.quant-gas},
}

@article{grover2012quantum,
  title = {Quantum phase transition between integer quantum {Hall} states of bosons},
  author = {Grover, Tarun and Vishwanath, Ashvin},
  year = 2013,
  month = jan,
  journal = {Physical Review B},
  publisher = {American Physical Society},
  volume = 87,
  pages = {045129},
  issue = 4,
  numpages = 10,
  doi = {10.1103/PhysRevB.87.045129},
  url = {https://arxiv.org/abs/1210.0907},
  eprint = {1210.0907},
  archiveprefix = {arXiv},
  primaryclass = {cond-mat.str-el},
}

@article{senthil2012integer,
  title = {Integer Quantum {Hall} Effect for Bosons},
  author = {Senthil, Todadri and Levin, Michael},
  year = 2013,
  month = jan,
  journal = {Physical Review Letters},
  publisher = {American Physical Society},
  volume = 110,
  pages = {046801},
  issue = 4,
  numpages = 5,
  doi = {10.1103/PhysRevLett.110.046801},
  url = {https://arxiv.org/abs/1206.1604},
  eprint = {1206.1604},
  archiveprefix = {arXiv},
  primaryclass = {cond-mat.str-el},
}

@article{chen2011symmetry,
  title = {Symmetry protected topological orders and the group cohomology of their symmetry group},
  author = {Chen, Xie and Gu, Zheng-Cheng and Liu, Zheng-Xin and Wen, Xiao-Gang},
  year = 2013,
  month = apr,
  journal = {Physical Review B},
  publisher = {American Physical Society},
  volume = 87,
  pages = 155114,
  issue = 15,
  numpages = 48,
  doi = {10.1103/PhysRevB.87.155114},
  url = {https://arxiv.org/abs/1106.4772},
  eprint = {1106.4772},
  archiveprefix = {arXiv},
  primaryclass = {cond-mat.str-el},
}

@article{wu2013quantum,
  title = {Quantum {Hall} effect of two-component bosons at fractional and integral fillings},
  author = {Wu, Ying-Hai and Jain, Jainendra K.},
  year = 2013,
  month = jun,
  journal = {Physical Review B},
  publisher = {American Physical Society},
  volume = 87,
  pages = 245123,
  issue = 24,
  numpages = 14,
  doi = {10.1103/PhysRevB.87.245123},
  url = {https://arxiv.org/abs/1304.7553},
  eprint = {1304.7553},
  archiveprefix = {arXiv},
  primaryclass = {cond-mat.str-el},
}

@article{furukawa2013integer,
  title = {Integer Quantum {Hall} State in Two-Component {Bose} Gases in a Synthetic Magnetic Field},
  author = {Furukawa, Shunsuke and Ueda, Masahito},
  year = 2013,
  month = aug,
  journal = {Physical Review Letters},
  publisher = {American Physical Society},
  volume = 111,
  pages = {090401},
  issue = 9,
  numpages = 5,
  doi = {10.1103/PhysRevLett.111.090401},
  url = {https://arxiv.org/abs/1304.5716},
  eprint = {1304.5716},
  archiveprefix = {arXiv},
  primaryclass = {cond-mat.quant-gas},
}

@article{cano2013bulkedge,
  title = {Bulk-edge correspondence in (2 + 1)-dimensional {Abelian} topological phases},
  author = {Cano, Jennifer and Cheng, Meng and Mulligan, Michael and Nayak, Chetan and Plamadeala, Eugeniu and Yard, Jon},
  year = 2014,
  month = mar,
  journal = {Physical Review B},
  publisher = {American Physical Society},
  volume = 89,
  pages = 115116,
  issue = 11,
  numpages = 25,
  doi = {10.1103/PhysRevB.89.115116},
  url = {https://arxiv.org/abs/1310.5708},
  eprint = {1310.5708},
  archiveprefix = {arXiv},
  primaryclass = {cond-mat.str-el},
}

@article{kapustin2014bosonic,
  title = {Bosonic Topological Insulators and Paramagnets: a view from cobordisms},
  author = {Anton Kapustin},
  year = 2014,
  month = apr,
  journal = {arXiv preprint},
  url = {https://arxiv.org/abs/1404.6659},
  eprint = {1404.6659},
  archiveprefix = {arXiv},
  primaryclass = {cond-mat.str-el},
}

@article{kong2014braided,
  title = {Braided fusion categories, gravitational anomalies, and the mathematical framework for topological orders in any dimensions},
  author = {Liang Kong and Xiao-Gang Wen},
  year = 2014,
  month = may,
  journal = {arXiv preprint},
  url = {https://arxiv.org/abs/1405.5858},
  eprint = {1405.5858},
  archiveprefix = {arXiv},
  primaryclass = {cond-mat.str-el},
}

@article{lu2012quantum,
  title = {Quantum phase transitions between bosonic symmetry-protected topological phases in two dimensions: Emergent {${\mathrm{QED}}_{3}$} and anyon superfluid},
  author = {Lu, Yuan-Ming and Lee, Dung-Hai},
  year = 2014,
  month = may,
  journal = {Physical Review B},
  publisher = {American Physical Society},
  volume = 89,
  pages = 195143,
  issue = 19,
  numpages = 11,
  doi = {10.1103/PhysRevB.89.195143},
  url = {https://arxiv.org/abs/1210.0909},
  eprint = {1210.0909},
  archiveprefix = {arXiv},
  primaryclass = {cond-mat.str-el},
}

@article{barkeshli2012continuous,
  title = {Continuous transition between fractional quantum {Hall} and superfluid states},
  author = {Barkeshli, Maissam and McGreevy, John},
  year = 2014,
  month = jun,
  journal = {Physical Review B},
  publisher = {American Physical Society},
  volume = 89,
  pages = 235116,
  issue = 23,
  numpages = 6,
  doi = {10.1103/PhysRevB.89.235116},
  url = {https://arxiv.org/abs/1201.4393},
  eprint = {1201.4393},
  archiveprefix = {arXiv},
  primaryclass = {cond-mat.str-el},
}

@article{freed2014shortrange,
  title = {Short-range entanglement and invertible field theories},
  author = {Daniel S. Freed},
  year = 2014,
  month = jun,
  journal = {arXiv preprint},
  url = {https://arxiv.org/abs/1406.7278},
  eprint = {1406.7278},
  archiveprefix = {arXiv},
  primaryclass = {cond-mat.str-el},
}

@article{dodooamoo2014nonuniversality,
  title = {Non-universality of scaling exponents in quantum {Hall} transitions},
  author = {Dodoo-Amoo, Nii A. and Saeed, Kashif and Mistry, Divyang and Khanna, Suraj P. and Li, Lianhe and Linfield, Edmund H. and Davies, A. Giles and Cunningham, John E.},
  year = 2014,
  month = oct,
  journal = {Journal of Physics: Condensed Matter},
  publisher = {IOP Publishing},
  volume = 26,
  number = 47,
  pages = 475801,
  doi = {10.1088/0953-8984/26/47/475801},
  url = {https://dx.doi.org/10.1088/0953-8984/26/47/475801},
}

@article{peterson2015abelian,
  title = {{Abelian} and non-{Abelian} states in $\nu=2/3$ bilayer fractional quantum {Hall} systems},
  author = {Peterson, Michael R. and Wu, Yang-Le and Cheng, Meng and Barkeshli, Maissam and Wang, Zhenghan and Das Sarma, Sankar},
  year = 2015,
  month = {JUL},
  journal = {Physical Review B},
  publisher = {American Physical Society},
  volume = 92,
  pages = {035103},
  issue = 3,
  numpages = 7,
  doi = {10.1103/PhysRevB.92.035103},
  url = {https://arxiv.org/abs/1502.02671},
  eprint = {1502.02671},
  archiveprefix = {arXiv},
  primaryclass = {cond-mat.str-el},
}

@article{geraedts2015competing,
  title = {Competing {Abelian} and non-{Abelian} topological orders in $\nu=1/3+1/3$ quantum {Hall} bilayers},
  author = {Scott Geraedts and Michael P. Zaletel and Zlatko Papić and Roger S. K. Mong},
  year = 2015,
  month = may,
  journal = {Physical Review B},
  publisher = {American Physical Society},
  volume = 91,
  pages = 205139,
  issue = 20,
  numpages = 16,
  doi = {10.1103/PhysRevB.91.205139},
  url = {https://arxiv.org/abs/1502.01340},
  eprint = {1502.01340},
  archiveprefix = {arXiv},
  primaryclass = {cond-mat.str-el},
}

@article{liu2015nonabelian,
  title = {Non-{Abelian} phases in two-component $\nu=2/3$ fractional quantum {Hall} states: Emergence of {Fibonacci} anyons},
  author = {Liu, Zhao and Vaezi, Abolhassan and Lee, Kyungmin and Kim, Eun-Ah},
  year = 2015,
  month = aug,
  journal = {Physical Review B},
  publisher = {American Physical Society},
  volume = 92,
  pages = {081102},
  issue = 8,
  numpages = 6,
  doi = {10.1103/PhysRevB.92.081102},
  url = {https://arxiv.org/abs/1502.05391},
  eprint = {1502.05391},
  archiveprefix = {arXiv},
  primaryclass = {cond-mat.str-el},
}

@article{son2015composite,
  title = {Is the Composite Fermion a {Dirac} Particle?},
  author = {Son, Dam Thanh},
  year = 2015,
  month = sep,
  journal = {Physical Review X},
  publisher = {American Physical Society},
  volume = 5,
  pages = {031027},
  issue = 3,
  numpages = 14,
  doi = {10.1103/PhysRevX.5.031027},
  url = {https://link.aps.org/doi/10.1103/PhysRevX.5.031027},
  eprint = {1502.03446},
  archiveprefix = {arXiv},
  primaryclass = {cond-mat.mes-hall},
}

@article{chester2016anomalous,
  title = {Anomalous dimensions of scalar operators in {QED}$_3$},
  author = {Chester, Shai M. and Pufu, Silviu S.},
  year = 2016,
  month = aug,
  journal = {Journal of High Energy Physics},
  volume = 2016,
  number = 8,
  pages = 69,
  doi = {10.1007/JHEP08(2016)069},
  isbn = {1029-8479},
  url = {https://doi.org/10.1007/JHEP08(2016)069},
  eprint = {1603.05582},
  archiveprefix = {arXiv},
  primaryclass = {hep-th},
}

@article{zucker2016phpfaffian,
  title = {Stabilization of the Particle-Hole {Pfaffian} Order by {Landau}-Level Mixing and Impurities That Break Particle-Hole Symmetry},
  author = {Zucker, Philip T. and Feldman, Dima E.},
  year = 2016,
  month = aug,
  journal = {Physical Review Letters},
  publisher = {American Physical Society},
  volume = 117,
  pages = {096802},
  issue = 9,
  numpages = 5,
  doi = {10.1103/PhysRevLett.117.096802},
  url = {https://link.aps.org/doi/10.1103/PhysRevLett.117.096802},
}

@article{hansson2016quantum,
  title = {Quantum {Hall} physics: Hierarchies and conformal field theory techniques},
  author = {Hansson, Thors H. and Hermanns, Maria and Simon, Steven H. and Viefers, Susanne F.},
  year = 2017,
  month = may,
  journal = {Reviews of Modern Physics},
  publisher = {American Physical Society},
  volume = 89,
  pages = {025005},
  issue = 2,
  numpages = 61,
  doi = {10.1103/RevModPhys.89.025005},
  url = {https://arxiv.org/abs/1601.01697},
  eprint = {1601.01697},
  archiveprefix = {arXiv},
  primaryclass = {cond-mat.str-el},
}

@article{thomson2017qed3,
  title = {Quantum electrodynamics in 2+1 dimensions with quenched disorder: Quantum critical states with interactions and disorder},
  author = {Thomson, Alex and Sachdev, Subir},
  year = 2017,
  month = jun,
  journal = {Physical Review B},
  publisher = {American Physical Society},
  volume = 95,
  pages = 235146,
  issue = 23,
  numpages = 32,
  doi = {10.1103/PhysRevB.95.235146},
  url = {https://arxiv.org/abs/1702.047236},
  eprint = {1702.04723},
  archiveprefix = {arXiv},
  primaryclass = {cond-mat.str-el},
}

@article{kane2017pairing,
  title = {Pairing in {Luttinger} Liquids and Quantum {Hall} States},
  author = {Kane, Charles L. and Stern, Ady and Halperin, Bertrand I.},
  year = 2017,
  month = jul,
  journal = {Physical Review X},
  publisher = {American Physical Society},
  volume = 7,
  pages = {031009},
  issue = 3,
  numpages = 17,
  doi = {10.1103/PhysRevX.7.031009},
  url = {https://link.aps.org/doi/10.1103/PhysRevX.7.031009},
  eprint = {1701.06200},
  archiveprefix = {arXiv},
  primaryclass = {cond-mat.str-el},
}

@article{zibrov2016robust,
  title = {Tunable interacting composite fermion phases in a half-filled bilayer-graphene {Landau} level},
  author = {Alexander A. Zibrov and Carlos. R. Kometter and Haoxin Zhou and Eric M. Spanton and Takashi Taniguchi and Kenji Watanabe and Michael P. Zaletel and Andrea F. Young},
  year = 2017,
  month = sep,
  journal = {Nature},
  volume = 549,
  number = 7672,
  pages = {360--364},
  doi = {10.1038/nature23893},
  isbn = {1476-4687},
  url = {https://arxiv.org/abs/1611.07113},
  eprint = {1611.07113},
  archiveprefix = {arXiv},
  primaryclass = {cond-mat.str-el},
}

@article{li2017denominator,
  title = {Even-denominator fractional quantum {Hall} states in bilayer graphene},
  author = {Jia I. A. Li and Cheng Tan and Shaowen Chen and Yihang Zeng and Takashi Taniguchi and Kenji Watanabe and James Hone and Cory R. Dean},
  year = 2017,
  month = oct,
  journal = {Science},
  volume = 358,
  number = 6363,
  pages = {648--652},
  doi = {10.1126/science.aao2521},
  url = {https://arxiv.org/abs/1705.07846},
  eprint = {1705.07846},
  archiveprefix = {arXiv},
  primaryclass = {cond-mat.mes-hall},
}

@article{dipientro2017epsilonqed,
  title = {Scaling dimensions in {QED3} from the $\epsilon$-expansion},
  author = {Di Pietro, Lorenzo and Stamou, Emmanuel},
  year = 2017,
  month = dec,
  journal = {Journal of High Energy Physics},
  volume = 2017,
  number = 12,
  pages = 54,
  doi = {10.1007/JHEP12(2017)054},
  isbn = {1029-8479},
  url = {https://doi.org/10.1007/JHEP12(2017)054},
  eprint = {1708.03740},
  archiveprefix = {arXiv},
  primaryclass = {hep-th},
}

@article{wen2017colloquium,
  title = {Colloquium: Zoo of quantum-topological phases of matter},
  author = {Wen, Xiao-Gang},
  year = 2017,
  month = dec,
  journal = {Review of Modern Physics},
  publisher = {American Physical Society},
  volume = 89,
  pages = {041004},
  issue = 4,
  numpages = 17,
  doi = {10.1103/RevModPhys.89.041004},
  url = {https://arxiv.org/abs/1610.03911},
  eprint = {1610.03911},
  archiveprefix = {arXiv},
  primaryclass = {cond-mat.str-el},
}

@article{burnell2017anyoncondensation,
  title = {Anyon Condensation and Its Applications},
  author = {Burnell, Fiona J.},
  year = 2018,
  journal = {Annual Review of Condensed Matter Physics},
  volume = 9,
  number = 1,
  pages = {307--327},
  doi = {10.1146/annurev-conmatphys-033117-054154},
  url = {https://doi.org/10.1146/annurev-conmatphys-033117-054154},
  eprint = {1706.04940},
  archiveprefix = {arXiv},
  primaryclass = {cond-mat.str-el},
}

@article{lee2018emergent,
  title = {Emergent Multi-Flavor {$\mathrm{QED}_{3}$} at the Plateau Transition between Fractional {Chern} Insulators: Applications to Graphene Heterostructures},
  author = {Lee, Jong Yeon and Wang, Chong and Zaletel, Michael P. and Vishwanath, Ashvin and He, Yin-Chen},
  year = 2018,
  month = jul,
  journal = {Physical Review X},
  publisher = {American Physical Society},
  volume = 8,
  pages = {031015},
  issue = 3,
  numpages = 21,
  doi = {10.1103/PhysRevX.8.031015},
  url = {https://link.aps.org/doi/10.1103/PhysRevX.8.031015},
  eprint = {1802.09538},
  archiveprefix = {arXiv},
  primaryclass = {cond-mat.str-el},
}

@article{mross2017theory,
  title = {Theory of Disorder-Induced Half-Integer Thermal {Hall} Conductance},
  author = {Mross, David F. and Oreg, Yuval and Stern, Ady and Margalit, Gilad and Heiblum, Moty},
  year = 2018,
  month = jul,
  journal = {Physical Review Letters},
  publisher = {American Physical Society},
  volume = 121,
  pages = {026801},
  issue = 2,
  numpages = 6,
  doi = {10.1103/PhysRevLett.121.026801},
  url = {https://arxiv.org/abs/1711.06278},
  eprint = {1711.06278},
  archiveprefix = {arXiv},
  primaryclass = {cond-mat.mes-hall},
}

@article{wang2017topological,
  title = {Topological order from disorder and the quantized {Hall} thermal metal: Possible applications to the $\nu=5/2$ state},
  author = {Wang, Chong and Vishwanath, Ashvin and Halperin, Bertrand I.},
  year = 2018,
  month = jul,
  journal = {Physical Review B},
  publisher = {American Physical Society},
  volume = 98,
  pages = {045112},
  issue = 4,
  numpages = 12,
  doi = {10.1103/PhysRevB.98.045112},
  url = {https://arxiv.org/abs/1711.11557},
  eprint = {1711.11557},
  archiveprefix = {arXiv},
  primaryclass = {cond-mat.str-el},
}

@article{balram2018anomalfqhe,
  title = {Fractional Quantum {Hall} Effect at {$\nu=2+6/13$}: The Parton Paradigm for the Second {Landau} Level},
  author = {Balram, Ajit C. and Mukherjee, Sutirtha and Park, Kwon and Barkeshli, Maissam and Rudner, Mark S. and Jain, Jainendra K.},
  year = 2018,
  month = nov,
  journal = {Physical Review Letters},
  publisher = {American Physical Society},
  volume = 121,
  pages = 186601,
  issue = 18,
  numpages = 6,
  doi = {10.1103/PhysRevLett.121.186601},
  url = {https://arxiv.org/abs/1807.02997},
  eprint = {1807.02997},
  archiveprefix = {arXiv},
  primaryclass = {cond-mat.str-el},
}

@article{benvenuti2019easyplane,
  title = {Easy-plane {QED}$_3$'s in the large {$N_f$} limit},
  author = {Benvenuti, Sergio and Khachatryan, Hrachya},
  year = 2019,
  month = may,
  journal = {Journal of High Energy Physics},
  volume = 2019,
  number = 5,
  pages = 214,
  doi = {10.1007/JHEP05(2019)214},
  isbn = {1029-8479},
  url = {https://doi.org/10.1007/JHEP05(2019)214},
  eprint = {1902.05767},
  archiveprefix = {arXiv},
  primaryclass = {hep-th},
}

@article{kumar2018symmetry,
  title = {Symmetry-enforced fractonicity and two-dimensional quantum crystal melting},
  author = {Kumar, Ajesh and Potter, Andrew C.},
  year = 2019,
  month = jul,
  journal = {Physical Review B},
  publisher = {American Physical Society},
  volume = 100,
  pages = {045119},
  issue = 4,
  numpages = 13,
  doi = {10.1103/PhysRevB.100.045119},
  url = {https://arxiv.org/abs/1808.05621},
  eprint = {1808.05621},
  archiveprefix = {arXiv},
  primaryclass = {cond-mat.str-el},
}

@article{spnsltt2019topological,
  title = {Topological Classification of Shot Noise on Fractional Quantum {Hall} Edges},
  author = {Sp\aa{}nsl\"att, Christian and Park, Jinhong and Gefen, Yuval and Mirlin, Alexander D.},
  year = 2019,
  month = sep,
  journal = {Physical Review Letters},
  publisher = {American Physical Society},
  volume = 123,
  pages = 137701,
  issue = 13,
  numpages = 6,
  doi = {10.1103/PhysRevLett.123.137701},
  url = {https://arxiv.org/abs/1906.05623},
  eprint = {1906.05623},
  archiveprefix = {arXiv},
  primaryclass = {cond-mat.mes-hall},
}

@article{simon2020pfaph,
  title = {Energetics of {Pfaffian}--{anti-Pfaffian} domains},
  author = {Simon, Steven H. and Ippoliti, Matteo and Zaletel, Michael P. and Rezayi, Edward H.},
  year = 2020,
  month = jan,
  journal = {Physical Review B},
  publisher = {American Physical Society},
  volume = 101,
  pages = {041302(R)},
  issue = 4,
  numpages = 4,
  doi = {10.1103/PhysRevB.101.041302},
  url = {https://arxiv.org/abs/1909.12844},
  eprint = {1909.12844},
  archiveprefix = {arXiv},
  primaryclass = {cond-mat.mes-hall},
}

@article{ma2020emergent,
  title = {Emergent {$\mathrm{QCD}_{3}$} quantum phase transitions of fractional {Chern} insulators},
  author = {Ma, Ruochen and He, Yin-Chen},
  year = 2020,
  month = sep,
  journal = {Physical Review Research},
  publisher = {American Physical Society},
  volume = 2,
  pages = {033348},
  issue = 3,
  numpages = 19,
  doi = {10.1103/PhysRevResearch.2.033348},
  url = {https://link.aps.org/doi/10.1103/PhysRevResearch.2.033348},
  eprint = {2003.05954},
  archiveprefix = {arXiv},
  primaryclass = {cond-mat.str-el},
}

@article{hsin2020eftfqh,
  title = {Effective field theory for fractional quantum {Hall} systems near $\nu=5/2$},
  author = {Hsin, Po-Shen and Lin, Ying-Hsuan and Paquette, Natalie M. and Wang, Juven},
  year = 2020,
  month = nov,
  journal = {Physical Review Research},
  publisher = {American Physical Society},
  volume = 2,
  pages = {043242},
  issue = 4,
  numpages = 31,
  doi = {10.1103/PhysRevResearch.2.043242},
  url = {https://link.aps.org/doi/10.1103/PhysRevResearch.2.043242},
  eprint = {2005.10826},
  archiveprefix = {arXiv},
  primaryclass = {cond-mat.str-el},
}

@article{wang2020abelian,
  title = {In and around abelian anyon models},
  author = {Wang, Liang and Wang, Zhenghan},
  year = 2020,
  month = {nov},
  journal = {Journal of Physics A: Mathematical and Theoretical},
  publisher = {IOP Publishing},
  volume = 53,
  number = 50,
  pages = 505203,
  doi = {10.1088/1751-8121/abc6c0},
  url = {https://arxiv.org/abs/2004.12048},
  eprint = {2004.12048},
  archiveprefix = {arXiv},
  primaryclass = {math.QA},
}

@article{sbierski2020criticality,
  title = {Criticality of Two-Dimensional Disordered {Dirac} Fermions in the Unitary Class and Universality of the Integer Quantum {Hall} Transition},
  author = {Sbierski, Bj\"orn and Dresselhaus, Elizabeth J. and Moore, Joel E. and Gruzberg, Ilya A.},
  year = 2021,
  month = feb,
  journal = {Physical Review Letters},
  publisher = {American Physical Society},
  volume = 126,
  pages = {076801},
  issue = 7,
  numpages = 6,
  doi = {10.1103/PhysRevLett.126.076801},
  url = {https://arxiv.org/abs/2008.09025},
  eprint = {2008.09025},
  archiveprefix = {arXiv},
  primaryclass = {cond-mat.mes-hall},
}

@article{freed2021reflection,
  title = {Reflection positivity and invertible topological phases},
  author = {Freed, Daniel S. and Hopkins, Michael J.},
  year = 2021,
  month = may,
  journal = {Geometry \& Topology},
  publisher = {Mathematical Sciences Publishers},
  volume = 25,
  number = 3,
  pages = {1165--1330},
  doi = {10.2140/gt.2021.25.1165},
  issn = {1465-3060},
  eprint = {1604.06527},
  archiveprefix = {arXiv},
  primaryclass = {hep-th},
}

@article{haldane2021graviton,
  title = {Graviton chirality and topological order in the half-filled {Landau} level},
  author = {Haldane, F. Duncan M. and Rezayi, Edward H. and Yang, Kun},
  year = 2021,
  month = sep,
  journal = {Physical Review B},
  publisher = {American Physical Society},
  volume = 104,
  pages = {L121106},
  issue = 12,
  numpages = 6,
  doi = {10.1103/PhysRevB.104.L121106},
  url = {https://arxiv.org/abs/2103.11019},
  eprint = {2103.11019},
  archiveprefix = {arXiv},
  primaryclass = {cond-mat.mes-hall},
}

@article{yutushui2021identifying,
  title = {Identifying the $\nu=\frac{5}{2}$  Topological Order through Charge Transport Measurements},
  author = {Yutushui, Misha and Stern, Ady and Mross, David F.},
  year = 2022,
  month = jan,
  journal = {Physical Review Letters},
  publisher = {American Physical Society},
  volume = 128,
  pages = {016401},
  issue = 1,
  numpages = 6,
  doi = {10.1103/PhysRevLett.128.016401},
  url = {https://arxiv.org/abs/2106.07667},
  eprint = {2106.07667},
  archiveprefix = {arXiv},
  primaryclass = {cond-mat.str-el},
}

@article{pu2021anderson,
  title = {Anderson Localization in the Fractional Quantum {Hall} Effect},
  author = {Pu, Songyang and Sreejith, Ganesh J. and Jain, Jainendra K.},
  year = 2022,
  month = mar,
  journal = {Physical Review Letters},
  publisher = {American Physical Society},
  volume = 128,
  pages = 116801,
  issue = 11,
  numpages = 6,
  doi = {10.1103/PhysRevLett.128.116801},
  url = {https://link.aps.org/doi/10.1103/PhysRevLett.128.116801},
  eprint = {2109.00362},
  archiveprefix = {arXiv},
  primaryclass = {cond-mat.mes-hall},
}

@article{schiller2021extracting,
  title = {Extracting the scaling dimension of quantum {Hall} quasiparticles from current correlations},
  author = {Schiller, Noam and Oreg, Yuval and Snizhko, Kyrylo},
  year = 2022,
  month = apr,
  journal = {Physical Review B},
  publisher = {American Physical Society},
  volume = 105,
  pages = 165150,
  issue = 16,
  numpages = 17,
  doi = {10.1103/PhysRevB.105.165150},
  url = {https://arxiv.org/abs/2111.05399},
  eprint = {2111.05399},
  archiveprefix = {arXiv},
  primaryclass = {cond-mat.mes-hall},
}

@article{barkeshli2021classification,
  title = {Classification of $(2+1)${D} invertible fermionic topological phases with symmetry},
  author = {Barkeshli, Maissam and Chen, Yu-An and Hsin, Po-Shen and Manjunath, Naren},
  year = 2022,
  month = jun,
  journal = {Physical Review B},
  publisher = {American Physical Society},
  volume = 105,
  pages = 235143,
  issue = 23,
  numpages = 60,
  doi = {10.1103/PhysRevB.105.235143},
  url = {https://arxiv.org/abs/2109.11039},
  eprint = {2109.11039},
  archiveprefix = {arXiv},
  primaryclass = {cond-mat.str-el},
}

@article{huang2021valley,
  title = {Valley Isospin Controlled Fractional Quantum {Hall} States in Bilayer Graphene},
  author = {Huang, Ke and Fu, Hailong and Hickey, Danielle Reifsnyder and Alem, Nasim and Lin, Xi and Watanabe, Kenji and Taniguchi, Takashi and Zhu, Jun},
  year = 2022,
  month = jul,
  journal = {Physical Review X},
  publisher = {American Physical Society},
  volume = 12,
  pages = {031019},
  issue = 3,
  numpages = 14,
  doi = {10.1103/PhysRevX.12.031019},
  url = {https://link.aps.org/doi/10.1103/PhysRevX.12.031019},
  eprint = {2105.07058},
  archiveprefix = {arXiv},
  primaryclass = {cond-mat.mes-hall},
}

@article{kumar2020interaction,
  title = {Interaction effects on quantum {Hall} transitions: Dynamical scaling laws and superuniversality},
  author = {Kumar, Prashant and Nosov, Pavel A. and Raghu, Srinivas},
  year = 2022,
  month = aug,
  journal = {Physical Review Research},
  publisher = {American Physical Society},
  volume = 4,
  pages = {033146},
  issue = 3,
  numpages = 12,
  doi = {10.1103/PhysRevResearch.4.033146},
  url = {https://link.aps.org/doi/10.1103/PhysRevResearch.4.033146},
  eprint = {2006.11862},
  archiveprefix = {arXiv},
  primaryclass = {cond-mat.str-el},
}

@article{kaidi2021higher,
  title = {Higher central charges and topological boundaries in 2+1-dimensional  {TQFTs}},
  author = {Justin Kaidi and Zohar Komargodski and Kantaro Ohmori and Sahand Seifnashri and Shu-Heng Shao},
  year = 2022,
  month = sep,
  journal = {SciPost Physics},
  publisher = {SciPost},
  volume = 13,
  pages = {067},
  doi = {10.21468/SciPostPhys.13.3.067},
  url = {https://scipost.org/10.21468/SciPostPhys.13.3.067},
  eprint = {2107.13091},
  archiveprefix = {arXiv},
  primaryclass = {hep-th},
}

@article{lee2022nonabelian,
  title = {Non-{Abelian} anyon collider},
  author = {Lee, June-Young M. and Sim, Heung-Sun},
  year = 2022,
  month = nov,
  journal = {Nature Communications},
  volume = 13,
  number = 1,
  pages = 6660,
  doi = {10.1038/s41467-022-34329-y},
  isbn = {2041-1723},
  url = {https://doi.org/10.1038/s41467-022-34329-y},
  eprint = {2202.03649},
  archiveprefix = {arXiv},
  primaryclass = {cond-mat.mes-hall},
}

@article{wei2023thermalinterferometry,
  title = {Thermal interferometry of anyons},
  author = {Wei, Zezhu and Batra, Navketan and Mitrovi\'{c}, Vesna F. and Feldman, Dima E.},
  year = 2023,
  month = mar,
  journal = {Physical Review B},
  publisher = {American Physical Society},
  volume = 107,
  pages = 104406,
  issue = 10,
  numpages = 32,
  doi = {10.1103/PhysRevB.107.104406},
  url = {https://arxiv.org/abs/2209.06234},
  eprint = {2209.06234},
  archiveprefix = {arXiv},
  primaryclass = {cond-mat.mes-hall},
}

@article{hein2022thermal,
  title = {Thermal conductance and noise of {Majorana} modes along interfaced $\nu=\frac{5}{2}$ fractional quantum {Hall} states},
  author = {Hein, Michael and Sp\aa{}nsl\"att, Christian},
  year = 2023,
  month = jun,
  journal = {Physical Review B},
  publisher = {American Physical Society},
  volume = 107,
  pages = 245301,
  issue = 24,
  numpages = 25,
  doi = {10.1103/PhysRevB.107.245301},
  url = {https://arxiv.org/abs/2211.08000},
  eprint = {2211.08000},
  archiveprefix = {arXiv},
  primaryclass = {cond-mat.mes-hall},
}

@article{madathil2023delocalization,
  title = {Delocalization and Universality of the Fractional Quantum {Hall} Plateau-to-Plateau Transitions},
  author = {Madathil, Pranav T. and Villegas Rosales, Kevin A. and Tai, Tai T. and Chung, Yoon J. and Pfeiffer, Loren N. and West, Kirk W. and Baldwin, Kirk W. and Shayegan, Mansour},
  year = 2023,
  month = jun,
  journal = {Physical Review Letters},
  publisher = {American Physical Society},
  volume = 130,
  pages = 226503,
  issue = 22,
  numpages = 6,
  doi = {10.1103/PhysRevLett.130.226503},
  url = {https://arxiv.org/abs/2306.03704},
  eprint = {2306.03704},
  archiveprefix = {arXiv},
  primaryclass = {cond-mat.mes-hall},
}

@article{tantivasadakarn2022hierarchy,
  title = {Hierarchy of Topological Order From Finite-Depth Unitaries, Measurement, and Feedforward},
  author = {Tantivasadakarn, Nathanan and Vishwanath, Ashvin and Verresen, Ruben},
  year = 2023,
  month = jun,
  journal = {PRX Quantum},
  publisher = {American Physical Society},
  volume = 4,
  pages = {020339},
  issue = 2,
  numpages = 25,
  doi = {10.1103/PRXQuantum.4.020339},
  url = {https://link.aps.org/doi/10.1103/PRXQuantum.4.020339},
  eprint = {2209.06202},
  archiveprefix = {arXiv},
  primaryclass = {quant-ph},
}

@article{nguyen2023quantum,
  title = {On quantum melting of superfluid vortex crystals: from {Lifshitz} scalar to dual gravity},
  author = {Dung Xuan Nguyen and Sergej Moroz},
  year = 2023,
  month = oct,
  journal = {arXiv preprint},
  publisher = {SciPost},
  volume = 17,
  pages = 164,
  doi = {10.21468/SciPostPhys.17.6.164},
  url = {https://arxiv.org/abs/2310.13741},
  eprint = {2310.13741},
  archiveprefix = {arXiv},
  primaryclass = {cond-mat.quant-gas},
}

@article{assouline2023energy,
  title = {Energy gap of the even-denominator fractional quantum {Hall} state in bilayer graphene},
  author = {Assouline, Alexandre and Wang, Taige and Zhou, Haoxin and Cohen, Liam A. and Yang, Fangyuan and Zhang, Ruining and Taniguchi, Takashi and Watanabe, Kenji and Mong, Roger S. K. and Zaletel, Michael P. and Young, Andrea F.},
  year = 2024,
  month = jan,
  journal = {Physical Review Letters},
  publisher = {American Physical Society},
  volume = 132,
  pages = {046603},
  issue = 4,
  numpages = 6,
  doi = {10.1103/PhysRevLett.132.046603},
  url = {https://arxiv.org/abs/2308.05729},
  eprint = {2308.05729},
  archiveprefix = {arXiv},
  primaryclass = {cond-mat.mes-hall},
}

@article{sharma2023compositefermion,
  title = {Composite-fermion pairing at half-filled and quarter-filled lowest Landau level},
  author = {Sharma, Anirban and Balram, Ajit C. and Jain, Jainendra K.},
  year = 2024,
  month = jan,
  journal = {Physical Review B},
  publisher = {American Physical Society},
  volume = 109,
  pages = {035306},
  issue = 3,
  numpages = 22,
  doi = {10.1103/PhysRevB.109.035306},
  url = {https://arxiv.org/abs/2311.05083},
  eprint = {2311.05083},
  archiveprefix = {arXiv},
  primaryclass = {cond-mat.str-el},
}

@article{park2024fingerprints,
  title = {Fingerprints of anti-{Pfaffian} topological order in quantum point contact transport},
  author = {Park, Jinhong and Sp\aa{}nsl\"att, Christian and Mirlin, Alexander D.},
  year = 2024,
  month = jun,
  journal = {Physical Review Letters},
  publisher = {American Physical Society},
  volume = 132,
  pages = 256601,
  issue = 25,
  doi = {10.1103/PhysRevLett.132.256601},
  url = {https://arxiv.org/abs/2402.02157},
  eprint = {2402.02157},
  archiveprefix = {arXiv},
  primaryclass = {cond-mat.mes-hall},
}

@article{andrews2023localization,
  title = {Localization renormalization and quantum {Hall} systems},
  author = {Andrews, Bartholomew and Reiss, Dominic and Harper, Fenner and Roy, Rahul},
  year = 2024,
  month = mar,
  journal = {Physical Review B},
  publisher = {American Physical Society},
  volume = 109,
  pages = 125132,
  issue = 12,
  numpages = 10,
  doi = {10.1103/PhysRevB.109.125132},
  url = {https://arxiv.org/abs/2310.14074},
  eprint = {2310.14074},
  archiveprefix = {arXiv},
  primaryclass = {cond-mat.mes-hall},
}

@article{manna2022classification,
  title = {Full Classification of Transport on an Equilibrated $5/2$ Edge via Shot Noise},
  author = {Manna, Sourav and Das, Ankur and Goldstein, Moshe and Gefen, Yuval},
  year = 2024,
  month = mar,
  journal = {Physical Review Letters},
  publisher = {American Physical Society},
  volume = 132,
  pages = 136502,
  issue = 13,
  numpages = 7,
  doi = {10.1103/PhysRevLett.132.136502},
  url = {https://arxiv.org/abs/2212.05732},
  eprint = {2212.05732},
  archiveprefix = {arXiv},
  primaryclass = {cond-mat.mes-hall},
}

@article{kumar2024quarter,
  title = {Quarter- and half-filled quantum {Hall} states and their competing interactions in bilayer graphene},
  author = {Ravi Kumar and André Haug and Jehyun Kim and Misha Yutushui and Konstantin Khudiakov and Vishal Bhardwaj and Alexey Ilin and Kenji Watanabe and Takashi Taniguchi and David F. Mross and Yuval Ronen},
  year = 2024,
  month = may,
  journal = {arXiv preprint},
  volume = 16,
  number = 1,
  pages = 7255,
  doi = {10.1038/s41467-025-62650-9},
  isbn = {2041-1723},
  url = {https://arxiv.org/abs/2405.19405},
  eprint = {2405.19405},
  archiveprefix = {arXiv},
  primaryclass = {cond-mat.mes-hall},
}

@article{singh2023topological,
  title = {Topological phase transition between {Jain} states and daughter states of the $\nu = 1/2$  fractional quantum {Hall} state},
  author = {Siddharth Kumar Singh and Chengyu Wang and Chia-Tse Tai and Casey S. Calhoun and Adbhut Gupta and Kirk W. Baldwin and Loren N. Pfeiffer and Mansour Shayegan},
  year = 2024,
  month = may,
  journal = {Nature Physics},
  volume = 20,
  number = 8,
  pages = {1247--1252},
  doi = {10.1038/s41567-024-02517-w},
  isbn = {1745-2481},
  url = {https://arxiv.org/abs/2309.00111},
  eprint = {2309.00111},
  archiveprefix = {arXiv},
  primaryclass = {cond-mat.mes-hall},
}

@article{zhang2024hierarchy,
  title = {Hierarchy construction for non-abelian fractional quantum {Hall} states via anyon condensation},
  author = {Carolyn Zhang and Ashvin Vishwanath and Xiao-Gang Wen},
  year = 2024,
  month = jun,
  journal = {arXiv preprint},
  publisher = {American Physical Society},
  volume = 112,
  pages = 125116,
  issue = 12,
  numpages = 25,
  doi = {10.1103/jndb-435f},
  url = {https://arxiv.org/abs/2406.12068},
  eprint = {2406.12068},
  archiveprefix = {arXiv},
  primaryclass = {cond-mat.str-el},
}

@article{du2023quantum,
  title = {Quantum vortex lattice: {Lifshitz} duality, topological defects, and multipole symmetries},
  author = {Du, Yi-Hsien and Lam, Ho Tat and Radzihovsky, Leo},
  year = 2024,
  month = jul,
  journal = {Physical Review B},
  publisher = {American Physical Society},
  volume = 110,
  pages = {035164},
  issue = 3,
  numpages = 16,
  doi = {10.1103/PhysRevB.110.035164},
  url = {https://arxiv.org/abs/2310.13794},
  eprint = {2310.13794},
  archiveprefix = {arXiv},
  primaryclass = {cond-mat.str-el},
}

@article{han2024interferometry,
  title = {Anyon Interferometry to Detect Braiding Statistics of Neutral Modes},
  author = {Han, Cheolhee and Lee, June-Young M. and Sim, Heung-Sun},
  year = 2024,
  month = oct,
  journal = {Physical Reviev Letters},
  publisher = {American Physical Society},
  volume = 133,
  pages = 186603,
  issue = 18,
  numpages = 7,
  doi = {10.1103/PhysRevLett.133.186603},
  url = {https://link.aps.org/doi/10.1103/PhysRevLett.133.186603},
}

@article{kaur2024qhtransition,
  title = {Universality of quantum phase transitions in the integer and fractional quantum {Hall} regimes},
  author = {Kaur, Simrandeep and Chanda, Tanima and Amin, Kazi Rafsanjani and Sahani, Divya and Watanabe, Kenji and Taniguchi, Takashi and Ghorai, Unmesh and Gefen, Yuval and Sreejith, Ganesh Jaya and Bid, Aveek},
  year = 2024,
  month = oct,
  journal = {Nature Communications},
  volume = 15,
  number = 1,
  pages = 8535,
  doi = {10.1038/s41467-024-52927-w},
  isbn = {2041-1723},
  url = {https://doi.org/10.1038/s41467-024-52927-w},
  eprint = {2312.06194},
  archiveprefix = {arXiv},
  primaryclass = {cond-mat.mes-hall},
}

@article{yutushui2024paired,
  title = {Paired fermions in strong magnetic fields and daughters of even-denominator {Hall} plateaus},
  author = {Yutushui, Misha and Hermanns, Maria and Mross, David F.},
  year = 2024,
  month = oct,
  journal = {Physical Review B},
  publisher = {American Physical Society},
  volume = 110,
  pages = 165402,
  issue = 16,
  numpages = 19,
  doi = {10.1103/PhysRevB.110.165402},
  url = {https://arxiv.org/abs/2405.03753},
  eprint = {2405.03753},
  archiveprefix = {arXiv},
  primaryclass = {cond-mat.str-el},
}

@article{zheltonozhskii2024identifying,
  title = {Identifying the topological order of quantized half-filled {Landau} levels through their daughter states},
  author = {Evgenii Zheltonozhskii and Stern, Ady and Lindner, Netanel H.},
  year = 2024,
  month = dec,
  journal = {Physical Review B},
  publisher = {American Physical Society},
  volume = 110,
  pages = 245140,
  issue = 24,
  numpages = 7,
  doi = {10.1103/PhysRevB.110.245140},
  url = {https://link.aps.org/doi/10.1103/PhysRevB.110.245140},
  eprint = {2405.03780},
  archiveprefix = {arXiv},
  primaryclass = {cond-mat.mes-hall},
  code = {https://github.com/Randl/daughter-states-half-filling},
  arxiv = {2405.03780},
}

@article{hu2024studying,
  title = {High-resolution tunnelling spectroscopy of fractional quantum {Hall} states},
  author = {Hu, Yuwen and Tsui, Yen-Chen and He, Minhao and Kamber, Umut and Wang, Taige and Mohammadi, Amir S. and Watanabe, Kenji and Taniguchi, Takashi and Papi{\'c}, Zlatko and Zaletel, Michael P. and Yazdani, Ali},
  year = 2025,
  month = may,
  journal = {Nature Physics},
  volume = 21,
  number = 5,
  pages = {716--723},
  doi = {10.1038/s41567-025-02830-y},
  isbn = {1745-2481},
  url = {https://arxiv.org/abs/2308.05789},
  eprint = {2308.05789},
  archiveprefix = {arXiv},
  primaryclass = {cond-mat.mes-hall},
}

@article{musser2024fractionalization,
  title = {Fractionalization as an alternate to charge ordering in electronic insulators},
  author = {Musser, Seth and Cheng, Meng and Senthil, Todadri},
  year = 2025,
  month = jun,
  journal = {Physical Review B},
  publisher = {American Physical Society},
  volume = 111,
  pages = 235108,
  issue = 23,
  numpages = 26,
  doi = {10.1103/PhysRevB.111.235108},
  url = {https://arxiv.org/abs/2408.03984},
  eprint = {2408.03984},
  archiveprefix = {arXiv},
  primaryclass = {cond-mat.str-el},
}

@article{lotric2025reconstructionnu52,
  title = {Majorana edge reconstruction and the $\nu=5/2$ non-{Abelian} thermal {Hall} puzzle},
  author = {Tevž Lotrič and Taige Wang and Michael P. Zaletel and Steven H. Simon and S. A. Parameswaran},
  year = 2025,
  month = jul,
  journal = {arXiv preprint},
  url = {https://arxiv.org/abs/2507.07161},
  eprint = {2507.07161},
  archiveprefix = {arXiv},
  primaryclass = {cond-mat.mes-hall},
}

@article{singh2025nonabeliantransition,
  title = {Fractional Quantum {Hall} State at $\nu=1/2$ with Energy Gap Up to 6 {K} and Possible Transition from the One- to Two-Component State},
  author = {Singh, Siddharth Kumar and Wang, Chengyu and Gupta, Adbhut and Baldwin, Kirk W. and Pfeiffer, Loren N. and Shayegan, Mansour},
  year = 2025,
  month = dec,
  journal = {Physical Review Letters},
  publisher = {American Physical Society},
  volume = 135,
  pages = 246603,
  issue = 24,
  numpages = 8,
  doi = {10.1103/ywpx-qm7d},
  url = {https://arxiv.org/abs/2510.03983},
  eprint = {2510.03983},
  archiveprefix = {arXiv},
  primaryclass = {cond-mat.mes-hall},
}

@article{cheng2025ordering,
  title = {Ordering the topological order in the fractional quantum {Hall} effect},
  author = {Meng Cheng and Seth Musser and Amir Raz and Nathan Seiberg and Todadri Senthil},
  year = 2026,
  month = mar,
  journal = {Physical Review B},
  publisher = {American Physical Society},
  volume = 113,
  pages = 115103,
  issue = 11,
  numpages = 45,
  doi = {10.1103/7gtq-y9d9},
  url = {https://arxiv.org/abs/2505.14767},
  eprint = {2505.14767},
  archiveprefix = {arXiv},
  primaryclass = {cond-mat.str-el},
}

@article{lotri2025paired,
  title = {Paired Parton Trial States for the Superfluid-Fractional {Chern} Insulator Transition},
  author = {Tevž Lotrič and Steven H. Simon},
  year = 2026,
  month = mar,
  journal = {Physical Review Letters},
  publisher = {American Physical Society},
  volume = 136,
  pages = {096601},
  issue = 9,
  numpages = 6,
  doi = {10.1103/l9rp-y68y},
  url = {https://arxiv.org/abs/2504.20139},
  eprint = {2504.20139},
  archiveprefix = {arXiv},
  primaryclass = {cond-mat.str-el},
}

@article{chanda2025denominator,
  title = {Even Denominator Fractional Quantum {Hall} States in the Zeroth {Landau} Level of {ABA} Trilayer Graphene},
  author = {Chanda, Tanima and Kaur, Simrandeep and Singh, Harsimran and Watanabe, Kenji and Taniguchi, Takashi and Jain, Manish and Khanna, Udit and Balram, Ajit C. and Bid, Aveek},
  year = 2026,
  month = jul,
  journal = {Physical Review Letters},
  publisher = {American Physical Society},
  volume = 137,
  pages = {036601},
  issue = 3,
  numpages = 8,
  doi = {10.1103/tsnc-4jjl},
  url = {https://arxiv.org/abs/2502.06245},
  eprint = {2502.06245},
  archiveprefix = {arXiv},
  primaryclass = {cond-mat.mes-hall},
}

@book{cardy1996scaling,
  title = {Scaling and Renormalization in Statistical Physics},
  author = {Cardy, John},
  year = 1996,
  publisher = {Cambridge University Press},
  series = {Cambridge Lecture Notes in Physics},
  doi = {10.1017/CBO9781316036440},
  place = {Cambridge},
  collection = {Cambridge Lecture Notes in Physics},
}

@book{difrancesco1997cft,
  title = {Conformal Field Theory},
  author = {Di Francesco, Philippe and Mathieu, Pierre and S{\'e}n{\'e}chal, David},
  year = 1997,
  publisher = {Springer Science \& Business Media},
  doi = {10.1007/978-1-4612-2256-9},
  isbn = 9781461222569,
}

@book{jain2007cfbook,
  title = {Composite Fermions},
  author = {Jain, Jainendra K.},
  year = 2007,
  publisher = {Cambridge University Press},
  doi = {10.1017/CBO9780511607561},
  isbn = 9780511607561,
  place = {Cambridge},
}

@book{carr2010understanding,
  title = {Understanding quantum phase transitions},
  author = {Carr, Lincoln},
  year = 2010,
  publisher = {CRC press},
  doi = {10.1201/b10273},
}

@book{sachdev2011qpt,
  title = {Quantum Phase Transitions},
  author = {Sachdev, Subir},
  year = 2011,
  month = may,
  publisher = {Cambridge University Press},
  doi = {10.1017/CBO9780511973765},
  place = {Cambridge},
  edition = 2,
}

@book{tao2012topics,
  title = {Topics in random matrix theory},
  author = {Tao, Terence},
  year = 2012,
  publisher = {American Mathematical Society},
  volume = 132,
  url = {https://bookstore.ams.org/gsm-132/},
}

@incollection{ma2022fractional,
  title = {Fractional quantum {Hall} effect at the filling factor $\nu$ = 5/2},
  author = {Ken K.W. Ma and Michael R. Peterson and Vito W. Scarola and Kun Yang},
  year = 2024,
  booktitle = {Encyclopedia of Condensed Matter Physics},
  publisher = {Academic Press},
  address = {Oxford},
  pages = {324--365},
  doi = {https://doi.org/10.1016/B978-0-323-90800-9.00135-9},
  isbn = {978-0-323-91408-6},
  url = {https://arxiv.org/abs/2208.07908},
  eprint = {2208.07908},
  archiveprefix = {arXiv},
  primaryclass = {cond-mat.mes-hall},
  editor = {Tapash Chakraborty},
  edition = {Second},
}

@article{rosenow2010neutral,
  title = {Signatures of neutral quantum {Hall} modes in transport through low-density constrictions},
  author = {Rosenow, Bernd and Halperin, Bertrand I.},
  year = 2010,
  month = apr,
  journal = {Physical Review B},
  publisher = {American Physical Society},
  volume = 81,
  pages = 165313,
  issue = 16,
  doi = {10.1103/PhysRevB.81.165313},
  url = {https://arxiv.org/abs/0806.0869},
  eprint = {0806.0869},
  archiveprefix = {arXiv},
  primaryclass = {cond-mat.mes-hall},
}

@article{takei2011neutral,
  title = {Neutral mode heat transport and fractional quantum {Hall} shot noise},
  author = {Takei, So and Rosenow, Bernd},
  year = 2011,
  month = dec,
  journal = {Physical Review B},
  publisher = {American Physical Society},
  volume = 84,
  pages = 235316,
  issue = 23,
  doi = {10.1103/PhysRevB.84.235316},
  url = {https://arxiv.org/abs/1012.0315},
  eprint = {1012.0315},
  archiveprefix = {arXiv},
  primaryclass = {cond-mat.mes-hall},
}

@article{spnsltt2020conductance,
  title = {Conductance plateaus and shot noise in fractional quantum {Hall} point contacts},
  author = {Sp\aa{}nsl\"att, Christian and Park, Jinhong and Gefen, Yuval and Mirlin, Alexander D.},
  year = 2020,
  month = feb,
  journal = {Physical Review B},
  publisher = {American Physical Society},
  volume = 101,
  pages = {075308},
  issue = 7,
  doi = {10.1103/PhysRevB.101.075308},
  url = {https://arxiv.org/abs/1911.11821},
  eprint = {1911.11821},
  archiveprefix = {arXiv},
  primaryclass = {cond-mat.mes-hall},
}

@article{park2020noise,
  title = {Noise on the non-{Abelian} $\nu=5/2$ Fractional Quantum {Hall} Edge},
  author = {Park, Jinhong and Sp\aa{}nsl\"att, Christian and Gefen, Yuval and Mirlin, Alexander D.},
  year = 2020,
  month = oct,
  journal = {Physical Review Letters},
  publisher = {American Physical Society},
  volume = 125,
  pages = 157702,
  issue = 15,
  doi = {10.1103/PhysRevLett.125.157702},
  url = {https://arxiv.org/abs/2006.06018},
  eprint = {2006.06018},
  archiveprefix = {arXiv},
  primaryclass = {cond-mat.mes-hall},
}

@article{das2022current,
  title = {dc electrical current generated by upstream neutral modes},
  author = {Das, Ankur and Rao, Sumathi and Gefen, Yuval and Murthy, Ganpathy},
  year = 2022,
  month = apr,
  journal = {Physical Review B},
  publisher = {American Physical Society},
  volume = 105,
  pages = 165154,
  issue = 16,
  doi = {10.1103/PhysRevB.105.165154},
  url = {https://arxiv.org/abs/2103.06200},
  eprint = {2103.06200},
  archiveprefix = {arXiv},
  primaryclass = {cond-mat.mes-hall},
}

@article{spnsltt2026bosonization,
  title = {Nonequilibrium bosonization of fractional quantum {Hall} edges},
  author = {Sp\aa{}nsl\"att, Christian and Park, Jinhong and Mirlin, Alexander D.},
  year = 2026,
  month = aug,
  journal = {Physical Review B},
  publisher = {American Physical Society},
  volume = 114,
  pages = 105418,
  issue = 10,
  doi = {10.1103/3js1-qd4m},
  url = {https://arxiv.org/abs/2603.05088},
  eprint = {2603.05088},
  archiveprefix = {arXiv},
  primaryclass = {cond-mat.mes-hall},
}

@article{vinjamuri2025shotnoise,
  title = {Shot noise signatures of candidate states for the fractional quantum {Hall} $\nu=12/5$ state},
  author = {Vinjamuri, Goutham and Das, Ankur},
  year = 2025,
  month = oct,
  journal = {Physical Review B},
  publisher = {American Physical Society},
  volume = 112,
  pages = 155117,
  issue = 15,
  doi = {10.1103/tp7s-nf8b},
  url = {https://arxiv.org/abs/2504.06226},
  eprint = {2504.06226},
  archiveprefix = {arXiv},
  primaryclass = {cond-mat.mes-hall},
}

\clearpage

\appendix
\setcounter{secnumdepth}{4}
\crefalias{section}{appendix}
\onecolumngrid

% %%% FIGURE NUMBERING IN APPENDIX
% \renewcommand\thefigure{A.\arabic{figure}} 
% \renewcommand\thetable{A.\arabic{table}} 
% \renewcommand\theequation{A.\arabic{equation}}  

% % % \numberwithin{equation}{section}
% % % \numberwithin{figure}{section}
% % % \numberwithin{table}{section}

% \setcounter{figure}{0}  
% \setcounter{table}{0}
% \setcounter{equation}{0}

\section{Parton construction in presence $C=0$ parton}
\label{app:c0}
Consider $N_{\text{p}}$ partons coupled to $N_{\text{g}}$ emergent gauge fields $a_j$ and to the external electromagnetic field $A$. The gauge field seen by parton $i$ is
\begin{align}
B_i = Q_{ij}a_j+t_i A ,
\end{align}
where $Q$ is an $N_{\text{p}}\times N_{\text{g}}$ charge matrix and $t$ is the electromagnetic charge vector (which dictates the coupling of the partons to the external field $A$).

Throughout this appendix, we distinguish between $(2+1)$D spacetime components, denoted by Greek indices (e.g., $\mu, \nu, \rho \in \{0, x, y\}$), and 2D spatial components, denoted by Latin indices (e.g., $m, n, p \in \{x, y\}$). Let $\Pi_i^{mn}(\omega)$ be the spatial current-current correlator at $\vb{q}=0$ for parton $i$, defined via linear response as
\begin{align}
\Pi_i^{mn}(\omega) = \expval {j_i^m(\omega) j_i^n(-\omega) }.
\end{align}
The current response of a gapped state at $\mathbf{q}=0$ has the form
\begin{align}
\Pi_i^{mn}(\omega) = \chi_i\omega^2\delta^{mn} + \frac{i\omega}{2\pi}\mathcal{K}_{ii}\epsilon^{mn} + \order{\omega^3}, \qquad m,n=x,y.
\end{align}
Here, $\mathcal{K}_{ii} \in \qty{\pm  1, 0}$ is the Chern number of the parton $f_i$. A parton is termed trivial if $\mathcal{K}_{ii}=0$, and topological otherwise. The $\chi_i\omega^2$ term is the ordinary insulating response, where $\chi_i \in \mathbb{R}$ is a real number representing the electric susceptibility.

We introduce a hydrodynamic field $\beta_i$ by
\begin{align}
j_i^\mu = \frac{1}{2\pi} \epsilon^{\mu\nu\rho}\partial_\nu \beta_{i\rho}.
\end{align}
At $\mathbf{q}=0$ (i.e., for spatially uniform fields), the spatial components of the current are:
\begin{align}
j_i^m(\omega) = -\frac{i\omega}{2\pi} \epsilon^{mn}\beta_{in}(\omega).
\end{align}

The correlator for the $\beta$ field is similarly defined as $\expval{ \beta_i^m(\omega) \beta_i^n(-\omega) } = [D_{\beta,i}^{-1}(\omega)]^{mn}$. Its inverse kernel, $D_{\beta,i}^{mn}(\omega)$, is therefore related to the inverse current response by
\begin{align}
D_{\beta,i}^{mn}(\omega) = \frac{\omega^2}{(2\pi)^2} \epsilon^{mp} \qty[\Pi_i^{-1}(\omega)]^{pq} \epsilon^{qn}.
\end{align}

If for some $i$, $\mathcal{K}_{ii}\neq0$ (a non-trivial parton), then (omitting spatial indices for brevity, where $I$ is the $2\times 2$ spatial identity matrix and $\epsilon$ is the $2\times 2$ spatial antisymmetric tensor)
\begin{align}
\Pi_i(\omega) = \frac{i\omega}{2\pi}\mathcal{K}_{ii}\,\epsilon + \chi_i\omega^2 I + \dots,
\end{align}
so its inverse is 
\begin{align}
D_{\beta,i}(\omega) = \frac{i\omega}{2\pi}\mathcal{K}_{ii}^{-1}\,\epsilon + \order{\omega^2}.
\end{align}
Since $\mathcal{K}_{ii} \in \qty{\pm 1}$, we have $\mathcal{K}_{ii}^{-1} = \mathcal{K}_{ii}$. Thus, a Chern parton gives a Chern--Simons term for $\beta_i$.

For $\mathcal{K}_{ii}=0$ (a trivial parton), the response and its inverse are 
\begin{align}
\Pi_i(\omega) &= \chi_i\omega^2 I + \dots, \\
D_{\beta,i}(\omega) &= \frac{1}{(2\pi)^2\chi_i} I + \order{\omega^2}.
\end{align}
Thus, a trivial parton gives an $\order{\omega^0}$ term in the $\beta$ inverse kernel, while a topological parton does not.

Therefore the low-energy $\beta$-field Lagrangian is
\begin{align}
\mathcal{L}_{\beta} &= \frac{1}{4\pi} \mathcal{K}_{ij} \beta_i \partial \beta_j + \frac{1}{2\pi} Q_{ij} \beta_i \partial a_j + \frac{1}{2\pi} t_i \beta_i \partial A + \frac{1}{2} \beta_{i\mu} M_{ij} \mathcal{P}_T^{\mu\nu} \beta_{j\nu} + \order{\partial^2}.
\end{align}
Here $\mathcal{P}_T^{\mu\nu} = \eta^{\mu\nu} - \frac{\partial^\mu \partial^\nu}{\partial^2}$ is the transverse projector, where $\eta^{\mu\nu}$ is the Minkowski metric; gauge invariance requires that any mass-like term only applies to the transverse components of the field. We choose to work in the temporal gauge, setting $\beta_{i0} = 0$. In this gauge, when evaluating fields at $\mathbf{q}=0$ (where spatial derivatives vanish), the transverse projector reduces to simply the spatial identity matrix, $\mathcal{P}_T^{mn}=\delta^{mn}$. 

For decoupled partons, the $K$-matrix $\mathcal{K}$ and mass matrix $M$ are diagonal:
\begin{align}
\mathcal{K}_{ij} = \delta_{ij}\mathcal{K}_{ii},
\qquad
M_{ij} = \delta_{ij}
\begin{cases}
0, & \mathcal{K}_{ii}\neq0,\\
\frac{1}{(2\pi)^2\chi_i}, & \mathcal{K}_{ii}=0.
\end{cases}
\end{align}

We now solve the constraint equation for gauge-invariant fields satisfying
\begin{align}
Q^\transp \partial \beta = 0.
\end{align}
Let $P$ be a full-rank matrix whose rows span the neutral subspace $\ker Q^\transp$, so that
\begin{align}
P Q = 0.
\end{align}
To solve the constraint, we define a basis change for $\beta$ into gauge-invariant fields $\varphi$ and gauge-dependent fields $\xi$:
\begin{align}
\beta = P^\transp \varphi + Q \xi.
\end{align}
The gauge-dependent $\xi$ fields couple directly to the emergent gauge fields $a_j$. When $a_j$ are integrated out, they impose the constraint that entirely removes $\xi$ from the low-energy effective action. Locally, after solving this constraint, we are left strictly with the physical degrees of freedom:
\begin{align}
 \beta = P^\transp \varphi .
\end{align}
Substituting into the action gives the physical quadratic Lagrangian
\begin{align}
\mathcal{L}_{\text{phys}} &= \frac{1}{4\pi} K^{(0)}_{ab} \varphi_a \partial \varphi_b + \frac{1}{2\pi} \tau_a \varphi_a \partial A  + \frac{1}{2} \varphi_{a\mu} G_{ab} \mathcal{P}_T^{\mu\nu} \varphi_{b\nu} + \order{\partial^2},
\end{align}
where
\begin{align}
K^{(0)} &= P\mathcal{K} P^\transp,\\
G &= P M P^\transp,\\
\tau &= P t.
\end{align}
The matrix $K^{(0)}$ is only the coefficient of the Chern--Simons-like part of the physical inverse kernel. It is not necessarily the final topological $K$-matrix $K_{\text{e}}$.

We define the correlator of these physical fields as $\langle \varphi_a^m(\omega) \varphi_b^n(-\omega) \rangle = [D_{\text{phys}}^{-1}(\omega)]_{ab}^{mn}$. At $\mathbf{q}=0$, the physical inverse kernel (the matrix operator in the action) is
\begin{align}
D_{\text{phys}}^{mn}(\omega) = G\,\delta^{mn} + \frac{i\omega}{2\pi}K^{(0)}\epsilon^{mn} + \order{\omega^2}.
\end{align}

The physical electromagnetic current is given by
\begin{align}
J^\mu_{\text{em}} = \frac{1}{2\pi} \tau_a \epsilon^{\mu\nu\rho}\partial_\nu\varphi_{a\rho}.
\end{align}
Moving to momentum space at $\vb{q}=0$ (where spatial derivatives vanish, meaning only the temporal derivative $\partial_0 \to -i\omega$ survives), the spatial components of the current $J^m_{\text{em}}$ become $J_{\text{em}}^m(\omega) = -\frac{i\omega}{2\pi} \tau_a \epsilon^{mp} \varphi_{ap}(\omega)$. 

The spatial electromagnetic current correlator is then defined as a $2\times 2$ spatial matrix: $\Pi_{\text{em}}^{mn}(\omega) = \langle J_{\text{em}}^m(\omega) J_{\text{em}}^n(-\omega) \rangle$. To calculate this, we integrate out the physical fields $\varphi_a$ from the action. In momentum space, the effective quadratic action involves the term $\frac{1}{2} \varphi(-\omega) D_{\text{phys}}(\omega) \varphi(\omega)$ and the source coupling $A(-\omega) J_{\text{em}}(\omega)$. By substituting the spatial components of the current $J_{\text{em}}^m(\omega)$ derived above, we perform the Gaussian integral over $\varphi$. This yields the effective action for $A$:
\begin{align}
S_{\text{eff}}[A] = \frac{1}{2} \int_\omega A_m(-\omega) \Pi_{\text{em}}^{mn}(\omega) A_n(\omega),
\end{align}
where inserting the completed square directly gives the correlator in terms of the $\varphi$ field correlator $D_{\text{phys}}^{-1}(\omega)$:
\begin{align}
\Pi_{\text{em}}^{mn}(\omega) = - \frac{\omega^2}{(2\pi)^2} \tau_a \epsilon^{mp} \qty[D_{\text{phys}}^{-1}(\omega)]_{ab}^{pq} \epsilon^{qn} \tau_b .
\end{align}
The finite-frequency conductivity is
\begin{align}
\sigma^{mn}(\omega) = \frac{1}{i\omega} \Pi_{\text{em}}^{mn}(\omega).
\end{align}

If $G$ is invertible, then (omitting spatial indices for brevity, with $\delta$ representing the $2\times 2$ spatial identity matrix $\delta^{mn}$) the contribution of $K$ to $D_{\text{phys}}^{-1}$ can be neglected:
\begin{align}
D_{\text{phys}}^{-1}(\omega) = G^{-1}\delta + \order{\omega},
\end{align}
and therefore the correlator and conductivity are
\begin{align}
\Pi_{\text{em}}(\omega) = \order{\omega^2}, \qquad \sigma(\omega) = \order{\omega}.
\end{align}
This is a trivial insulating response.

If $G=0$ and $K^{(0)}$ is invertible, then 
\begin{align}
D_{\text{phys}}^{-1}(\omega) = -\frac{2\pi}{i\omega} \qty(K^{(0)})^{-1}\epsilon + \order{1},
\end{align}
so we obtain a tensor with full spatial indices restored:
\begin{align}
\Pi_{\text{em}}^{mn}(\omega) = \frac{i\omega}{2\pi} \tau^\transp \qty(K^{(0)})^{-1} \tau\, \epsilon^{mn} + \order{\omega^2},
\end{align}
and hence
\begin{align}
\sigma_{xy} = \frac{1}{2\pi} \tau^\transp \qty(K^{(0)})^{-1} \tau.
\end{align}
In this case, we obtain the regular Hall response of a quantum Hall state properly described by the topological $K$-matrix $K^{(0)}$.

In the general mixed case, the mass matrix $G$ is non-zero but not invertible ($G \neq 0$ but $\abs{G} = 0$), i.e., the system contains a mixture of topological and massive trivial modes. We extract the topological sector by defining
\begin{align}
T_\varphi = \ker G.
\end{align}
Choose a matrix $L$ whose rows form a basis for $T_\varphi$. Write
\begin{align}
\varphi = L^\transp\psi + R^\transp\zeta,
\end{align}
where the rows of $R$ span a complementary subspace on which $G$ is invertible. The $\zeta$ fields are trivial insulating fields. Their kernel has an $\order{\omega^0}$ piece, so integrating them out changes the $\psi$ kernel only at order $\order{\omega^2}$ (with spatial indices omitted for brevity):
\begin{align}
D_{\text{eff}}^\psi(\omega) &= L D_{\text{phys}}L^\transp - L D_{\text{phys}}R^\transp \qty(R D_{\text{phys}}R^\transp)^{-1} R D_{\text{phys}}L^\transp \nonumber\\
&= \frac{i\omega}{2\pi} L K^{(0)}L^\transp\,\epsilon + \order{\omega^2}.
\end{align}
Therefore, the topological $K$-matrix $K_{\text{e}}$ and charge vector $t_{\text{e}}$ are
\begin{align}
K_{\text{e}} &= L P\mathcal{K} P^\transp L^\transp, \\
t_{\text{e}} &= L P t.
\end{align}
Only after restricting to $\ker G$, we acquire the physical $K$-matrix.

If the effective topological $K$-matrix vanishes, the leading term in the effective inverse kernel $D_{\text{eff}}^\psi(\omega)$ for the remaining gapless field $\psi$ is an ordinary Maxwell term of order $\order{\omega^2}$. If this mode couples to the electromagnetic field, substituting this into the physical electromagnetic response yields a $1/i\omega$ pole in the conductivity, indicating a finite Meissner weight, i.e., a superconductor.

\section{Details of the basis transformation and parton construction}
\label{app:matrices}
Here, we provide the full definitions of the matrices used in our construction. First, to map between the $I_8 \oplus \sigma_z \oplus \sigma_z$ fermionic state and $\sigma_x \oplus E_8 \oplus \sigma_z $ state, we use the $\mathrm{SL}(\mathbb{Z})$ transformation that maps between $K_{\text{f}}$ and $K'_{\text{f}}$ as defined in \cref{eq:maptoe8}:
\begin{align}
    W &= \begin{pmatrix}
 1 & 0 & 0 & 0 & 0 & 0 & 0 & 0 & -1 & 0 & 1 & 2 \\
 0 & 1 & 0 & 0 & 0 & 0 & 0 & 0 & 0 & 0 & -1 & 0 \\
 0 & 1 & 0 & 0 & 0 & 0 & 0 & 0 & 1 & -1 & 0 & 1 \\
 0 & 0 & -1 & 1 & 0 & 0 & 0 & 0 & 0 & 0 & 0 & -1 \\
 0 & 0 & 0 & -1 & 1 & 0 & 0 & 0 & 0 & 0 & 0 & -1 \\
 0 & 0 & 0 & 0 & -1 & 1 & 0 & 0 & 0 & 0 & 0 & -1 \\
 0 & 0 & 0 & 0 & 0 & -1 & 1 & 0 & 0 & 0 & 0 & -1 \\
 0 & 0 & 0 & 0 & 0 & 0 & -1 & 1 & 0 & 1 & 0 & -1 \\
 0 & 0 & 0 & 0 & 0 & 0 & 0 & -1 & 1 & 1 & 0 & -1 \\
 1 & -1 & 0 & 0 & 0 & 0 & 0 & 0 & -1 & 0 & 1 & 2 \\
 0 & 0 & 1 & 0 & 0 & 0 & 0 & 0 & 0 & 0 & 0 & -1 \\
 0 & 1 & 0 & 0 & 0 & 0 & 0 & 0 & 0 & -1 & 0 & 3 \\
    \end{pmatrix}
\end{align}

To construct the physical state, we use 12 total partons, defined by matrices $P$ and $Q$:
\begin{align}
    P &= 
\begin{pmatrix}
 1 & -1 & 0 & 0 & 0 & 0 & 0 & 0 & 0 & 0 & 0 & 0 \\
 0 & 1 & -1 & 0 & 0 & 0 & 0 & 0 & 0 & 0 & 0 & 0 \\
 0 & 0 & 1 & -1 & 0 & 0 & 0 & 0 & 0 & 0 & 0 & 0 \\
 0 & 0 & 0 & 1 & -1 & 0 & 0 & 0 & 0 & 1 & 1 & 0 \\
 0 & 0 & 0 & 0 & 1 & -1 & 0 & 0 & 0 & -1 & -1 & 0 \\
 0 & 0 & 0 & 0 & 0 & 1 & 0 & -1 & -1 & 0 & 0 & 1 \\
 0 & 0 & 0 & 0 & 0 & 0 & 1 & 1 & 0 & 0 & 0 & 0 \\
 0 & 0 & 0 & 0 & 0 & 1 & 0 & 0 & 1 & 0 & 0 & 0 \\
\end{pmatrix} \label{appeq:mainexamplep}\\
    Q^\transp &= 
\begin{pmatrix} 
 0 & 0 & 0 & 0 & 1 & 0 & 0 & 0 & 0 & 0 & 1 & 0 \\
 0 & 0 & 0 & 0 & 0 & 0 & -1 & 1 & 0 & 0 & 0 & 1 \\
 1 & 1 & 1 & 1 & 0 & 1 & 0 & 0 & -1 & 0 & -1 & -2 \\
 0 & 0 & 0 & 0 & 0 & 0 & 0 & 0 & 0 & 1 & -1 & 0 \\
\end{pmatrix}. \label{appeq:mainexampleq}
\end{align}

An alternative construction is
\begin{align}
    P&= 
\begin{pmatrix}
 1 & -1 & 0 & 0 & 0 & 0 & 0 & 0 & 0 & 1 & 1 & 0 \\
 0 & 1 & -1 & 0 & 0 & 0 & 0 & 0 & 0 & -1 & -1 & 0 \\
 0 & 0 & 1 & -1 & 0 & 0 & 0 & 0 & 0 & -1 & -1 & 0 \\
 0 & 0 & 0 & 1 & -1 & 0 & 0 & 0 & 0 & 0 & 0 & 0 \\
 0 & 0 & 0 & 0 & 1 & -1 & 0 & 0 & 0 & 0 & 0 & 0 \\
 0 & 0 & 0 & 0 & 0 & 1 & 1 & 0 & -1 &1 & 1 & 1 \\
 0 & 0 & 0 & 0 & 0 & 0 & 0 & -1 & 1 & -1 & -1 & 0 \\
 0 & 0 & 0 & 0 & 0 & 1 & -1 & 0 & 0 & 0 & 0 & 0 \\
\end{pmatrix} \label{appeq:secondexamplep}\\
Q^\transp &=
\begin{pmatrix}
 1 & 1 & 1 & 1 & 1 & 1 & 1 & 0 & 0 & 0 & 0 & -2 \\
 1 & 2 & 1 & 0 & 0 & 0 & 0 & -1 & 0 & 0 & 1 & -1 \\
 1 & 2 & 1 & 0 & 0 & 0 & 0 & 0 & 1 & 0 & 1 & 0 \\
 0 & 0 & 0 & 0 & 0 & 0 & 0 & 0 & 0 & 1 & -1 & 0 \\
\end{pmatrix} \label{appeq:secondexampleq}
\end{align}

\section{Properties of intermediate phases}
\label{app:intermediate}
\subsection{Rank-1 updates}
A change of the single parton $f_i$ Chern number maps the physical $K$-matrix as 
\begin{align}
    K_{\text{e}} = P \mathcal{K} P^\transp \mapsto P (\mathcal{K} + E) P^\transp
\end{align}
where $E$ is a matrix with a single non-zero integer element $E_{ii}$. We can rewrite the new $K_{\text{e}}$ as  $P \mathcal{K}P^\transp  + P E P^\transp$. Since $E$ has rank one, $P E P^\transp$ has rank at most one, and has rank one when the corresponding column of $P$ is nonzero. We now use a variation of the Cauchy interlacing theorem~\cite{golub1973matrixreview}: for a symmetric $n\times n$ matrix $D$ with eigenvalues $\lambda_i$, the eigenvalues $\mu_i$ of $D+\sigma \vb{u}\vb{u}^\transp$ for $\sigma>0$ satisfy the following inequality
\begin{align}
    \lambda_1 \leqslant \mu_1 \leqslant \lambda_2 \leqslant \dots \leqslant \lambda_n \leqslant \mu_n. 
\end{align}
If $\sigma<0$ an equivalent inequality is obtained by changing the roles of $D$ and $D+\sigma \vb{u}\vb{u}^\transp$.

Now consider a concrete case of $K_{E_8}$, which has eight positive eigenvalues. After a rank-1 update arising from the one parton Chern number changing sign, only the smallest eigenvalue can become negative; as such, the signature of the new $K$-matrix differs at most by $2$. After repeating the process four times, we arrive at a matrix with signature $0$; the change must be exactly $2$ at each mass change.

The same argument shows that there cannot be linear dependence between the columns of $P$ corresponding to partons that change sign. We can write the trivial phase $K$-matrix as $K_{E_8} - P \tilde{E} P^\transp$, where $\tilde{E}_{ii} = 2$ for all partons that change sign; and since the difference between the signature of $K_{E_8}$ and the trivial one is $8$, the rank of $P \tilde{E} P^\transp$ is at least 4, meaning all four columns of $P$ corresponding to partons that change sign are linearly independent.

\subsection{\texorpdfstring{$K$}{K}-matrix determinant parity}
We first show that two phases differing by a single parton Chern number sign have the same determinant parity.
Let $A$ and $B$ be two matrices of size $m\times n$ and $n \times m$ correspondingly. Denote by $\binom{[n]}{m}$ the set of all subsets of $\qty{1,\dots,n}$ of size $m$. Now for $S \in \binom{[n]}{m}$ we denote  by $A_{[m], S}$ a matrix with $m$ columns of $A$ at indices from $S$ and by $B_{S, [m]}$ a matrix with $m$ rows of $B$ at indices from $S$. The Cauchy–Binet formula \cite{tao2012topics} then gives the determinant of $AB$ as
\begin{align}
    \det(AB) = \sum_{S\in \binom{[n]}{m}} \det(A_{[m],S})\det(B_{S,[m]}). \label{appeq:cauchybinet}
\end{align}

We apply it to $K_e$ by choosing $A=P$ and $B=\mathcal{K}P^\transp$, with $m=N_{\text{e}}$ and $n=N_{\text{p}}$.
Since $\mathcal{K}$ is diagonal, changing the sign of some diagonal elements of $\mathcal{K}$ changes the signs of the corresponding rows of $B$. As a result, $\det(B_{S,[m]})$ is multiplied by $-1$ if an odd number of rows change sign in $S$, and unchanged otherwise. This means the determinant of the new $K_e$ is the same sum (\ref{appeq:cauchybinet}), but with some elements having a different sign, so the parity of the determinant is unchanged.

\subsection{Minimal degeneracy and different \texorpdfstring{$\kappa_{\mathrm{xy}}$}{κxy}}
Following the previous result, minimal nontrivial degeneracy is at least $3$ for even $\kappa_{\mathrm{xy}}$ and $2$ for odd ones along these parton paths. The even-$\kappa_{\mathrm{xy}}$ phases have odd determinant; the odd-$\kappa_{\mathrm{xy}}$ phases require projection and can have even determinant. We now show that the minimal absolute determinant for $\kappa_{\mathrm{xy}}=4$ under this odd-determinant restriction is 5, and for $\kappa_{\mathrm{xy}}\in\qty{3,5}$ is 4, using the decomposition into prime anyon models \cite{wang2020abelian}.

Specifically, we show that the bosonic phase with ground state degeneracy 3 cannot have $\kappa_{\mathrm{xy}}=4$. We denote the absolute value of the determinant as $\abs{K}$ in what follows.
Since the determinant of the $K$-matrix is multiplicative, the prime decomposition of the state with $\abs{K}=3$ has to have one component with $\abs{K}=3$ and some number of components with $\abs{K}=1$. The latter cannot change $\kappa_{\mathrm{xy}} \mod 8$ and thus can be ignored. 

The only prime anyonic models with $\abs{K}=3$ are $A_3$ or $B_3$, with $K$-matrices being Cartan matrix of $E_6$ ($\kappa_{\mathrm{xy}}=6$) and $K_{B_3}=\begin{pmatrix}
    2&1\\1&2
\end{pmatrix}$ ($\kappa_{\mathrm{xy}}=2$), correspondingly. As a result, $\abs{K}\neq 3$  for $\kappa_{\mathrm{xy}}=4$.

Similarly, only $A_{2^r}$, $B_{2^r}$, $C_{2^r}$, and $D_{2^r}$ can have odd $\kappa_{\mathrm{xy}}$, and for  $\kappa_{\mathrm{xy}}\in\qty{3,5}$ the only candidates are $C_{2^r}$ and $D_{2^r}$ with even $r$, i.e., minimal degeneracy for these values of $\kappa_{\mathrm{xy}}$ are $4$ for $C_4$ and $D_4$, with explicit $K$-matrices given by \citet{wang2020abelian}.

\section{RG analysis of \texorpdfstring{$U(1)^k$}{U(1)ᵏ} Chern--Simons--\texorpdfstring{QED$_{3}$}{QED3}}
\label{app:rg}
In this section, we perform an RG analysis of a $U(1)^k$ Chern--Simons--QED$_3$ with large-$N$ expansion. Specifically, we compute the scaling dimension of the mass term for a generic Chern--Simons coupling and fermionic charges, and then specialize the results to the theory obtained in the main text.
To calculate the scaling dimension, we are interested in the logarithmic divergence of the leading-order (in $1/N$) correction to the mass term.

\subsection{RG analysis of the critical point}

We first remind the reader how the logarithmic divergence of the two-point correlator is related to the scaling dimension of the operator. 
For some operator $\mathcal{O}$ define
\begin{align}
    \mathcal{M}(x) &= \expval{\mathcal{O}(x) \bar{\mathcal{O}}(0)} = \mathcal{M}^{(0)}(x) + \frac{1}{N} \mathcal{M}^{(1)}(x) + \dots
\end{align}
Expanding conformal correlator around $\Delta=\Delta_0$, we get
\begin{align}
    \mathcal{M}(x) = \frac{M_0}{\abs{x}^{2\Delta}} = \frac{M_0}{\abs{x}^{2\qty(\Delta^{(0)} + \frac{1}{N}\Delta^{(1)})}} = \frac{M_0}{\abs{x}^{2 \Delta^{(0)}}}\qty(1-\frac{1}{N}\Delta^{(1)} \log(\abs{x}^2\Lambda^2) + \order{\frac{1}{N^2}}),  \label{appeq:correlator}
\end{align}
where $\Lambda$ is the momentum cutoff.
By calculating the logarithmic divergence of the correlator, we can extract the scaling dimension correction
\begin{align}
    \Delta^{(1)} &= \frac{M_1}{M_0}\\
    \Delta &= \Delta^{(0)} + \Delta^{(1)} \cdot \frac{1}{N} +\order{\frac{1}{N^2}}.
\end{align}  

The logarithmic divergence (\ref{appeq:correlator}) is the same both in momentum and position space. Starting from
\begin{align}
    \int \dd[d]{x} e^{ipx} \abs{x}^{-\alpha} = C(\alpha) \abs{p}^{-(d-\alpha)}
\end{align}
for $0<\alpha<d$, and differentiating this equation with respect to $\alpha$, we get
\begin{align}
    \int \dd[d]{x} e^{ipx} \abs{x}^{-\alpha} \log(\abs{x}) = C(\alpha)\abs{p}^{-(d-\alpha)} \log(1/\abs{p}) + \order{\abs{p}^{-(d-\alpha)} }.
\end{align}

\subsection{Useful identities}
Gamma matrix algebra in 3D:
\begin{align}
    \gamma^\mu\gamma^\eta\gamma^\nu &= \delta^{\mu\eta}\gamma^\nu - \delta^{\mu\nu}\gamma^\eta + \delta^{\eta\nu}\gamma^\mu + i\epsilon^{\mu\eta\nu} \label{eq:3gammas}\\
    \gamma^\mu\gamma^\nu\gamma^\rho\gamma^\sigma &= \delta^{\mu\nu}\delta^{\rho\sigma} - \delta^{\mu\rho}\delta^{\nu\sigma} + \delta^{\mu\sigma}\delta^{\nu\rho} + i\qty[\delta^{\mu\nu}\epsilon^{\rho\sigma\lambda}+\delta^{\rho\sigma}\epsilon^{\mu\nu\lambda}-\delta^{\nu\lambda}\epsilon^{\rho\sigma\mu}+\delta^{\mu\lambda}\epsilon^{\rho\sigma\nu}]\gamma_\lambda. \label{eq:4gammas}
\end{align}
We will use the fact that $\delta^{\mu\nu}\delta^{\mu\nu} = d$ in $d$ dimensions. As we are working in Euclidean space, $\gamma^{\mu\nu} = \delta^{\mu\nu}\gamma_{\mu\nu}$.

Radial integrals
\begin{align}
    \int \frac{\dd[3]{k}}{(2\pi)^3}  \frac{1}{\abs{k}^3} &=  \int_0^\Lambda \frac{\dd{k}}{2\pi^2}  \frac{1}{k} = \frac{1}{4\pi^2}\log(\Lambda^2 x^2)\\
    \int \frac{\dd[3]{k}}{(2\pi)^3}  \frac{k_\mu k_\nu}{\abs{k}^5} &=  \frac{\delta_{\mu\nu}}{3}\int \frac{\dd[3]{k}}{(2\pi)^3}  \frac{k^2}{\abs{k}^5} = \frac{\delta_{\mu\nu}}{12\pi^2}\log(\Lambda^2 x^2),
\end{align}
where in the second integral we used the reflection symmetry to establish that the integral vanishes if $\mu\neq \nu$ and the rotational symmetry to replace $k^\mu k^\nu$ by $\frac{\delta_{\mu\nu}}{3}k^2$.

Momentum integrals
\begin{align}
    \int \frac{\dd[3]{q}}{(2\pi)^3}  \frac{1}{q^2(q-k)^2} &= \frac{1}{8\abs{k}} \label{appeq:momint1}\\
       \int \frac{\dd[3]{q}}{(2\pi)^3}  \frac{q^\mu}{q^2(q-k)^2} &= \frac{k^\mu}{16\abs{k}} \label{appeq:momint2}
\end{align}
(the second can be obtained by contracting it with $k_\mu$ and observing that it equals half the first one when completing a square to $(q-k)^2$). 

For $\abs{k}\gg \abs{p}$ we can expand
\begin{align}
    \frac{1}{(p-k)^2} = \frac{1}{k^2} \cdot \frac{1}{1 - 2\frac{p\vdot k}{k^2} + \qty(\frac{p}{k})^2} = \frac{1}{k^2}\qty[1 + 2\frac{p\vdot k}{k^2} - \qty(\frac{p}{k})^2  + 4\qty(\frac{p\vdot k}{k^2})^2 + \order{\qty(\frac{p}{k})^3}] \label{eq:frac_expansion}
\end{align}

Fermion propagator and vertex for the Lagrangian (\ref{eq:LtwophasesNf})
\begin{align}
    \mathcal{L} &= \sum_{k=1}^{N}\sum_{i=1}^4\sum_{s=\pm}\bar{f}_{i,k,s}\qty[\gamma_\mu (\partial_\mu - i{R}_{ij} a_{j\mu})+m]f_{i,k,s}+\frac{iN}{4\pi}\mathcal{K}^{\text{eff}}_{ij}a^i\partial a^j
\end{align}
are given by (with Fourier factors of $i$ suppressed in the diagram notation)
\begin{align}
G^0(p)=\begin{tikzpicture}[baseline=(x.base)]
\begin{feynman}
    \vertex (x) at (0,0);    
    \vertex (w)  at (3,0);  
    \diagram*{
      (x) -- [fermion, momentum'={$p$}]     (w)
    };
  \end{feynman}
  \end{tikzpicture} &=  \frac{\gamma^\mu p_\mu}{p^2}\\
G^0(x_1,x_2)=\begin{tikzpicture}[baseline=(x.base)]
\begin{feynman}
    \vertex  (x) at (0,0) {$x_1$};    
    \vertex (w)  at (3,0) {$x_2$};  
    \diagram*{
      (x) -- [fermion]     (w) 
    };
  \end{feynman}
  \end{tikzpicture} &=  \frac{\gamma^\mu (x_1-x_2)_\mu}{4\pi \abs{x_1-x_2}^3}\\
\begin{tikzpicture}[baseline=(w.base)]
\begin{feynman}
    \vertex (x) at (-1,-1); 
    \vertex (y) at (-1,1);
    \vertex (c) at (0,0);
    \vertex (w)  at (1,0);  
    \diagram*{
      (x) -- [fermion, edge label'=i]     (c),
      (y) -- [fermion, edge label=i]     (c),
      (c) -- [photon, edge label=j]     (w),
    };
  \end{feynman}
  \end{tikzpicture} &=  R_{ij} \gamma_\mu.
\end{align}

\subsection{Photon propagator}
There are two methods for calculating the photon propagator. We can either calculate the bare photon propagator and use it to evaluate geometric series that include polarization bubbles, or we can integrate out the fermions first and then calculate the photon propagator in the effective theory.

We start from a simpler calculation involving one gauge field, and then generalize it to the case of multiple $U(1)$ gauge fields.
\subsubsection{\texorpdfstring{QED$_3$}{QED3} photon propagator}
As a warm-up, we calculate the propagator for the QED$_3$ \cite{chester2016anomalous}.
The Lagrangian density for photons is
\begin{align}
    \mathcal{L} &= \frac{1}{2}A^\mu  O^{\mathrm{QED}}_{\mu\nu} A^\nu\\
    O^{\mathrm{QED}}_{\mu\nu} &= q^2\delta_{\mu\nu} - \frac{N}{16(\zeta-1)\abs{q}}q_{\mu}q_{\nu},
\end{align}
where $\zeta$ is a gauge-fixing parameter.
The bare photon propagator $D^{\mathrm{QED},0}_{\rho\nu}$ is given by 
\begin{align}
        O^{\mathrm{QED}}_{\mu\rho} D^{\mathrm{QED},0}_{\rho\nu} &= \delta_{\mu\nu}
\end{align}

Since the fermion loop contributions are not suppressed for the photon, we need to sum the geometric series.
The polarization bubble (fermion loop) of massless Dirac fermions is given by
\begin{align}
    \Pi^{\mu \nu}(q) &= N\int \frac{\dd[d]{p}}{(2\pi)^d} \Tr[i\gamma^\mu \frac{ p_\alpha \gamma^\alpha}{p^2} i \gamma^\nu \frac{ (p+q)_\beta \gamma^\beta}{(p+q)^2}]= \frac{N \abs{q}}{16}\qty(\delta^{\mu\nu} - \frac{q^\mu q^\nu}{q^2}).
\end{align}
The full photonic propagator given by the geometric series sum
\begin{align}
    D^{\mathrm{QED}}_{\mu\nu}(p)  &= D_{\mu\nu}^{\mathrm{QED},0}(p) - D_{\mu\alpha}^{\mathrm{QED},0}(p) \Pi^{\alpha \beta}(p) D_{\beta\nu}^{\mathrm{QED},0}(p) + \dots \nonumber\\&= \qty\Big(\qty[D_{\mu\nu}^{\mathrm{QED},0}(p)]^{-1}+\Pi^{\mu \nu}(p)  )^{-1} = \qty\Big(O^{\mathrm{QED}}_{\mu\nu}(p) +\Pi^{\mu \nu}(p)  )^{-1}. \label{appeq:geomser}
\end{align}

The same expression can be obtained by integrating out the fermions. 
In this case, an additional term of the form $\Pi^{\mu \nu}(p)$
appears in the effective Lagrangian for photons, and the photon propagator again acquires the form of \cref{appeq:geomser}.
We guess an ansatz of the form
\begin{align}
    D^{\mathrm{QED}}_{\rho\nu} &= a(q) \delta_{\rho\nu} + b(q) q_\rho q_\nu,
\end{align}
which gives us
\begin{align}
    &\qty(O^{\mathrm{QED}}_{\mu\rho}(q) +\Pi^{\mu \rho}(q)  ) D^{\mathrm{QED}}_{\rho\nu} =  \qty[q^2\delta_{\mu\rho} - \frac{N}{16(\zeta-1)\abs{q}}q_{\mu}q_{\rho} + \frac{N \abs{q}}{16}\qty(\delta^{\mu\rho} - \frac{q^\mu q^\rho}{q^2})]\qty[a(q) \delta_{\rho\nu} + b(q) q_\rho q_\nu] = \nonumber \\&=
    q^2\delta_{\mu\nu} a(q)   - \frac{N}{16(\zeta-1)\abs{q}}q_{\mu}q_{\nu} a(q)   + \frac{N \abs{q}}{16}\qty(\delta^{\mu\nu} - \frac{q^\mu q^\nu}{q^2})a(q)  +  q^2  b(q) q_\mu q_\nu - \frac{N\abs{q}}{16(\zeta-1)}q_{\mu} q_\nu b(q) = \nonumber \\&=
        \qty(q^2 + \frac{N \abs{q}}{16})\delta^{\mu\nu}a(q) +\qty[- \frac{N}{16(\zeta-1)\abs{q}} a(q)    -\frac{N}{16 \abs{q}}  a(q)  +  q^2  b(q)  - \frac{N\abs{q}}{16(\zeta-1)} b(q) ] q_\mu q_\nu = \nonumber \\&=
        \qty(q^2 + \frac{N \abs{q}}{16}) \delta^{\mu\nu}a(q) +\qty[
        - \frac{N}{16\abs{q}} \frac{\zeta}{\zeta-1}   a(q)  +  \qty(q^2   - \frac{N\abs{q}}{16(\zeta-1)}) b(q)  ] q_\mu q_\nu,
\end{align}
meaning $a(q) = \qty(q^2 + \frac{N \abs{q}}{16})^{-1} = \frac{16}{N\abs{q}} + \order{1/N^2}$ and
\begin{align}
  \qty(q^2   - \frac{N\abs{q}}{16(\zeta-1)}) b(q) &=  \frac{N\zeta}{16(\zeta-1)\abs{q}}a(q)   \\
  b(q) &= -\frac{16\zeta}{N\abs{q}^3} + \order{1/(q^2 N^2)}, 
\end{align}
and resulting in a photon propagator
\begin{align}
    D^{\mathrm{QED}}_{\rho\nu} &= \frac{16}{N\abs{q}} \qty(\delta_{\rho\nu} -\zeta \frac{q_\rho q_\nu}{q^2}).
\end{align}
\subsubsection{Chern--Simons--\texorpdfstring{QED$_3$}{QED3}}
We now turn to the Chern--Simons--QED$_3$  with a single $U(1)$ gauge field.
The action is
\begin{align}
   S_E[a, \psi] = \int \dd[3]{x} \qty[ \sum_{i=1}^{N} 
   \bar{\psi}_{i}\gamma_\mu (\partial_\mu - i a_\mu)\psi_{i}
   + \frac{i k_{\text{eff}}}{4\pi} \epsilon_{\mu\nu\rho} a_\mu \partial_\nu a_\rho ] \label{appeq:cs_action}
\end{align}
Here the charge $e=1$. Note that since the Maxwell term is less relevant than the Chern--Simons term, it can be neglected.

After integrating out the fermions, which adds a polarization bubble term to the action and switching to momentum space, we get an effective Lagrangian
\begin{align}
\mathcal{L}[a(q)] &= \mathcal{L}_{\text{fermion}} +  \mathcal{L}_{\text{CS}} + \mathcal{L}_{\text{gauge fixing}}\\
    \mathcal{L}_{\text{fermion}} &= \frac{1}{2} a_\mu(-q) \qty[ \frac{N}{16}\abs{q}\qty(\delta^{\mu\nu} - \frac{q^\mu q^\nu}{q^2}) ] a_\nu(q)\\
    \mathcal{L}_{\text{CS}} &= \frac{k_{\text{eff}}}{4\pi} \epsilon_{\rho\mu\nu} q_\rho a_\mu(-q) a_\nu(q) \label{appeq:cslagrangian}\\
    \mathcal{L}_{\text{gauge fixing}} &= -\frac{N}{32(\zeta-1)}a_\mu(-q) \frac{q_\mu q_\nu} {\abs{q}}a_\nu(q).
\end{align}
We write the quadratic part of the action in the form
\begin{align}
    \mathcal{L} &= \frac{1}{2} a_\mu O_{\mu\nu} a_\nu\\
    O_{\mu\nu} &= \alpha \abs{q} \delta_{\mu\nu} + \beta \epsilon_{\mu\nu\rho} q_\rho + \gamma q_\mu q_\nu\\
    \alpha &= \frac{N}{16}\\
    \beta &= \frac{k_{\text{eff}}}{2\pi}\\
    \gamma &= \frac{N}{\xi \abs{q}}
\end{align}
with $\xi = \frac{16(1-\zeta)}{\zeta}$. 

We introduce projections on transverse and longitudinal directions
\begin{align}
    P^T_{\mu\nu} &= \delta_{\mu\nu} - \frac{q_\mu q_\nu}{q^2}\\
    P^L_{\mu\nu} &= \frac{q_\mu q_\nu}{q^2}
\end{align}
that satisfy $P_T+P_L = \delta_{\mu\nu}$, $P_T^2 = P_T$, $P_L^2=P_L$, and $P_L P_T = 0$.
In addition, we introduce an antisymmetric transversal tensor 
\begin{align}
    S_{\mu\nu} &= \epsilon_{\mu\nu\rho} \frac{q_\rho}{\abs{q}}
\end{align}
satisfying
\begin{align}
    S^2 &=  S_{\mu\sigma} S_{\sigma\nu} = -\qty(\delta_{\mu\nu}\delta_{\rho\alpha} - \delta_{\mu\alpha}\delta_{\nu\rho})\frac{q_\rho q_\alpha}{q^2} = \frac{q_\mu q_\nu}{q^2} - \delta_{\mu\nu} = -P^T\\
    S P_T &= \epsilon_{\mu\sigma\rho} \frac{q_\rho}{\abs{q}} \qty( \delta_{\sigma\nu} - \frac{q_\sigma q_\nu}{q^2}) = \epsilon_{\mu\nu\rho} \frac{q_\rho}{\abs{q}}  = S.
\end{align}

We rewrite $O$ as
\begin{align}
     O_{\mu\nu} &= \alpha \abs{q} \delta_{\mu\nu} + \beta \epsilon_{\mu\nu\rho} q_\rho + \gamma q_\mu q_\nu = \alpha \abs{q} P^T + \beta \abs{q} S + \alpha \abs{q}P^L + \gamma q^2 P^L
\end{align}
and we write the propagator as
\begin{align}
    D_{\mu\nu} &= a(q) P^T + b(q) P^L + c(q) S. \label{appeq:prop_generic}
\end{align}
Note that $a$ and $b$ have a similar meaning as in the previous section, but different numerical values.

Multiplying the two, we get
\begin{align}
      O_{\mu\eta} D_{\eta\nu} &= \qty(\alpha \abs{q} P^T + \beta \abs{q} S + \alpha \abs{q}P^L + \gamma q^2 P^L)\qty(a(q) P^T + b(q) P^L + c(q) S) = \nonumber \\&= \qty(\alpha \abs{q} P^T + \beta \abs{q} S)a(q) P^T + \qty(\alpha \abs{q}P^L + \gamma q^2 P^L) b(q) P^L  + \qty(\alpha \abs{q} P^T + \beta \abs{q} S)c(q) S = \nonumber\\&=
     \qty(\alpha \abs{q} a(q) - \beta \abs{q}c(q)) P^T + \qty(\alpha \abs{q} + \gamma q^2 ) b(q) P^L + \qty(\alpha \abs{q} c(q)  + \beta \abs{q}a(q)) S
\end{align}
which gives us three equations
\begin{align}
    \alpha \abs{q} a(q) - \beta \abs{q}c(q) &= 1\\
    \qty(\alpha \abs{q} + \gamma q^2 ) b(q) &= 1\\
    \alpha \abs{q} c(q)  + \beta \abs{q}a(q) &= 0
\end{align}
with solution
\begin{align}
    b(q) &= \frac{1}{\alpha \abs{q} + \gamma q^2}\\
     a(q) &= -  \frac{\alpha}{\beta}  c(q) \label{appeq:acs}\\
c(q) &= -\frac{\beta}{  \abs{q}  (\alpha^2 + \beta^2 )}, \label{appeq:ccs}
\end{align}
i.e.,
\begin{align}
    D_{\mu\nu} &= -c(q)  \qty(\frac{\alpha}{\beta}    P^T  -  S) + b(q) P^L.
\end{align}
The form of \cref{appeq:acs,appeq:ccs} can be understood analogously to \cref{appeq:geomser}. However, for action (\ref{appeq:cs_action}), the transversal part of the bare propagator is acquired by inverting \cref{appeq:cslagrangian} and as such is antisymmetric, i.e., $D^{0}_{\mu\nu} \propto S$, while $\Pi^{\mu\nu}$ is symmetric; $D_{\mu\nu}$ has both symmetric and antisymmetric components. This means the symmetric part of $D_{\mu\nu}$ must contain an odd number of $D^{0}_{\mu\alpha}\Pi^{\alpha\nu}$ factors, and the antisymmetric part must contain an even number. This amounts to summing geometric series with the term squared relative to one in \cref{appeq:geomser}; as a result, the denominator in \cref{appeq:ccs} contains the $\alpha^2$ and $\beta^2$ terms. 

We define $\lambda = \frac{8k_{\text{eff}} }{\pi N}$, i.e., $k_{\text{eff}} = \frac{\pi N \lambda }{8}$
\begin{align}
    %\alpha &= \frac{N}{16}\\
    \beta &= \frac{N \lambda }{16} = \alpha \lambda %\\
    %\gamma &= \frac{N}{\xi \abs{q}}.
\end{align}
Substituting these, we get
\begin{align}
    b(q) &= \frac{16\xi}{N (\xi  + 16 ) \abs{q}}\\
c(q) &= -\frac{\lambda }{  \abs{q} \alpha  (1 + \lambda ^2 )}
\end{align}
to acquire the propagator
\begin{align}
      D_{\mu\nu} &= \frac{16 }{  \abs{q} N  (1 + \lambda ^2 )} (   P^T  -   \lambda S) +  \frac{16\xi}{N (\xi  + 16 ) \abs{q}} P^L.
\end{align}
A good choice of $\xi$ is such that $\frac{q_\mu q_\nu}{q^2}$ cancels, i.e., $\xi=\frac{16}{\lambda^2}$:
\begin{align}
      D_{\mu\nu} &= \frac{16 }{  \abs{q} N  (1 + \lambda ^2 )} \qty( \delta_{\mu\nu}   -   \lambda \frac{q_\rho}{\abs{q}} \epsilon_{\rho\mu\nu})
\end{align}

\subsubsection{Propagator of \texorpdfstring{$U(1)^k$}{U(1)ᵏ} QED--Chern--Simons}
For more general $U(1)^k$ case we have Lagrangian given by
\begin{align}
    \mathcal{L}_{\text{crit}} &= \frac{1}{2}a_{i \mu}(-q)\qty[  \frac{\abs{q}}{16}
                G_{ij} \delta^{\mu\nu}
            +       \frac{1}{2 \pi} \epsilon^{\mu\nu\rho}q_\rho K^{\text{eff}}_{ij}
            +       \frac{\abs{q}}{16} G_{ij} \frac{\zeta}{1-\zeta}
                             \frac{q^\mu q^\nu}{q^2}
              ] a_{j \nu}(q). 
\end{align}
with $G = \bar{R}^\transp\bar{R}=2N R^\transp R$, where $N$ now counts replicas and $\bar{R}$ is the $8N\times N_{\text{g}}$ matrix acquired from $R$ by repeating each charge row $2N$ times.
For the example in the main text, defined by \cref{appeq:mainexamplep,appeq:mainexampleq}, after integration out we get \cref{eq:Keff,eq:Rcrit}:
\begin{align}
    \mathcal{K}^{\text{eff}} &= 
\begin{pmatrix}
 0 & 0 & 1 & 1 \\
 0 & -1 & 2 & 0 \\
 1 & 2 & 0 & -1 \\
 1 & 0 & -1 & -1 \\
\end{pmatrix} \label{appeq:Keff} \\
R &= \begin{pmatrix} 
0 & -1 & 0 & 0 \\ 
0 & 1 & 0 & 0 \\ 
0 & 0 & -1 & 0 \\ 
0 & 0 & 0 & 1 \end{pmatrix} \label{appeq:Rone}. 
\end{align}
For convenience we redefine $K^{\text{eff}} =  N   \mathcal{K}^{\text{eff}}$.  

We define the matrices
 \begin{align}
     \alpha &= \frac{G}{16}\\  
     \beta &= \frac{K^{\text{eff}}}{2\pi}\\
 \gamma &= \frac{G}{\xi \abs{q}} 
 \end{align}
and repeat the same process with a generic (\ref{appeq:prop_generic}), where $a$,  $b$, $c$ are now matrices too. Specifically, we get the matrix equations
\begin{align}
    \alpha \abs{q} a(q) - \beta \abs{q}c(q) &= 1\\
    \qty(\alpha \abs{q} + \gamma q^2 ) b(q) &= 1\\
    \alpha \abs{q} c(q)  + \beta \abs{q}a(q) &= 0
\end{align}
with solution
\begin{align}
 a(q) &= - \beta^{-1}  \alpha  c(q)  \\
    b(q) &=  \qty(\alpha \abs{q} + \gamma q^2 )^{-1}\\
  c(q) &= -\frac{1}{ \abs{q} }\qty[ \alpha \beta^{-1}  \alpha   + \beta]^{-1}
\end{align}
or equivalently
\begin{align}
a(q)  &= \frac{ 1}{\abs{q}}\qty[\alpha   + \beta  \alpha^{-1}\beta ]^{-1}\\
b(q) &=  \qty(\alpha \abs{q} + \gamma q^2 )^{-1}\\
    c(q)  &=- \alpha^{-1}\beta a(q).
\end{align} 
Note that the expression for $b$ is well-defined only if $G$ is invertible; for singular $G$, we impose transverse gauge ($\zeta=1$) before inversion and omit the longitudinal sector.
The two different expressions for $a$ and $c$ are well-defined if either $K^{\text{eff}}$ or $G$  is invertible, correspondingly. In the case in the main text, $K^{\text{eff}}$ is invertible and thus we use the first version.
Thus
\begin{align}
     D_{\mu\nu} &=  \frac{1}{ \abs{q} }\frac{\pi}{8}\qty(K^{\text{eff}})^{-1} G   \qty[ \frac{\pi }{128} G\qty(K^{\text{eff}})^{-1} G   + \frac{1}{2\pi}K^{\text{eff}}]^{-1} P^T +  \frac{16\xi}{(16+\xi)\abs{q}}G^{-1}  P^L -\frac{1}{ \abs{q} }\qty[ \frac{\pi }{128} G\qty(K^{\text{eff}})^{-1} G   + \frac{1}{2\pi}K^{\text{eff}}]^{-1} S
\end{align}
Substituting $\frac{16\xi}{(16+\xi)} = 16(1-\zeta)$,
this gives us \cref{eq:final_propagator}
\begin{align}
    \tilde{D}_{\mu\nu}(p)  = R D_{\mu\nu}(p) R^\transp &= \frac{\tilde{A}}{\abs{p}} \delta_{\mu\nu} + \tilde{B}\frac{p_\mu p_\nu}{\abs{p}^3}+\tilde{C} \frac{p^\rho}{p^2}\epsilon_{\rho\mu\nu}  \\
    V &= \qty[\frac{\pi}{128} G {K^{\text{eff}}}^{-1}G + \frac{1}{2\pi}K^{\text{eff}}]^{-1} \\ 
       \tilde{A}&=\frac{\pi}{8} R   {K^{\text{eff}}}^{-1}G V  R^{\transp} \\
   \tilde{B}&= 16(1-\zeta) R  G^{-1}R^{\transp} - \tilde{A}\\
     \tilde{C}&= -R  V R^{\transp},
\end{align}
where $\tilde{A}$, $\tilde{B}$, $\tilde{C}$ are all $\order{1/N}$  independent of $p$, and we can make convenient choice of $\zeta=1$ in what follows. As a result, for the diagrams we plan to calculate, each bosonic propagator contributes a factor of $1/N$, and each fermionic loop contributes a factor of $N$, which allows us to calculate correlators order-by-order in $1/N$.

For the example in the main text, using \cref{appeq:Keff,appeq:Rone}  we get
\begin{align}
    \tilde{A} &= \frac{4\pi^2 }{(20  + \pi^2)N} \begin{pmatrix}
        1&-1&0&0\\
        -1&1&0&0\\
        0&0&1&1\\
        0&0&1&1
    \end{pmatrix} \label{appeq:tildea1}\\
    \tilde{C} &= \frac{4\pi   }{(20   + \pi^2)N} \begin{pmatrix}
        2&-2&-4&-4\\
        -2&2&4&4\\
        -4&4&-2&-2\\
        -4&4&-2&-2\\
    \end{pmatrix} \label{appeq:tildec1}
\end{align}
These are $4\times4$ matrices in pair space. In the full species space, all blocks are repeated, since all replicas couple to the same gauge fields.

\subsection{Fermion self-energy}
We start by calculating the self-energy of the fermion.
It is given by a diagram 
\begin{align}
\Sigma(p)&=\begin{tikzpicture}[baseline=(x.base)]
\begin{feynman}
    \vertex (x) at (0,0);     
    \vertex (y)  at (1,0);   
    \vertex (z)  at (2,0);  
    \vertex (w)  at (3,0);  
    \diagram*{
      (y)  -- [fermion] (z),
      (y)  -- [photon,  half left, looseness=1.5]        (z),
    };
  \end{feynman}
  \end{tikzpicture} =  - \int \frac{\dd[3]{k}}{(2\pi)^3}  \gamma^\nu G^0(p-k) \gamma^\mu  \tilde{D}_{\mu\nu}(k).
\end{align}
Substituting the propagators and using \cref{eq:3gammas}, we get
\begin{align}
\Sigma(p)&= -  \int \frac{\dd[3]{k}}{(2\pi)^3} \gamma^\mu \frac{\gamma^\eta (p-k)_\eta}{(p-k)^2} \gamma^\nu  \qty[\frac{\tilde{A}}{\abs{k}} \delta_{\mu\nu} + \tilde{B}\frac{k_\mu k_\nu}{\abs{k}^3}+\tilde{C} \frac{k^\rho}{k^2}\epsilon_{\rho\mu\nu}] =\\&=  -  \int \frac{\dd[3]{k}}{(2\pi)^3}  \frac{(p-k)_\eta}{(p-k)^2} \qty[\delta^{\mu\eta}\gamma^\nu - \delta^{\mu\nu}\gamma^\eta + \delta^{\eta\nu}\gamma^\mu + i\epsilon^{\mu\eta\nu}]
\qty[\frac{\tilde{A}}{\abs{k}} \delta_{\mu\nu} + \tilde{B}\frac{k_\mu k_\nu}{\abs{k}^3}+\tilde{C} \frac{k^\rho}{k^2}\epsilon_{\rho\mu\nu}] =\\&=  -  \int \frac{\dd[3]{k}}{(2\pi)^3} \qty[ \frac{\gamma^\nu  (p-k)^\mu}{(p-k)^2} - \delta^{\mu\nu}  \frac{\gamma^\eta (p-k)_\eta}{(p-k)^2} +  \frac{\gamma^\mu (p-k)^\nu}{(p-k)^2} + i\epsilon^{\mu\eta\nu} \frac{(p-k)_\eta}{(p-k)^2} ]
\qty[\frac{\tilde{A}}{\abs{k}} \delta_{\mu\nu} + \tilde{B}\frac{k_\mu k_\nu}{\abs{k}^3}+\tilde{C} \frac{k^\rho}{k^2}\epsilon_{\rho\mu\nu}]=\\&=  -  \int \frac{\dd[3]{k}}{(2\pi)^3} \frac{1}{(p-k)^2}\qty[ -  \tilde{A}\frac{\gamma^\eta (p-k)_\eta}{\abs{k}}  +\tilde{B}\qty[ 2\frac{(p-k)^\mu k_\mu \gamma^\nu k_\nu }{ \abs{k}^3}  -  \frac{\gamma^\eta (p-k)_\eta}{\abs{k}} ]  -2  \tilde{C}  \frac{(p-k)_\eta k^\eta}{k^2 }]
=\\&= 
-  \int \frac{\dd[3]{k}}{(2\pi)^3} \frac{1}{(p-k)^2}\qty[ -  (\tilde{A}+\tilde{B})\frac{\gamma^\eta p_\eta}{\abs{k}}+ 2\tilde{B} \frac{p^\mu k_\mu \gamma^\nu k_\nu }{ \abs{k}^3}  +  (\tilde{A}-   \tilde{B})\frac{\gamma^\eta  k_\eta}{\abs{k}}  -2 i \tilde{C}  \frac{p_\eta k^\eta}{k^2 }    +2 i \tilde{C} ]
\end{align}
We are interested in logarithmic divergence, i.e., terms proportional to $k^{-3}$. We thus split the integrand by powers of $k$ and keep only logarithmic divergent parts:
\begin{align}
    \mathcal{I}_1 = -  \int \frac{\dd[3]{k}}{(2\pi)^3} \frac{1}{(p-k)^2}\qty[ -  (\tilde{A}+\tilde{B})\frac{\gamma^\eta p_\eta}{\abs{k}}+ 2\tilde{B} \frac{p^\mu k_\mu \gamma^\nu k_\nu }{ \abs{k}^3}-2 i \tilde{C}  \frac{p_\eta k^\eta}{k^2 } ] &=    \int \frac{\dd[3]{k}}{(2\pi)^3} \frac{1}{k^2}\qty[   (\tilde{A}+\tilde{B})\frac{\gamma^\eta p_\eta}{\abs{k}} - 2\tilde{B} \frac{p^\mu k_\mu \gamma^\nu k_\nu }{ \abs{k}^3}]\\
      \mathcal{I}_2 = -  \int \frac{\dd[3]{k}}{(2\pi)^3} \frac{1}{(p-k)^2}\qty[  (\tilde{A}-   \tilde{B})\frac{\gamma^\eta  k_\eta}{\abs{k}} +2 i \tilde{C}   ] &=  - 2  p^\mu \int \frac{\dd[3]{k}}{(2\pi)^3} \frac{ k_\mu}{\abs{k}^5}  (\tilde{A}-   \tilde{B}) \gamma^\eta  k_\eta,      
\end{align}
where we drop numerators with $k$-odd terms.
The first integral gives 
\begin{align}
    \mathcal{I}_1 =  \frac{3\tilde{A}+\tilde{B}}{12\pi^2}   \gamma^\mu p_\mu \log(\Lambda^2 / p^2)   
\end{align}
and the second gives
\begin{align}
    \mathcal{I}_2 =  -      \frac{2\tilde{A} -   2\tilde{B}}{12\pi^2}\gamma^\mu p_\mu \log(\Lambda^2 / p^2)   
\end{align}
giving in total
\begin{align}
   \mathcal{I}_1 +  \mathcal{I}_2 =  \frac{\tilde{A}+3\tilde{B}}{12\pi^2}   \gamma^\mu p_\mu \log(\Lambda^2 / p^2).
\end{align}
% As usual we calculate the full diagram by summing the geometric series to get
As a result, the fermion propagator is given by
\begin{align}
   \begin{tikzpicture}[baseline=(x.base)]
\begin{feynman}
    \vertex (x) at (0,0);     
    \vertex (y)  at (1,0);   
    \vertex (z)  at (2,0);  
    \vertex (w)  at (3,0);  
    \diagram*{
      (x) -- [fermion]      (y)  -- [fermion] (z)
            -- [fermion]    (w),
      (y)  -- [photon,  half left, looseness=1.5]        (z),
    };
  \end{feynman}
  \end{tikzpicture} &= G^0(p) \qty[1-\frac{\Delta^{(1)}_\psi}{N} \log(\Lambda^2 / p^2) + \order{1/N^2}]\\
   \frac{\Delta^{(1)}_\psi}{N} &= \frac{\tilde{A}+3\tilde{B}}{12\pi^2} 
\end{align}
In real space, we get
\begin{align}
   \begin{tikzpicture}[baseline=(x.base)]
\begin{feynman}
    \vertex (x) at (0,0);     
    \vertex (y)  at (1,0);   
    \vertex (z)  at (2,0);  
    \vertex (w)  at (3,0);  
    \diagram*{
      (x) -- [fermion]      (y)  -- [fermion] (z)
            -- [fermion]    (w),
      (y)  -- [photon,  half left, looseness=1.5]        (z),
    };
  \end{feynman}
  \end{tikzpicture}  &= G^0(x) \qty[1-\frac{\Delta^{(1)}_\psi}{N} \log(\Lambda^2 x^2) + \order{1/N^2}] \label{appeq:selfenergy}
\end{align}

\subsection{Mass correction}
Now we will calculate the leading order correction to the scaling dimension of the mass operator in the form of \cref{appeq:correlator}. For that, we are interested in logarithmic divergences of the correlator  $\expval{\mathcal{M}^\alpha(x)\mathcal{M}^\beta(0)}$ proportional to $\log(\Lambda^2 x^2)$, where 
 \begin{align}
     \mathcal{M}^\alpha= \frac{1}{\sqrt{2N}}\sum_{k=1}^N\sum_{s=\pm}\bar f_{\alpha,k,s}f_{\alpha,k,s}
 \end{align} is a mass operator and  $\Lambda$ is the momentum cutoff.

The final values of all the diagrams contributing to the correlator $\expval{\mathcal{M}^\alpha(x)\mathcal{M}^\beta(0)}$ (calculated below) are
\begin{align}
   \begin{tikzpicture}[baseline=(e.base)]
\begin{feynman}
    \draw[postaction={decorate,
decoration={markings,
        mark=at position .0 with {\coordinate (e);},
        mark=at position 0.25 with {\arrowreversed{Latex}},
        mark=at position .5 with {\coordinate (a);},
        mark=at position 0.75 with {\arrowreversed{Latex}}
        }}] (0,0) circle (1); 
    \vertex[right=2cm of a] (d);
    \fill (e) circle (2pt);
    \fill (a) circle (2pt);
\end{feynman}
  \end{tikzpicture}  &= \delta_{\alpha\beta} \Gamma^{(0)}(x) = \frac{2}{4\pi \abs{x}^4} \delta_{\alpha\beta} \label{appeq:mass_o1}\\
   \begin{tikzpicture}[baseline=(e.base)]
\begin{feynman}
    \draw[postaction={decorate,
decoration={markings,
        mark=at position .0 with {\coordinate (e);},
        mark=at position 0.08 with {\arrowreversed{Latex}},
        mark=at position .166 with {\coordinate (b);},
        mark=at position 0.25 with {\arrowreversed{Latex}},
        mark=at position .333 with {\coordinate (c);},
        mark=at position 0.413 with {\arrowreversed{Latex}},
        mark=at position .5 with {\coordinate (a);},
        mark=at position 0.75 with {\arrowreversed{Latex}}
        }}] (0,0) circle (1); 
    \vertex[right=2cm of a] (d);
    \fill (e) circle (2pt);
    \fill (a) circle (2pt);
    \diagram*{
    (c)-- [photon, half right] (b)
    };
\end{feynman}
  \end{tikzpicture}  =
    \begin{tikzpicture}[baseline=(e.base)]
\begin{feynman}
    \draw[postaction={decorate,
decoration={markings,
        mark=at position .0 with {\coordinate (e);},
        mark=at position 0.25 with {\arrowreversed{Latex}},
        mark=at position .5 with {\coordinate (a);},
        mark=at position 0.58 with {\arrowreversed{Latex}},
        mark=at position .666 with {\coordinate (b);},
        mark=at position 0.75 with {\arrowreversed{Latex}},
        mark=at position .833 with {\coordinate (c);},
        mark=at position 0.913 with {\arrowreversed{Latex}},
        }}] (0,0) circle (1); 
    \vertex[right=2cm of a] (d);
    \fill (e) circle (2pt);
    \fill (a) circle (2pt);
    \diagram*{
    (c)-- [photon, half right] (b)
    };
\end{feynman}
  \end{tikzpicture}  &= -\frac{\Delta^{(1)}_\psi}{N} \delta_{\alpha\beta} \Gamma^{(0)}(x) \log(x^2 \Lambda^2 ) \label{appeq:mass_selfenergy}\\
   \begin{tikzpicture}[baseline=(e.base)]
\begin{feynman}
    \draw[postaction={decorate,
decoration={markings,
        mark=at position .0 with {\coordinate (e);},
        mark=at position 0.125 with {\arrowreversed{Latex}},
        mark=at position .25 with {\coordinate (b);},
        mark=at position 0.375 with {\arrowreversed{Latex}},
        mark=at position .5 with {\coordinate (a);},
        mark=at position 0.625 with {\arrowreversed{Latex}},
        mark=at position .75 with {\coordinate (c);},
        mark=at position 0.875 with {\arrowreversed{Latex}},
        }}] (0,0) circle (1); 
    \vertex[right=2cm of a] (d);
    \fill (e) circle (2pt);
    \fill (a) circle (2pt);
    \diagram*{
    (b)-- [photon] (c)
    };
\end{feynman}
  \end{tikzpicture} &=   \qty[\frac{3\tilde{A}+\tilde{B}}{2\pi^2 }]_{\alpha\beta} \delta_{\alpha\beta} \Gamma^{(0)}(x) \log(x^2 \Lambda^2 ) \label{appeq:mass_crossphoton}\\
   \begin{tikzpicture}[baseline=(e.base)]
\begin{feynman}
    \draw[postaction={decorate,
decoration={markings,
        mark=at position .0 with {\coordinate (e);},
        mark=at position 0.0 with {\arrowreversed{Latex}},
        mark=at position .1 with {\coordinate (b);},
        mark=at position 0.25 with {\arrowreversed{Latex}},
        mark=at position .5 with {\coordinate (a);},
        mark=at position 0.75 with {\arrowreversed{Latex}},
        mark=at position .9 with {\coordinate (c);},
        }}] (0,0) circle (1); 
    \draw[postaction={decorate,
decoration={markings,
        mark=at position .0 with {\coordinate (e2);},
        mark=at position 0.25 with {\arrowreversed{Latex}},
        mark=at position .4 with {\coordinate (b2);},
        mark=at position .5 with {\coordinate (a2);},
        mark=at position 0.5 with {\arrowreversed{Latex}},
        mark=at position .6 with {\coordinate (c2);},
        mark=at position 0.75 with {\arrowreversed{Latex}},
        }}] (2.5,0) circle (1); 
    \fill (e2) circle (2pt);
    \fill (a) circle (2pt);
    \diagram*{
    (b)-- [photon] (b2),
    (c2)-- [photon] (c)
    };
\end{feynman}
  \end{tikzpicture} =
   \begin{tikzpicture}[baseline=(e.base)]
\begin{feynman}
    \draw[postaction={decorate,
decoration={markings,
        mark=at position .0 with {\coordinate (e);},
        mark=at position 0.0 with {\arrowreversed{Latex}},
        mark=at position .1 with {\coordinate (b);},
        mark=at position 0.25 with {\arrowreversed{Latex}},
        mark=at position .5 with {\coordinate (a);},
        mark=at position 0.75 with {\arrowreversed{Latex}},
        mark=at position .9 with {\coordinate (c);},
        }}] (0,0) circle (1); 
    \draw[postaction={decorate,
decoration={markings,
        mark=at position .0 with {\coordinate (e2);},
        mark=at position 0.25 with {\arrow{Latex}},
        mark=at position .4 with {\coordinate (b2);},
        mark=at position .5 with {\coordinate (a2);},
        mark=at position 0.5 with {\arrow{Latex}},
        mark=at position .6 with {\coordinate (c2);},
        mark=at position 0.75 with {\arrow{Latex}},
        }}] (2.5,0) circle (1); 
    \fill (e2) circle (2pt);
    \fill (a) circle (2pt);
    \diagram*{
    (b)-- [photon] (b2),
    (c2)-- [photon] (c)
    };
\end{feynman}
 \end{tikzpicture}  &= -2N\qty[\frac{ \tilde{A}\odot\tilde{A}-\tilde{C}\odot\tilde{C}}{4\pi^2}]_{\alpha\beta}  \Gamma^{(0)}(x) \log(x^2 \Lambda^2 ) \label{appeq:mass_twoloop}
\end{align}
The dot denotes the insertion of the bilinear. Diagrams (\ref{appeq:mass_selfenergy}, \ref{appeq:mass_crossphoton}) represent the attraction between a particle and an anti-particle, and thus are expected to reduce the scaling dimension of the mass term \cite{hermele2005algebraic,hermele2005algebraicerratum}, while the diagram (\ref{appeq:mass_twoloop}) involves decay into two photons. 

We summarize the calculation of these diagrams below.

The first diagram (\ref{appeq:mass_o1}) gives $\order{1}$ value of the vertex and is given by
\begin{align}
  \expval{\mathcal{M}^\alpha(x)\mathcal{M}^\beta(0)} &= \expval{\bar{\psi}_i(x) m^\alpha_{ij} \psi_j(x) \bar{\psi}_{i'}(0) m^\beta_{i'j'} \psi_{j'}(0)} = G^0_{ij'}(x)m^\alpha_{ij}m^\beta_{i'j'}G^0_{i'j}(-x) =\\&=  \Tr(m^\alpha m^\beta) G^0(x)G^0(-x) =\frac{1}{8\pi^2\abs{x}^4} \Tr(m^\alpha m^\beta), 
\end{align} 
where we used explicit indices on fermion propagators for clarity. The trace is given by $\delta_{\alpha\beta}$ in our case. We denote 
\begin{align}
    \Gamma^{(0)}(x) =G^0(x)G^0(-x) =\frac{1}{8\pi^2\abs{x}^4}.
\end{align}

We now calculate the $\order{1/N}$ diagrams.
The second two diagrams (\ref{appeq:mass_selfenergy})  are symmetric; they are acquired by replacing one of the fermion propagators in \cref{appeq:mass_o1} by the self-energy one (\ref{appeq:selfenergy}).

The third diagram (\ref{appeq:mass_crossphoton}) is given by
\begin{align}
    G^0_{i}(y-x)G^0_{i}(0-y)m^\alpha_{ij}G^0_{j}(w-0)G^0_{j}(x-w) m^\beta_{ji} \tilde{D}_{ij}(w-y)
\end{align}
The log divergence occurs when $y$ and $w$ are close to $0$ or $x$. We calculate one of the cases and multiply the result by 2.
In this case we can amputate $G^0(-x)G^0(x)$ to get
\begin{align}
     \Gamma^{(0)}(x)  G^0_{i}(-y)m^\alpha_{ij}G^0_{j}(w) m^\beta_{ji} \tilde{D}_{ij}(w-y)
    \end{align}
 We first ignore the indices $i$, $j$, and calculate matrix elements for arbitrary fermion indices. After the Fourier transform, we need to calculate the integral
\begin{align}
      \Gamma^{(1)}_c(x) &\equiv \Gamma^{(0)}(x) \int \frac{\dd[3]{k}}{(2\pi)^3}  \gamma^{\mu} G^0(k)G^0(-k) \gamma^\sigma \tilde{D}_{\mu\sigma}(k)
    =\\&=
   - \Gamma^{(0)}(x)\int \frac{\dd[3]{k}}{(2\pi)^3} \gamma^{\mu} \gamma^\nu \frac{k_\nu}{k^2} \gamma^\rho \frac{k_\rho }{k^2}\gamma^\sigma \qty[\frac{\tilde{A}}{\abs{k}} \delta_{\mu\sigma} + \tilde{B}\frac{k_\mu k_\sigma}{\abs{k}^3}+\tilde{C} \frac{k^\eta}{k^2}\epsilon_{\eta\mu\sigma}] 
\end{align}
We substitute \cref{eq:4gammas}:
\begin{align}
    \Gamma^{(1)}_c(x) &=
    -\Gamma^{(0)}(x)\int \frac{\dd[3]{k}}{(2\pi)^3}  \frac{k_\nu k_\rho }{k^4}\qty[\delta^{\mu\nu}\delta^{\rho\sigma} - \delta^{\mu\rho}\delta^{\nu\sigma} + \delta^{\mu\sigma}\delta^{\nu\rho} + i\qty[\delta^{\mu\nu}\epsilon^{\rho\sigma\lambda}+\delta^{\rho\sigma}\epsilon^{\mu\nu\lambda}-\delta^{\nu\lambda}\epsilon^{\rho\sigma\mu}+\delta^{\mu\lambda}\epsilon^{\rho\sigma\nu}]\gamma_\lambda]  \times \nonumber \\& \quad \quad \times  \qty[\frac{\tilde{A}}{\abs{k}} \delta_{\mu\sigma} + \tilde{B}\frac{k_\mu k_\sigma}{\abs{k}^3}+\tilde{C} \frac{k^\eta}{k^2}\epsilon_{\eta\mu\sigma}] =\\&= 
   -\Gamma^{(0)}_m(x)  \int \frac{\dd[3]{k}}{(2\pi)^3} \frac{\tilde{A}}{\abs{k}}  \frac{k_\nu k_\rho }{k^4} \qty[ 3\delta^{\nu\rho}   + i\epsilon^{\rho\nu\lambda} \gamma_\lambda]   +\tilde{B} \frac{k_\nu k_\rho }{k^4} \qty[  \frac{1}{\abs{k}}\delta^{\nu\rho} + i\frac{k_\mu }{\abs{k}^3}\qty[k^\nu \epsilon^{\rho\mu\lambda}+k^\rho \epsilon^{\mu\nu\lambda}+k^\lambda \epsilon^{\rho\mu\nu}]\gamma_\lambda]+ \nonumber \\&\quad \quad +\tilde{C}  \frac{k_\nu k_\rho }{k^4}  \frac{k^\eta}{k^2}\qty[\epsilon^{\nu\rho}_{\eta}- \epsilon^{\rho\nu}_{\eta}+ i\qty[
    2\delta^{\rho\nu}\delta^{\lambda}_{\eta}-\delta^{\lambda\nu}\delta^{\rho}_{\eta}]\gamma_\lambda] =\\&=
   -\Gamma^{(0)}_m(x)  \int \frac{\dd[3]{k}}{(2\pi)^3} \frac{3\tilde{A}+\tilde{B}}{\abs{k}^3}  
\end{align}
which gives 
\begin{align}
        \Gamma^{(1)}_c(x) &= (3\tilde{A}+\tilde{B})\Gamma^{(0)}(x) \int \frac{\dd[3]{k}}{(2\pi)^3}   \frac{1}{\abs{k}^3} = \frac{3\tilde{A}+\tilde{B}}{4\pi^2} \Gamma^{(0)}(x)  \log(\Lambda^2 x^2).
\end{align}
We now reinstate the fermionic indices.
We are interested in singlet mass term $\mathcal{M}_i(x)=\frac{1}{\sqrt{2N}}\sum_{k=1}^N\sum_{s=\pm}\bar f_{i,k,s}f_{i,k,s}$.  The product $m^\alpha_{ij} m^{\beta}_{ji}$ is zero if $\alpha\neq \beta$. Thus, the diagram is given by the diagonal part of $\Gamma^{(1)}_{c}$.

The sum of the first three diagrams is thus 
\begin{align}
  \qty[ -2\frac{\tilde{A} + 3\tilde{B}}{12 \pi^2} +  \frac{3\tilde{A} + \tilde{B}}{2 \pi^2} ] \delta_{\alpha \beta} \log(\Lambda^2 / p^2) =  \frac{4\tilde{A}}{3 \pi^2}  \delta_{\alpha \beta}\log(\Lambda^2 / p^2).
\end{align}
Note that the contribution of the $\tilde{B}$ term cancels out as expected since it contains a gauge-dependent parameter.

The last two diagrams (\ref{appeq:mass_twoloop}) are
\begin{align}
    G^0_i(w-x)  G^0_i(y-w)  G^0_i(x-y)  G^0_j(0-z)  G^0_j(u-0)  G^0_j(z-u) m^\alpha_{ii}  \tilde{D}_{ij}(z-w) m^\beta_{jj}   \tilde{D}_{ji}(y-u)
\end{align}
Taking again $u,w,y,z$ close to zero, factoring out $\Gamma^{(0)}$ we need to calculate the matrix 
\begin{align}
   \Gamma^{(1)}_2(x) = \Gamma^{(0)}(x)  \iint \dd{y}\dd{w}\dd{z}\dd{u}G^0_i(y-w)   G^0_j(-z)  G^0_j(u)  G^0_j(z-u)  \tilde{D}_{ij}(z-w) \tilde{D}_{ji}(y-u)
\end{align}
and after Fourier transform 
\begin{align}
    \Gamma^{(1)}_2(x) = \Gamma^{(0)}(x) \iint \dd{k}\dd{q} G^0_i(-k)   G^0_j(q)  G^0_j(-q)  G^0_j(q-k)  \tilde{D}_{ij}(k) \tilde{D}_{ji}(-k)
\end{align}
Each gauge vertex preserves the fermion species, so both gauge lines connect the same fixed pair labels $\alpha,\beta$. Their flavor factors therefore form a Hadamard product, $(X\odot Y)_{\alpha\beta}=X_{\alpha\beta}Y_{\alpha\beta}$. Dropping fermionic indices for convenience, we write
\begin{align}
       \Gamma^{(1)}_2(x)  &=  \Gamma^{(0)}(x) \iint \frac{\dd[3]{k}}{(2\pi)^3}  \frac{\dd[3]{q}}{(2\pi)^3}  \gamma^\mu G^0(-k) \big(\tilde{D}_{\mu\rho}(k)\odot\tilde{D}_{\nu\sigma}(-k)\big)\gamma^\nu  \gamma^\rho   G^0(q-k) \gamma^\sigma G^0(q) G^0(-q).
\end{align}
We define 
\begin{align}
    \Gamma^{(1)}_2(x) &=  \Gamma^{(0)}(x) \int \frac{\dd[3]{k}}{(2\pi)^3}   \gamma^\mu G^0(-k) \big(\tilde{D}^\transp_{\mu\rho}(k)\odot\tilde{D}_{\nu\sigma}(-k)\big)\gamma^\nu   \Pi^{\rho\sigma}(k) \label{appeq:gamma2oneint}
\end{align}
where $ \Pi^{\rho\sigma}(k) $, calculated below, is given by
\begin{align}
    \Pi^{\rho\sigma}(k) &= \int  \frac{\dd[3]{q}}{(2\pi)^3} \Tr[G^0(q)  \gamma^\rho   G^0(q-k) \gamma^\sigma G^0(-q)] = -\int  \frac{\dd[3]{q}}{(2\pi)^3}  \Tr[\frac{ \gamma^\beta q_\beta \gamma^\rho  \gamma^\alpha (q-k)_\alpha \gamma^\sigma \gamma^\delta q_\delta}{q^4 (q-k)^2}] = \nonumber\\&=  i\epsilon^{\rho \sigma \alpha } \frac{k_\alpha}{8\abs{k}}
\end{align}
To calculate
\begin{align}
    \int  \frac{\dd[3]{q}}{(2\pi)^3}  \Tr[\frac{ \gamma^\beta  \gamma^\rho  \gamma^\alpha \gamma^\sigma \gamma^\delta q_\beta(q-k)_\alpha  q_\delta}{q^4 (q-k)^2}] 
\end{align}
we use  
\begin{align}
    \gamma^\beta \gamma^\rho = \delta^{\beta\rho} + i \epsilon^{\beta\rho\kappa} \gamma_\kappa
\end{align}
and write
\begin{align}
    \Tr[\gamma^\beta  \gamma^\rho  \gamma^\alpha \gamma^\sigma \gamma^\delta] =  \delta^{\beta\rho} \Tr[  \gamma^\alpha \gamma^\sigma \gamma^\delta]+ i \epsilon^{\beta\rho\kappa} \Tr[\gamma_\kappa  \gamma^\alpha \gamma^\sigma \gamma^\delta]
\end{align}
Using
\begin{align}
    \Tr[\gamma^\mu \gamma^\nu \gamma^\rho] &= 2i \epsilon^{\mu\nu\rho}
\end{align}
we get
\begin{align}
    \Tr[\gamma^\beta  \gamma^\rho  \gamma^\alpha \gamma^\sigma \gamma^\delta] =  2i\delta^{\beta\rho}\epsilon^{\alpha\sigma\delta}+ i \epsilon^{\beta\rho\kappa} \Tr[\gamma_\kappa  \gamma^\alpha \gamma^\sigma \gamma^\delta]
\end{align}
Repeating the two-trace exercise
\begin{align}
    \Tr[\gamma_\kappa  \gamma^\alpha \gamma^\sigma \gamma^\delta] = \delta_{\kappa}^\alpha \Tr[\gamma^\sigma \gamma^\delta] + i{\epsilon_{\kappa}^{\alpha}}_{\eta} \Tr[\gamma^\eta \gamma^\sigma \gamma^\delta]
\end{align}
Substituting also
\begin{align}
    \Tr[\gamma^\sigma \gamma^\delta] &= 2\delta^{\sigma \delta}
\end{align}
we get
\begin{align}
    \Tr[\gamma_\kappa  \gamma^\alpha \gamma^\sigma \gamma^\delta] &= 2\delta_{\kappa}^\alpha \delta^{ \sigma \delta} -2{\epsilon_{\kappa}^{\alpha}}_{\eta} \epsilon^{   \sigma  \delta\eta} = 2\delta_{\kappa}^\alpha \delta^{ \sigma \delta} -2\delta_{\kappa}^{\sigma  }\delta^{\alpha \delta} + 2\delta_{\kappa}^{ \delta }\delta^{\alpha \sigma}
\end{align}
contracting we get
\begin{align}
    \epsilon^{\beta\rho\kappa} \Tr[\gamma_\kappa  \gamma^\alpha \gamma^\sigma \gamma^\delta] &= 2    \epsilon^{\beta\rho\alpha} \delta^{ \sigma \delta} -2    \epsilon^{\beta\rho\sigma} \delta^{\alpha \delta} + 2    \epsilon^{\beta\rho\delta}\delta^{\alpha \sigma}
\end{align}
which gives us
\begin{align}
     \Tr[\gamma^\beta  \gamma^\rho  \gamma^\alpha \gamma^\sigma \gamma^\delta] =  2i \qty[ \delta^{\beta\rho}\epsilon^{\alpha\sigma\delta} + \epsilon^{\beta\rho\alpha} \delta^{ \sigma \delta} -    \epsilon^{\beta\rho\sigma} \delta^{\alpha \delta} +   \epsilon^{\beta\rho\delta}\delta^{\alpha \sigma}]
\end{align}
Finally
\begin{align}
   2i &\int  \frac{\dd[3]{q}}{(2\pi)^3}  \frac{ q_\beta  (q-k)_\alpha  q_\delta}{q^4 (q-k)^2}  \qty[ \delta^{\beta\rho}\epsilon^{\alpha\sigma\delta} + \epsilon^{\beta\rho\alpha} \delta^{ \sigma \delta} -    \epsilon^{\beta\rho\sigma} \delta^{\alpha \delta} +   \epsilon^{\beta\rho\delta}\delta^{\alpha \sigma}] = \nonumber \\=-2i &\int  \frac{\dd[3]{q}}{(2\pi)^3}    \qty[ \epsilon^{\alpha\sigma\beta} \frac{ q^\rho  k_\alpha  q_\beta}{q^4 (q-k)^2} + \epsilon^{\beta\rho\alpha} \frac{ q_\beta  k_\alpha  q^\sigma}{q^4 (q-k)^2} +    \epsilon^{\beta\rho\sigma}  \frac{ q_\beta  }{q^2 (q-k)^2} -     \epsilon^{\beta\rho\sigma}  \frac{ q_\beta  k^\alpha  q_\alpha}{q^4 (q-k)^2} ]
\end{align}
Using rotational invariance, we get
\begin{align}
    \Pi^{\rho\sigma} &=-2i \epsilon^{\alpha\sigma\rho}  \int  \frac{\dd[3]{q}}{(2\pi)^3}     \frac{     k_\alpha -q_\alpha  }{q^2 (q-k)^2}   
\end{align}
We now use the integrals \cref{appeq:momint1,appeq:momint2} to get
\begin{align}
    \Pi^{\rho\sigma}(k) &= i \epsilon^{\rho\sigma\alpha}    \frac{k^\alpha}{8\abs{k}} \label{appeq:pi_result}
\end{align}
and thus, substituting $ \Pi^{\rho\sigma}(k)$ into \cref{appeq:gamma2oneint}
\begin{align}
    \Gamma^{(1)}_2(x) &=  \Gamma^{(0)}(x) \int \frac{\dd[3]{k}}{(2\pi)^3}   \gamma^\mu G^0(-k) \big(\tilde{D}^\transp_{\mu\rho}(k)\odot\tilde{D}_{\nu\sigma}(-k)\big)\gamma^\nu   \Pi^{\rho\sigma}  =\\&=  -i  \Gamma^{(0)}(x) \int \frac{\dd[3]{k}}{(2\pi)^3}   \gamma^\mu \frac{\gamma^\lambda k_\lambda}{k^2} \qty[\frac{\tilde{A}}{\abs{k}} \delta_{\mu\rho} + \tilde{B}\frac{k_\mu k_\rho}{\abs{k}^3}+\tilde{C} \frac{k^\alpha}{k^2}\epsilon_{\alpha\mu\rho}]^\transp \odot\qty[  \frac{\tilde{A}}{\abs{k}} \delta_{\nu\sigma} + \tilde{B}\frac{k_\nu k_\sigma}{\abs{k}^3}-\tilde{C} \frac{k^\beta}{k^2}\epsilon_{\beta\nu\sigma} ] \gamma^\nu\epsilon^{\rho \sigma \eta } \frac{k_\eta}{8\abs{k}}
    =\\&= -i  \Gamma^{(0)}(x) \int \frac{\dd[3]{k}}{(2\pi)^3}   \gamma^\mu \frac{\gamma^\lambda k_\lambda}{k^2} \qty[\frac{\tilde{A}}{\abs{k}} \delta_{\mu\rho} + \tilde{B}\frac{k_\mu k_\rho}{\abs{k}^3}+\tilde{C} \frac{k^\alpha}{k^2}\epsilon_{\alpha\mu\rho}]^\transp \odot\qty[  -\frac{\tilde{A}}{\abs{k}}\epsilon_{\nu}^{\rho  \eta } \frac{k_\eta}{8\abs{k}}+\tilde{C} \qty(\frac{k^\rho}{k^2}\frac{k_\nu}{8\abs{k}}-\frac{1}{8\abs{k}}\delta_{\nu}^{\rho }) ]\gamma^\nu 
    =\\&=  -i  \Gamma^{(0)}(x) \int \frac{\dd[3]{k}}{(2\pi)^3}   \gamma^\mu \frac{\gamma^\lambda k_\lambda}{k^2} \Big\{ -\qty[\frac{\tilde{A}\odot\tilde{A} k_\eta}{8\abs{k}^3 }  \epsilon_{\nu\mu}^{  \eta }  -\frac{\tilde{A}\odot\tilde{C}}{8k^2}\qty(\frac{k_\mu k_\nu}{k^2} - \delta_{\mu\nu}) ]\nonumber\\&\qquad{}+\frac{1}{8\abs{k}}\tilde{C}\odot\qty[\frac{\tilde{A}}{\abs{k}} \qty(\frac{k_\mu k_\nu}{k^2}-\delta_{\mu\nu})-\tilde{C} \frac{k^\alpha}{k^2}\epsilon_{\nu\alpha\mu}] \Big\}\gamma^\nu =\\&=
      i  \Gamma^{(0)}(x) \int \frac{\dd[3]{k}}{(2\pi)^3}   \gamma^\mu \frac{\gamma^\lambda k_\lambda}{k^2}  \qty[\frac{(\tilde{A}\odot\tilde{A}-\tilde{C}\odot\tilde{C}) k^\alpha}{8\abs{k}^3 }  \epsilon_{\nu\mu \alpha } - \frac{\tilde{C}\odot\tilde{A} + \tilde{A}\odot\tilde{C}}{8k^2}\qty(\frac{k_\mu k_\nu}{k^2}-\delta_{\mu\nu})]\gamma^\nu
\end{align} 
The second term is $k$-odd, so we can drop it.
We now substitute \cref{eq:3gammas}:
\begin{align}
   \gamma^\mu \gamma^\lambda \gamma^\nu \epsilon_{\nu\mu \alpha } = \qty[\delta^{\mu\lambda}\gamma^\nu  + \delta^{\lambda\nu}\gamma^\mu ]\epsilon_{\nu\mu \alpha }+ 2i\delta^{\lambda}_{\alpha }
\end{align}
giving us (the first two terms cancel) 
\begin{align}
    \Gamma^{(1)}_2(x) &=  - \Gamma^{(0)}(x)  \frac{\tilde{A}\odot\tilde{A}-\tilde{C}\odot\tilde{C}}{4} \int \frac{\dd[3]{k}}{(2\pi)^3}  \frac{  k_\alpha}{k^2}  \frac{ k^\alpha}{\abs{k}^3 }  =  -  \Gamma^{(0)}(x) \frac{\tilde{A}\odot\tilde{A}-\tilde{C}\odot\tilde{C}}{4} \int \frac{\dd[3]{k}}{(2\pi)^3}   \frac{ 1}{\abs{k}^3}  
\end{align} 
resulting in
\begin{align}
      \Gamma^{(1)}_2(x) &= - \Gamma^{(0)}(x)  \frac{\tilde{A}\odot\tilde{A}-\tilde{C}\odot\tilde{C}}{4} \int \frac{\dd[3]{k}}{(2\pi)^3}   \frac{ 1}{ \abs{k}^3}  = - \Gamma^{(0)}(x)  \frac{\tilde{A}\odot\tilde{A}-\tilde{C}\odot\tilde{C}}{16\pi^2} \log(\Lambda^2 / p^2)
\end{align}
For $N>1$, we can use \cref{appeq:tildea1,appeq:tildec1} to calculate the value and then multiply it by the number of species, $N$.

Finally, the second diagram in \cref{appeq:mass_twoloop} is identical to the first one, but with $q+k$ instead of $q-k$ in the fermion propagator. As a result, in \cref{appeq:pi_result} we get $ i \epsilon^{\alpha\sigma\rho}    \frac{3 k^\alpha}{8\abs{k}}$ for a total multiplicative factor of $4$.

In total, we get
\begin{align}
     \frac{\Gamma^{(1)}_{\alpha\beta}(x)}{\Gamma^{(0)}(x)} = \qty[\frac{4\tilde{A}_{\alpha\beta}}{3 \pi^2}\delta_{\alpha\beta}  -  2N \qty[\frac{ \tilde{A}\odot\tilde{A}-\tilde{C}\odot\tilde{C}}{4\pi^2}]_{\alpha\beta}] \log(\Lambda^2 x^2),\label{appeq:final_log_mass}
\end{align}
as claimed in \cref{eq:finalm}.

\subsection{Scaling dimensions}
To calculate the scaling dimensions, we need the eigenvalues of the negative logarithmic coefficient in \cref{appeq:final_log_mass}. We also include the calculations of the adjoint mass term, to verify our results against the literature \cite{chester2016anomalous,lee2018emergent}. For the adjoint mass term, the diagrams (\ref{appeq:mass_twoloop}) vanish. 
\subsubsection{\texorpdfstring{QED$_3$}{QED3}}
For QED$_3$, the $   \Gamma^{(1)}(x)$ is a scalar with $a = \frac{16}{N}$ and $c=0$  and thus we get
\begin{align}
    \Delta_{\text{scalar}}^{(1)} = -\frac{16a - 3a^2 N + 3c^2 N}{12 \pi^2} = \frac{128}{3N \pi^2}\\
    \Delta_{\text{adj}}^{(1)} = -\frac{16a  }{12 \pi^2} = -\frac{64}{3N \pi^2}
\end{align}
\subsubsection{Chern--Simons--\texorpdfstring{QED$_3$}{QED3}}
For Chern--Simons--QED$_3$, we substitute the scalars $a=\frac{16}{N(1+\lambda^2)}$ and $c= - a \lambda$

\begin{align}
    \Delta_{\text{scalar}}^{(1)} = -\frac{16a - 3a^2 N + 3c^2 N}{12 \pi^2} &= -\frac{64  }{3 N(1+\lambda^2)\pi^2}+\frac{ 64  (1-\lambda^2)}{  N(1+\lambda^2)^2 \pi^2}\\
    \Delta_{\text{adj}}^{(1)} &= -\frac{64  }{3 N(1+\lambda^2)\pi^2}
\end{align}

\subsubsection{\texorpdfstring{$U(1)^k$}{U(1)ᵏ} Chern--Simons--\texorpdfstring{QED$_3$}{QED3}}
In this case, $\Delta$ depends on flavor, so we calculate the scaling dimensions by taking the eigenvalues of the matrix in \cref{eq:finalm}.  
For the example used in the main text, defined by \cref{appeq:mainexamplep,appeq:mainexampleq}, we substitute the values from \cref{appeq:tildea1,appeq:tildec1} and acquire the corrections (with $N$ again counting replicas)\begin{align}
     \Delta^{(1)}_{1,2} &= -\frac{16}{3N\qty(20+\pi^2)} \approx -\frac{0.178554}{N},\\
     \Delta^{(1)}_3 &= \frac{32\qty(\pi^2-40)}{3N\qty(20+\pi^2)^2} \approx -\frac{0.360226}{N},\\
     \Delta^{(1)}_4 &= \frac{32\qty(\pi^2+8)}{3N\qty(20+\pi^2)^2} \approx \frac{0.213641}{N}.
 \end{align}
The four pair-difference masses have no two-loop contribution and share $\Delta^{(1)}_{1,2}$. Thus all eight dimensions $\Delta=2+\Delta^{(1)}+\order{N^{-2}}$ remain below $3$ at this order, although one correction is positive.

For the second example, given by \cref{appeq:secondexamplep,appeq:secondexampleq}, the pair-singlet corrections are $\qty{-0.581655,-0.378594,-0.142490,+0.188693}/N$, and the pair-difference corrections are $\qty{-0.284941,-0.256230,-0.143549,-0.316785}/N$. These likewise leave all eight mass perturbations relevant at this order.

\end{document}